%% file: ms.tex
\documentclass[10pt,preprint2]{aastex}
\usepackage{amsmath}

\def\p/{\mbox{$^1$}}
\def\pp/{\mbox{$^2$}}
\def\ppp/{\mbox{$^3$}}
\def\pppp/{\mbox{$^4$}}
\def\m/{\mbox{$^{-1}$}}
\def\mm/{\mbox{$^{-2}$}}
\def\mmm/{\mbox{$^{-3}$}}
\def\mmmm/{\mbox{$^{-4}$}}
\def\Ms/{\mbox{M$_\odot$}}

\newcommand{\HeII}[1]{\mbox{He\,{\sc ii}~$\lambda${#1}}}
\newcommand{\HeI}[1]{\mbox{He\,{\sc i}~$\lambda${#1}}}

\shorttitle{The Galactic O-Star Spectroscopic Survey. I}
\shortauthors{Sota et al.}

\slugcomment{Submitted for publication to the Astrophysical Journal Supplement Series}
\begin{document}

\title{The Galactic O-Star Spectroscopic Survey. I. Classification System and Bright Northern Stars in the Blue-Violet at 
R$\sim$2500\altaffilmark{1}}

\author{A. Sota\altaffilmark{2}}
\author{J. Ma\'{\i}z Apell\'aniz\altaffilmark{2,3,4,5,6}}
\affil{Instituto de Astrof\'{\i}sica de Andaluc\'{\i}a-CSIC, Glorieta de la Astronom\'{\i}a s/n, 18008 Granada, Spain}
%\email{sota@iaa.es, jmaiz@iaa.es}
\author{N. R. Walborn}
\affil{Space Telescope Science Institute, 3700 San Martin Drive, Baltimore, MD 21218, USA}
%\email{walborn@stsci.edu}
\author{E. J. Alfaro}
\affil{Instituto de Astrof\'{\i}sica de Andaluc\'{\i}a-CSIC, Glorieta de la Astronom\'{\i}a s/n, 18008 Granada, Spain}
%\email{emilio@iaa.es}
\author{R. H. Barb\'a\altaffilmark{4}}
\affil{Instituto de Ciencias Astron\'omicas, de la Tierra y del Espacio, Casilla 467, 5400 San Juan, Argentina}
\affil{Departamento de F\'{\i}sica, Universidad de La Serena, Av. Cisternas 1200 Norte, La Serena, Chile}
%\email{rbarba@dfuls.cl}
\author{N. I. Morrell}
\affil{Las Campanas Observatory, Observatories of the Carnegie Institution of Washington, La Serena, Chile}
%\email{nmorrell@lco.cl}
\author{R. C. Gamen}
\affil{Instituto de Astrof\'{\i}sica de La Plata (CCT La Plata-CONICET, Universidad Nacional de La Plata), Paseo del Bosque s/n, 1900 La Plata, Argentina}
%\email{rgamen@fcaglp.unlp.edu.ar}
\author{J. I. Arias}
\affil{Departamento de F\'{\i}sica, Universidad de La Serena, Av. Cisternas 1200 Norte, La Serena, Chile}
%\email{julia@dfuls.cl}

% ---------------affiliations of the science group -----------------------

\altaffiltext{1}{The spectroscopic data in this article were gathered with three facilities: the 1.5 m telescope at the 
\facility{Observatorio de Sierra Nevada} (OSN), the 3.5 m telescope at \facility{Calar Alto Observatory} (CAHA), 
and the du Pont 2.5 m telescope at \facility{Las Campanas Observatory} (LCO). Some of the supporting imaging data 
were obtained with the 2.2 m telescope at CAHA and the NASA/ESA \facility{Hubble Space Telescope} (HST). The rest
were retrieved from the DSS2 and 2MASS surveys. The HST data were obtained at the Space Telescope Science 
Institute, which is operated by the Association of Universities for Research in Astronomy, Inc., under NASA 
contract NAS 5-26555.}

\altaffiltext{2}{Visiting Astronomer, CAHA, Spain.}
\altaffiltext{3}{Visiting Astronomer, OSN, Spain.}
\altaffiltext{4}{Visiting Astronomer, LCO, Chile.}
\altaffiltext{5}{e-mail contact: {\tt jmaiz@iaa.es}.}
\altaffiltext{6}{Ram\'on y Cajal fellow.}

\begin{abstract}
We present the first installment of a massive spectroscopic survey of
Galactic O stars, based on new, high signal-to-noise ratio, $R \sim 2500$ digital observations 
from both hemispheres selected from the Galactic O-Star Catalog of \citet{Maizetal04b} and
\citet{Sotaetal08}. 
The spectral classification system is rediscussed and 
a new atlas is presented, which supersedes previous versions.  Extensive
sequences of exceptional objects are given, including types Ofc, ON/OC,
Onfp, Of?p, Oe, and double-lined 
spectroscopic binaries. The remaining normal spectra bring this first sample 
to 184 stars, which is close to complete to $B=8$ and north of $\delta = -20^{\circ}$ and 
includes all of the northern objects in \citet{Maizetal04b} that are still classified as O stars.
The systematic and random accuracies of 
these classifications are substantially higher than previously attainable, 
because of the quality, quantity, and homogeneity of the data and analysis 
procedures. These results will enhance subsequent investigations in 
Galactic astronomy and stellar astrophysics. 
In the future we will publish the rest of the survey, beginning with a 
second paper that will include most of the southern stars in \citet{Maizetal04b}.
\end{abstract}


\section{Introduction}

	In \citet{Maizetal04b} we presented the first version of the Galactic O-Star Catalog (GOSC), a collection of 
spectral classifications for 378 Galactic O stars accompanied by astrometric, photometric, group membership, and 
multiplicity information. Most of the stars in that first version had been classified by one of us (N.R.W.) two or three 
decades earlier using photographic spectrograms. GOSC was subsequently expanded (version 2) by \citet{Sotaetal08}, 
who added $\sim 1000$ stars that had at least one spectral classification in the literature that identified them as O stars. 
As a quick look at the online version\footnote{\tt http://gosc.iaa.es} of GOSC v2 reveals, there 
is an unfortunately large disparity in the literature spectral classifications for the stars there. Some of the discrepancies 
are due to different spectral resolutions or signal-to-noise ratios (S/N), others to variability in the stars (spectroscopic
binaries being the major culprit here), and still others to errors or different criteria among classifiers. We believe it
is important to correct this situation, not only for the sake of the analysis of individual stars but also because
the use of inconsistent or incorrect spectral classifications may lead to errors in the derivation of statistically based
parameters such as the massive-star IMF or the overall number of ionizing photons in the Galaxy. 

	Thus was born in 2007 the idea for the Galactic O-Star Spectroscopic Survey (GOSSS), a project whose primary goal is to 
obtain new spectral classifications of at least all Galactic O stars brighter than $B$~=~13. Since then, we are deriving
classifications using new, uniform quality, high S/N spectrograms homogeneously processed and classified according to 
well-defined standards. The survey is described in \citet{Maizetal10b}.

	How opportune and feasible is such a project?  On the one hand, we are in a better 
position to do it than when similar surveys were attempted in the 1960s and 1970s: there are more telescopes, better 
detectors, improved data reduction software, and much larger reference databases. Furthermore, many of the targets are
relatively bright, making the project accessible to 1-4 m class telescopes. On the other hand, such a project still represents a 
large and complicated endeavor, with the targets scattered along the Galactic Plane in two hemispheres and requiring
hundreds of observation nights. Also, since most fields include none or only a few additional O stars within 
$\sim 10\arcmin$ of the primary target, the use of
a fiber spectrograph would be a waste of resources and a complication for harmonizing the data from different
observatories. Hence, the project is being conducted using long-slit spectrographs.

	The earliest results from GOSSS were presented in a letter \citep{Walbetal10a} that discussed the presence
of the C\,{\sc iii}~$\lambda$4650 blend in Of spectra. In this first paper we present [a] an overview of the project, 
[b] an atlas of the blue-violet spectral classification standards at $R\sim 2500$ from both hemispheres that will be the 
basis of the rest of the survey, and [c] a spectral library of 184 O stars without WR companions and 
with declinations larger than $-20\arcdeg$. The majority of the stars in this paper are from \citet{Maizetal04b}; a few 
have been added to achieve completeness\footnote{As described in this paper, some stars previously classified as B0 V to III 
have been assigned new spectral types O9.7 V to III (previously, the O9.7 spectral type was defined only for luminosity classes
II to Ia). Since we have only observed a small fraction of the stars with $B < 8.0$ previously classified as B0, it is possible that we have missed
some O9.7 stars within that magnitude range.} to $B = 8.0$, because of their presence in the same slit as other O stars, or 
because of their inclusion in \citet{Walbetal10a}. The declination limit is fixed by the accessibility from our northern 
observatories but it turns out to be a useful value because it splits the numbers in the original catalog into two 
nearly equal parts. Paper II will be the complement of part [c] of this one for declinations smaller than 
$-20\arcdeg$. Future papers will extend the O-star sample and the wavelength coverage; in both cases we already have
abundant data taken\footnote{We also point out out the existence of two related projects at higher spectral resolutions, OWN 
(southern hemisphere, \citealt{Barbetal10}) and IACOB (northern hemisphere, \citealt{SimDetal10}), 
which are obtaining $R\sim 40\,000$ optical spectrograms of hundreds of Galactic O, B, and WN stars.}. We may also publish the 
spectrograms of the hundreds of non-O and low-mass stars (B\footnote{Among the stars we are classifying as early-B there are some stars
that had previously considered to be O stars, e.g. RY Sct and HD 194\,280. The latter, the propotype late-OC supergiant, has been
reclassified as BC0 Iab.}, WR - including WNh stars\footnote{Hydrogen-rich WN stars appear to be relatively unevolved very massive stars
\citep{Crowetal10}.} -, hot subdwarfs) and the handful of O + WR systems that we 
are obtaining as byproducts of our search.

\section{Survey description}

\subsection{Blue-violet spectroscopy with $R$~$\sim$~2500}

	The primary goal of GOSSS is to obtain high S/N (200-300) blue-violet spectrograms of all O stars with $B < 13$ at
a high degree of uniformity and $R\sim 2500$. Given those conditions, our first step was to select the telescopes and 
instruments with which to carry on the survey. For the northern part of the survey, we settled on the Albireo 
spectrograph\footnote{\tt http://www.osn.iaa.es/Albireo/albireo.html .} at the 1.5~m telescope of the Observatorio de 
Sierra Nevada (OSN), which can reach stars down to $\delta = -20\arcdeg$. For the southern part of the survey 
($\delta < -20\arcdeg$), we chose the Boller \& Chivens 
spectrograph\footnote{\tt http://www.lco.cl/telescopes-information/irenee-du-pont/ telescopes-information/irenee-du-pont/instruments .}
at the 2.5~m du Pont telescope at Las Campanas Observatory (LCO). 
The du Pont telescope can reach the desired S/N values for the dimmest stars in the sample within a 
reasonable total integration time (approximately one hour), but in the north the 1.5~m at OSN requires significantly longer
exposure times, which compromise the quality of the spectra due to the required instrument stability. Therefore, the dimmer
stars ($B > 11$) in the northern part of the survey were observed with the TWIN 
spectrograph\footnote{\tt http://www.caha.es/pedraz/Twin/index.html .} at the 3.5 m telescope of Calar Alto Observatory 
(CAHA, Centro Astron\'omico Hispano Alem\'an). Also, since the image quality (seeing+telescope+instrument) is usually better with TWIN at CAHA 
than with Albireo at OSN, some of the bright northern stars
with close companions were observed from CAHA in order to better spatially separate the two spectra. 

	The characteristics of the three setups are shown in Table~\ref{settings}. We used observations of the same stars 
with two or three of the telescopes to check the uniformity of the data\footnote{Note that from LCO it is possible to access 
declinations much farther north than $\delta = -20\arcdeg$, thus providing a large overlap region of the sky for the three
observatories.}. The spectral resolution of our OSN and LCO observations as measured from the arc spectra turned out 
to be very similar and stable from night to night. $R_{4500} = 4500$ \AA$/\Delta\lambda = 2500 \pm 100$ with $\Delta\lambda$, the
FWHM of the calibration lamp emission lines, being nearly constant over the full wavelength range with a value of 1.8 \AA.
For our CAHA data the spectral resolution was somewhat higher ($R_{4500} \sim 3000$, $\Delta\lambda \sim 1.5$ \AA) and with a different 
dependence on wavelength. In order to provide a uniform spectral library, a smoothing filter was applied to the CAHA
data to achieve a constant $\Delta\lambda = 1.8$ \AA\ for the full spectral range.

\begin{table*}
\caption{Telescopes, instruments, and settings used.}
\centerline{
\begin{tabular}{lccccc}
\\
\hline
\multicolumn{1}{c}{Telescope} & Spectrograph      & Grating & Spectral scale & Spatial scale & Wav. range  \\
                              &                   & (l/mm)  & (\AA/px)       & (\arcsec/px)  & (\AA)       \\
\hline
OSN 1.5 m                     & Albireo           & 1800    & 0.66           & 0.85          & 3740$-$5090 \\
LCO 2.5 m (du Pont)           & Boller \& Chivens & 1200    & 0.80           & 0.56          & 3900$-$5510 \\
CAHA 3.5 m                    & TWIN (blue arm)   & 1200    & 0.54           & 0.69          & 3930$-$5020 \\
\hline
\end{tabular}
}
\label{settings}
\end{table*}

	In this paper we present mostly OSN and CAHA data, since the majority of the results here correspond to 
the northern part of the survey. Nevertheless, the atlas includes LCO data because for some spectral types southern
standards are better than northern ones\footnote{For some spectral types there are no northern standards in
\citet{Maizetal04b}.}. Our goal is to maintain our telescope triad for at least paper II of the survey. If we eventually
include new telescopes and/or instruments, we will first check for uniformity with the existing data.

	The data in this paper were obtained between 2007 and 2010. In some cases, observations were repeated due to 
focus and other instrument issues detected after the fact. For SB2 and SB3 spectroscopic binaries, multiple epochs were
obtained to observe the different orbit phases. In most cases with known orbits, observations near quadrature were attempted. 

\begin{figure*}
%\centerline{\includegraphics*[width=1.0\linewidth]{HD_190_429_2d.ps}}
\centerline{\includegraphics*[width=1.0\linewidth]{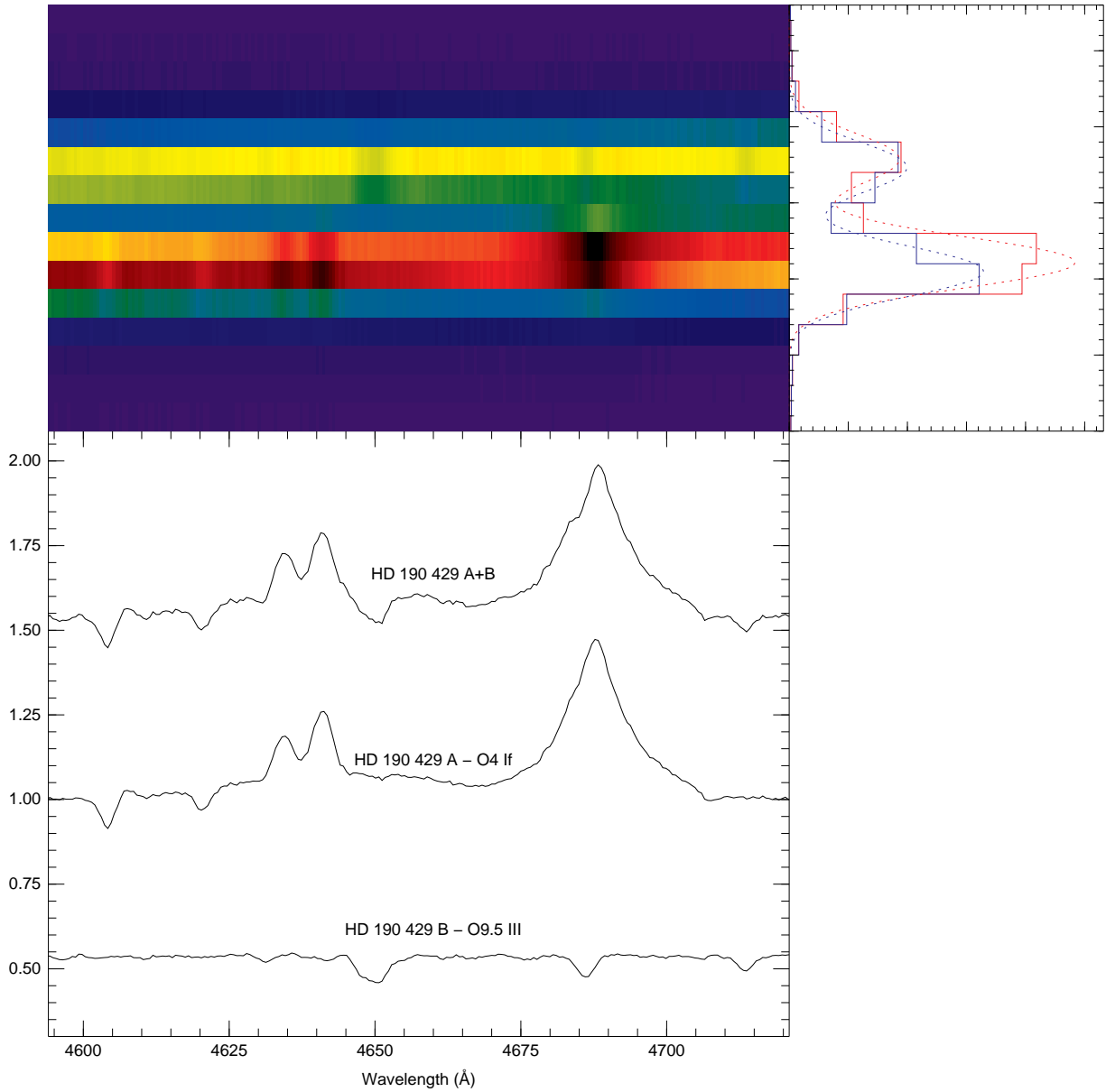}}
\caption{[Top left] False-color representation of a portion of a GOSSS long-slit exposure of HD~190\,429~A+B. 
The spectral direction is nearly parallel to the $x$ axis. The bottom 
(brighter) component is A and the top (weaker) component is B. [Top right] Spatial intensity cuts for two different wavelengths (one in red and 
one in blue) for the data on the top left panel. The dotted lines show the two-component fit to the data. [Bottom] Rectified extracted spectrum for
each component and for the sum of the two. The continua are all normalized to the value of the A component ($\Delta B_{\rm Ty}$ = 0.679 mag). Note
the appearance of C\,{\sc iii}~$\lambda$4647-50-51 \AA\ and \HeI{4713} absorptions and the change in the \HeII{4686}
profile for the A+B spectrum when compared to that of the A component.}
\label{2dplot}
\end{figure*}	

	In order to reduce the large amount of data in GOSSS, one of us (A.S.) wrote a pipeline in IDL. The pipeline first applies the bias and flat
and calculates a mask to eliminate cosmic rays and cosmetic defects. Second, the data are calibrated in wavelength and placed into the star rest frame.
Third, the star(s) in each long-slit exposure is/are identifed and extracted. Then, the spectra from different exposures (three or four per target) are combined 
and the final spectrogram is finally rectified. The pipeline can be run in either [a] a fully automated mode that is usually good enough for a quick 
look at the telescope or [b] an interactive mode that allows for the tweaking of some parameters such as the mask calculation or the spectrum rectification.

	A special case is that of close pairs with small magnitude differences ($\Delta m$). For those systems, we aligned the slit parallel to the line
joining the two stars to include both of them and we used a custom-made IDL fitting routine derived from the MULTISPEC code \citep{Maiz05a}
to deconvolve the two spatial profiles and extract the spectra for the two stars. An example is shown in Fig.~\ref{2dplot}.
The procedure works very well for large separations but becomes increasingly harder for small values, especially if $\Delta m$ is large or the seeing is
degraded. The closest pair for which we were able to extract separate spectra thanks to excellent seeing conditions was HD 17\,520 AB ($\Delta m \approx 0.7$
mag) with a separation of 0\farcs316 \citep{Maiz10}. On the other hand, we were unable to separate $\sigma$ Ori AB, which currently has a slightly lower 
separation (0\farcs260) but a significantly larger $\Delta m$ ($\approx 1.6$ mag, \citealt{Maiz10}). As will be shown later, the use of such a deconvolution 
technique is the reason for the largest changes in the spectral classifications in this paper with respect to previous works.

	The data from each observatory covers slightly different wavelength ranges (Table~\ref{settings}). The spectrograms shown in this paper have been 
cut to show the same spectral range.

\subsection{Complementary data, nomenclature, and cataloguing}

	Two problems that have complicated the spectral classification of massive stars in the past are [a] the presence of
nearby resolved companions that may or may not contribute to the observed primary spectrum depending on the magnitude 
difference, separation, slit orientation, and seeing; and [b] the misidentification of components in multiple systems. Both 
issues are known to be the sources of some discrepancies between literature spectral classifications of the same target.

	In order to correct those two issues as much as possible, we used two strategies. On the one hand, we analyzed
high-resolution imaging to identify and measure the magnitude differences of nearby companions. For the northern part of the
survey, this was done with Lucky-Imaging AstraLux observations at the 2.2 m telescope of CAHA and HST imaging (GO 
programs 10602 and 11981, P.I. Ma\'{\i}z Apell\'aniz, and archival data). The first results appeared in \citet{Maiz10} and will be
used here. For the southern part of the survey, we will use, among others, 
HST imaging from programs 10205 (P.I. Walborn), 10602, and 10898 (P.I. Ma\'{\i}z Apell\'aniz). On the other hand, we searched the
literature for results similar to those obtained with AstraLux 
(e.g. \citealt{McCaStau94,Duchetal01,Masoetal98,Turnetal08,Bouyetal08,Masoetal09}) and we plotted information from Simbad 
using Aladin images to ensure the correct identification of sources. In order to minimize possible future confusions, we 
provide charts for some specific cases. We followed the component nomenclature of the Washington Double Star Catalog 
\citep{Masoetal01}.

	In some cases, the information derived from the sources above allowed us to determine whether two or more visual 
components are spatially unresolved in our data. We considered that a secondary component is capable of significantly modifying 
the spectral type if $|\Delta B| \le 2.0$. In such cases we included in the name of the star the two components 
(e.g. Pismis~24-1~AB or HD~93\,129~AaAb); for larger values of $|\Delta B|$ the secondary component was not included in the name.
Note that when we are able to spatially resolve a nearby component and extract its spectrum independently from the primary, 
we do include the component name in each case (e.g. HD 218\,195 A) even if $|\Delta B|$ is larger than 2.0.

	As previously mentioned, the GOSSS sample was drawn from version 2.3 of GOSC \citep{Sotaetal08}. Our plans for the
future include using the new spectral classification to produce a new (3.0) version of the catalog. That version will
include not only the spectral classifications but the spectroscopic data themselves as well as the new distances \citep{Maizetal08c}
derived from the new Hipparcos data reduction \citep{vanL07a}.

\subsection{Spectral classification methodology}

	Spectral classification according to the MK process is carried out by [a] selecting a two-dimensional grid (in spectral 
type and luminosity class) of standard stars; [b] comparing the unknown spectrum with that grid, in terms of the line ratios 
that define the different subtypes; and [c] choosing the standard spectrum that most resembles the unknown spectrum,
if appropriate noting any anomalies such as broad lines or discrepancies among different line ratios compared to the standards. The
classification categories are discrete, whereas the phenomena are continuous, so interpolations or compromises may be required in some
cases, which should be noted.

\begin{figure*}
%\centerline{\includegraphics*[width=1.2\linewidth]{DH_Cep_090701_O_B2500.ps}}
%\centerline{\includegraphics*[width=1.2\linewidth]{Y_Cyg_090905_O_B2500.ps}}
\centerline{\includegraphics*[width=1.2\linewidth]{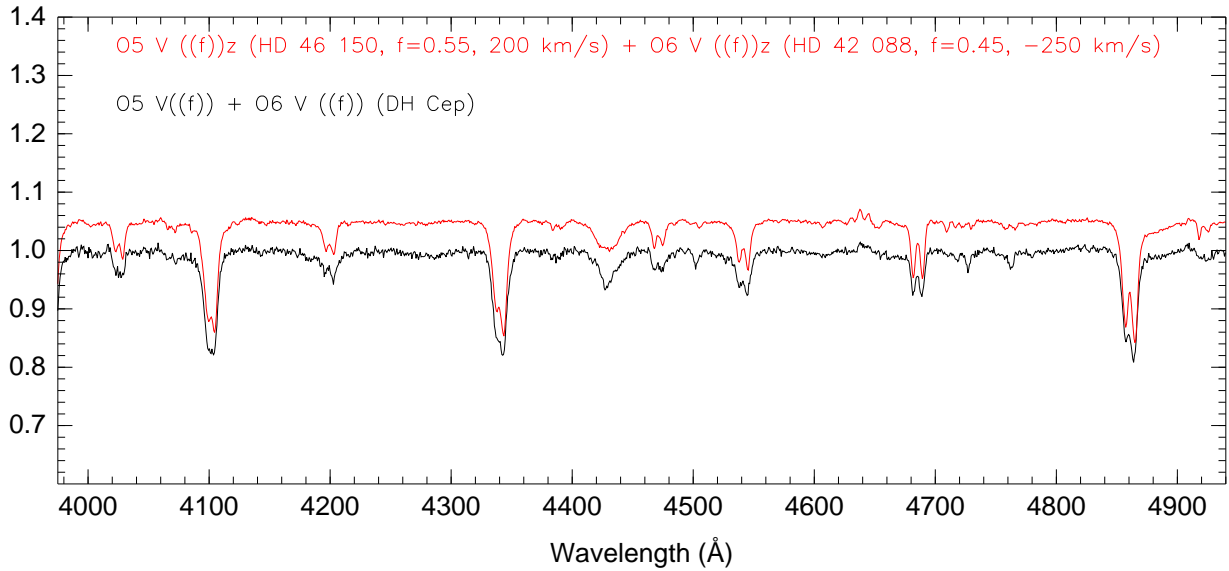}}
\centerline{\includegraphics*[width=1.2\linewidth]{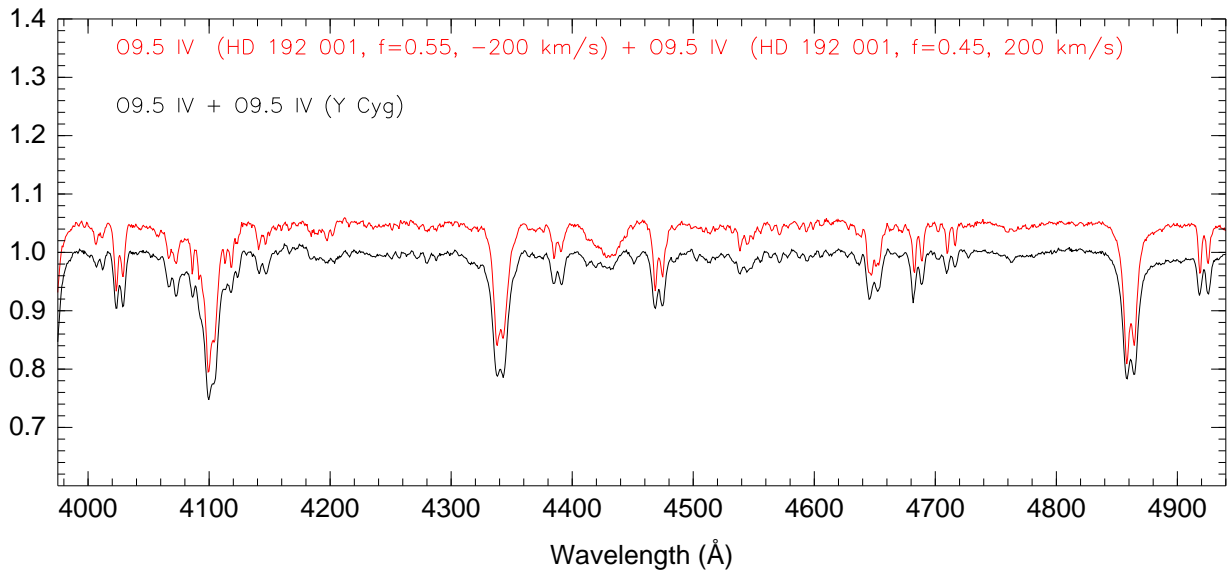}}
\caption{Two examples of spectral classifications of double-lined spectroscopic binaries using MGB. The bottom (black) line shows the spectrum to be
classified and the top (red or grey) line the linear combination of the two standards. The flux fraction and velocity of each standard are indicated along with 
its spectral type. [See the electronic version of the journal for a color version of this figure.]}
\label{SB2}
\end{figure*}	

	Many of the stars we selected as standard stars for this paper have been previously used as such, in some cases going back 
to the original definition of the O subtypes. Nevertheless, in some cases we noted inconsistencies that made us revise the spectral classification,
or we found other, superior definitions of the category among our expanded sample, as detailed in the next section. For
the comparison between the unknown spectra and the standards we used MGB\footnote{Marxist Ghost Buster.}, 
an IDL code developed by one of us (J.M.A.) that overplots the two
and allows the user to easily change from one standard to another.
MGB also allows the user to artificially broaden the standard spectra to measure the line broadening (see next section) and
also to combine two standard spectra adjusting their velocities and flux fraction in order to analyze spectroscopic binaries (see Fig.~\ref{SB2} for
two examples). The software was independently used by two of the authors and the results compared. In most cases there was an excellent agreement
in the classifications; discrepancies were subsequently analyzed in more detail.

	One important aspect is that spectral classification is subject to the effects of spectral, spatial, and temporal resolution as well as S/N. 
For example, a SB2 may remain undetected without adequate resolution or temporal coverage, possibly yielding anomalously wide lines due to blends;
in other cases some absorption lines may be too weak to be detected, e.g., \HeII{4542} at B0. In other cases, a close visual binary may 
have historical composite spectra (hence, intermediate spectral classifications and/or peculiarities) that cannot be separated until spatially resolved 
spectroscopy can be obtained. Such limitations are a major reason for discrepant spectral classifications in the literature. 
As previously described, we are taking steps to minimize such effects (e.g., obtaining multiple-epoch spectroscopy 
for known SB2s and to discover new ones), but it is impossible to eliminate them completely. That is one of the reasons why we publish not only the 
spectral types, but also the original spectrograms, since that enables comparison with past or future results. In that regard, we have searched the 
literature for spectrograms that may be in conflict with our classifications (because of, e.g., better 
temporal or spectral resolution) and analyzed those cases. We plan to continually update the GOSC whenever new data justify it in the future.

\section{Results}

	This section constitutes the main body of this paper and is divided in three parts. First, we present the new atlas 
of standard O stars and the associated spectral classification developments. Second, we briefly present the noteworthy
characteristics of some of the members of the
peculiar categories (Ofc, ON/OC, Onfp, Of?p, Oe, SB2+SB3) of the full Northern sample (atlas and non-atlas stars) in the
paper. Finally, we do the same with the normal O stars in the full Northern sample. 

\subsection{Atlas and spectral classification system developments}

\begin{table*}
\caption{Spectral classification standards.}
\scriptsize
\input{standards}
\label{standards}
\end{table*}

\begin{table*}
\caption{Qualifiers used for spectral classification in this work and in others.}
\label{qualifiers}
\centerline{
\begin{tabular}{cl}
	& \\
\hline
((f))   & Weak N\,{\sc iii}~$\lambda$4634-40-42 emission, strong \HeII{4686} absorption \\
(f)     & Medium N\,{\sc iii}~$\lambda$4634-40-42 emission, neutral or weak \HeII{4686} absorption \\
f       & Strong N\,{\sc iii}~$\lambda$4634-40-42 emission, \HeII{4686} emission above continuum \\
\hline
((f*))  & N\,{\sc iv}~$\lambda$4058 emission $\ge$ N\,{\sc iii}~$\lambda$4640 emission, strong \HeII{4686} absorption (O2-3.5) \\
(f*)    & N\,{\sc iv}~$\lambda$4058 emission $\ge$ N\,{\sc iii}~$\lambda$4640 emission, weaker \HeII{4686} absorption (O2-3.5) \\
f*      & N\,{\sc iv}~$\lambda$4058 emission $\ge$ N\,{\sc iii}~$\lambda$4640 emission, \HeII{4686} emission (O2-3.5) \\
\hline
((fc))  & As ((f)) plus C\,{\sc iii}~$\lambda$4647-50-51 emission equal to N\,{\sc iii}~$\lambda$4634\\
(fc)    & As (f) plus C\,{\sc iii}~$\lambda$4647-50-51 emission equal to N\,{\sc iii}~$\lambda$4634 \\
fc      & As f plus C\,{\sc iii}~$\lambda$4647-50-51 emission  equal to N\,{\sc iii}~$\lambda$4634 \\
f?p     & Variable C\,{\sc iii}~$\lambda$4647-50-51 emission $\ge$ N\,{\sc iii}~$\lambda$4634-40-42 at maximum; variable \\
        & $\;\;$ sharp absorption, emission, and/or P Cygni features at H and He\,{\sc i} lines \\
\hline
((f+))  & As ((f)) plus Si\,{\sc iv}~$\lambda$4089-4116 emission (O4-8, obsolete, see subsection 3.1.3) \\
(f+)    & As (f) plus Si\,{\sc iv}~$\lambda$4089-4116 emission (O4-8, obsolete, see subsection 3.1.3) \\
f+      & As f plus Si\,{\sc iv}~$\lambda$4089-4116 emission (O4-8, obsolete, see subsection 3.1.3) \\
\hline
(e)     & Probable H$\alpha$ emission but no red spectrogram available \\
e       & Emission components in H lines \\
pe      & As e with emission components in He\,{\sc i} and/or continuum veiling \\
$[$e$]$ & Emission spectrum including Fe forbidden lines \\
e+      & Fe\,{\sc ii} and H emission lines (subcategories in \citealt{Lesh68}) \\
\hline
((n))   & Broadened lines (not applied here, marginal) \\
(n)     & More broadened lines ($v\sin i\sim 200$ km/s) \\
n       & Even more broadened lines ($v\sin i\sim 300$ km/s) \\
nn      & Yet even more broadened lines ($v\sin i\sim 400$ km/s) \\
$[$n$]$ & H lines more broadened than He lines \\
nfp     & He\,{\sc ii} centrally reversed emission, broadened absorption lines (Conti Oef) \\
\hline
N       & N absorption enhanced, C and O deficient \\
Nstr    & Moderate case of above (e.g. N\,{\sc iii}~$\lambda$4640 enhanced but not $>$ C\,{\sc iii}~$\lambda$4650) \\
C       & C absorption enhanced, N deficient \\
Nwk     & Moderate case of above \\
\hline
var     & Variation in line spectrum intensities or content \\
p       & Peculiar spectrum \\
z       & \HeII{4686} in absorption and $>$ than both \HeI{4471} and \HeII{4542} \\
\hline
\end{tabular}
}
\end{table*}

\begin{figure}
%\centerline{\includegraphics*[width=1.0\linewidth]{broad_O9.ps}}
\centerline{\includegraphics*[width=1.0\linewidth]{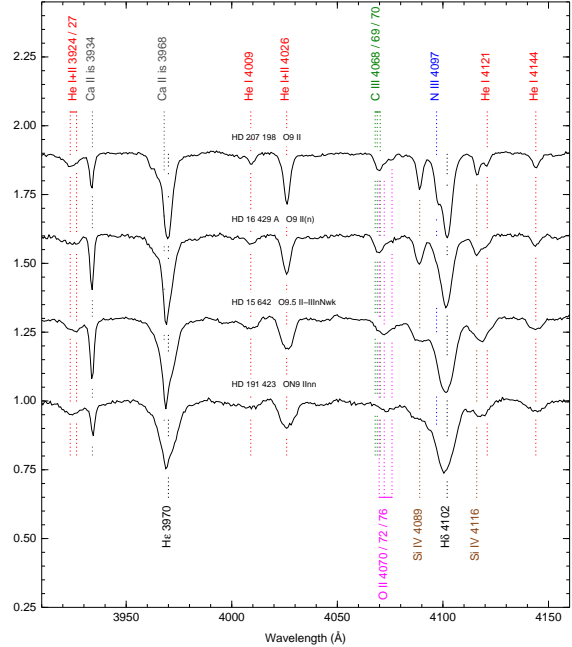}}
\caption{Broadening sequence [normal, (n), n, nn] for four O stars of similar spectral type. 
[See the electronic version of the journal for a color version of this figure.]}
\label{broadening}
\end{figure}	

A historical and technical review of the current spectral classification
system for the OB stars was given by \citet{Walb09a}. Because of the
unprecedented quality and quantity of the present dataset, several systemic 
developments and revisions for the O stars are introduced in the present 
work, which supersede previous procedures and are described here.  
Classification standards are listed in Table~\ref{standards}, and an extensive new
spectral atlas is presented in Figures~\ref{LC_Ia}--\ref{ST_O9}; the first four figures
provide spectral-type sequences at fixed luminosity classes, while the
latter five are luminosity-class sequences at fixed spectral types (with
a few exceptions because of positions unrepresented in the current sample)\footnote{It is important to note that the printed atlas plots
are necessarily very reduced. They must be enlarged online to reveal the full definition of the classification criteria, especially those
involving weak lines. As previously mentioned, with v3.0 of GOSC we plan to make the data themselves available to the astronomical
community.}. This atlas replaces that of \citet{WalbFitz90} for the O spectral types. 
A list of qualifiers for O spectral types is provided in Table~\ref{qualifiers}.

With regard to line broadening, we have consistently distinguished the
three degrees (n), n, and nn in this work. The ((n)) qualifier of
\citet{Walb71a} has not been applied, as it was judged too marginal and      
close to the slight resolution differences among the different
instruments involved. Figure~\ref{broadening} shows the sequence from normal to nn stars for stars around type O9 II.
See Table~\ref{qualifiers} for the approximate velocities that correspond to 
(n), n, and nn, respectively\footnote{Note that thos evalues are only approximate: spectral classification is done
by comparing data with standard spectra, such as those in Fig.~\ref{broadening}.}.

\begin{figure*}
%\centerline{\includegraphics*[width=1.01\linewidth]{LC_Ia.ps}}
\centerline{\includegraphics*[width=1.01\linewidth]{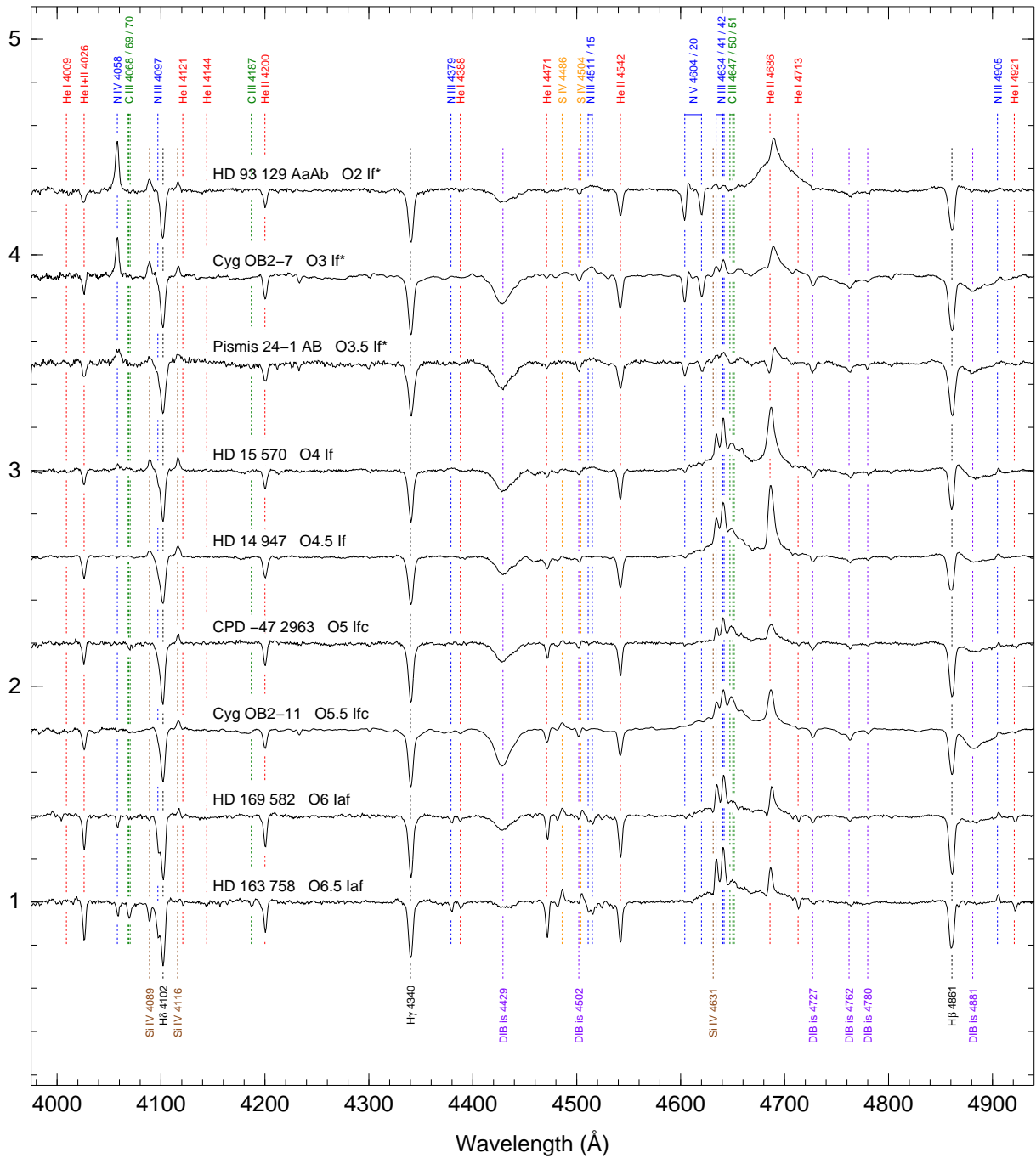}}
\caption{Atlas of rectified digital, linear-intensity spectrograms for luminosity class I, Galactic O stars. In this and 
subsequent figures, the $y$ axis is labeled in continuum units and the spectrograms are vertically displaced for display 
purposes. [See the electronic version of the journal for a color version of this figure.]}
\label{LC_Ia}
\end{figure*}	

\addtocounter{figure}{-1}

\begin{figure*}
%\centerline{\includegraphics*[width=1.01\linewidth]{LC_Ib.ps}}
\centerline{\includegraphics*[width=1.01\linewidth]{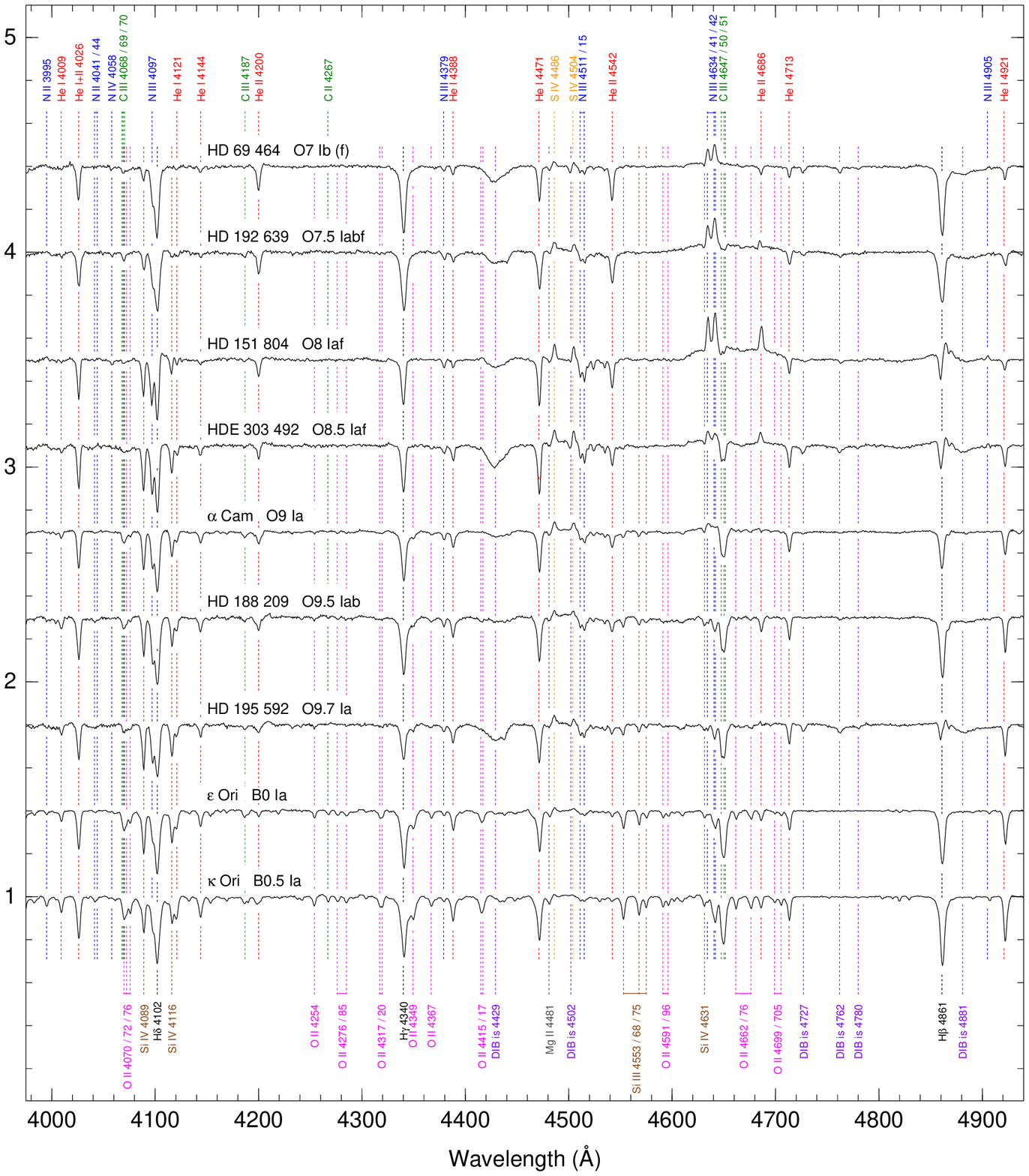}}
\caption{(continued).}
\label{LC_Ib}
\end{figure*}	

\begin{figure*}
%\centerline{\includegraphics*[width=1.01\linewidth]{LC_II.ps}}
\centerline{\includegraphics*[width=1.01\linewidth]{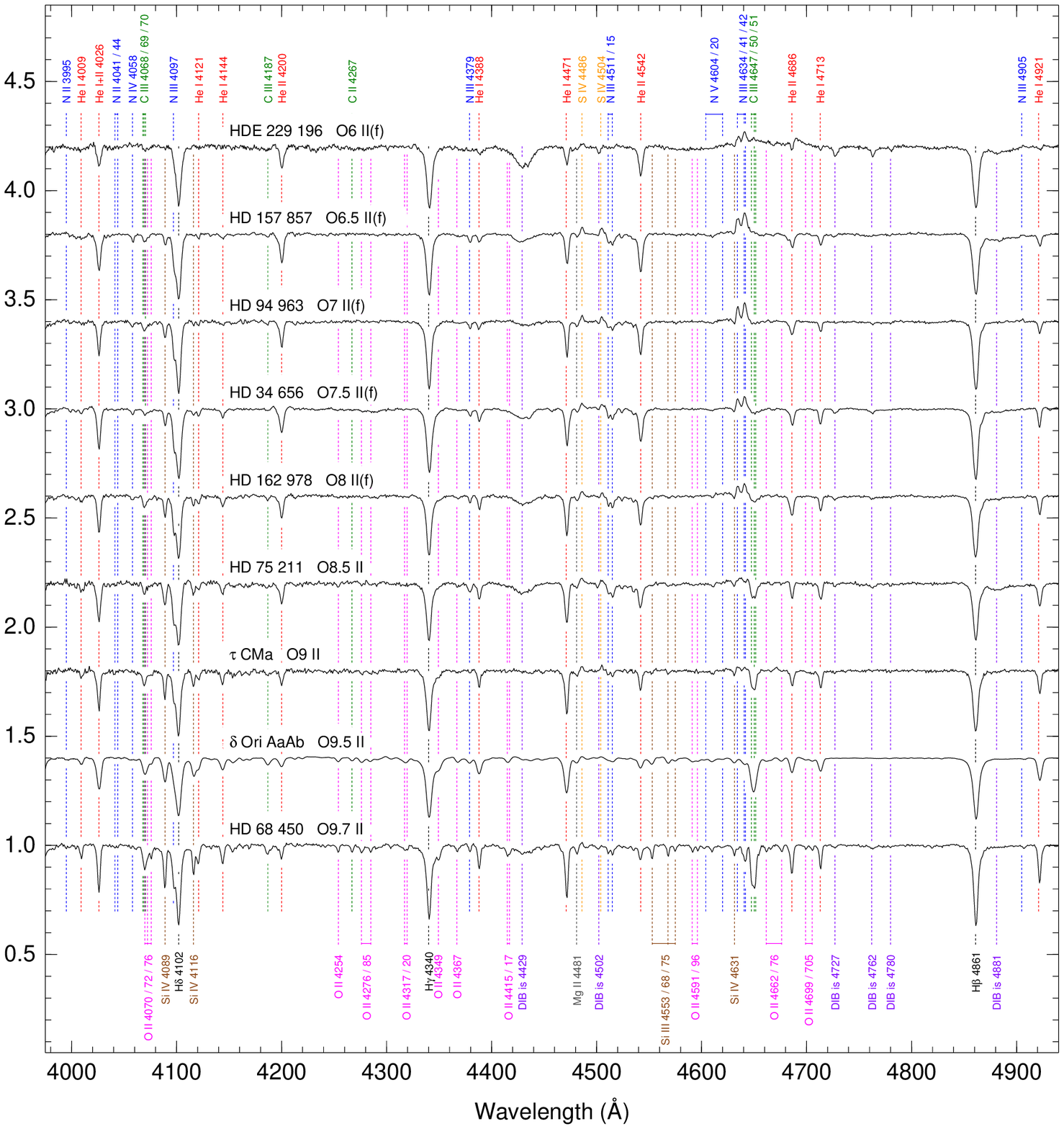}}
\caption{Same as Fig.~\ref{LC_Ia} for luminosity class II. [See the electronic version of the journal for a color version of this figure.]}
\label{LC_II}
\end{figure*}	

\begin{figure*}
%\centerline{\includegraphics*[width=1.01\linewidth]{LC_IIIa.ps}}
\centerline{\includegraphics*[width=1.01\linewidth]{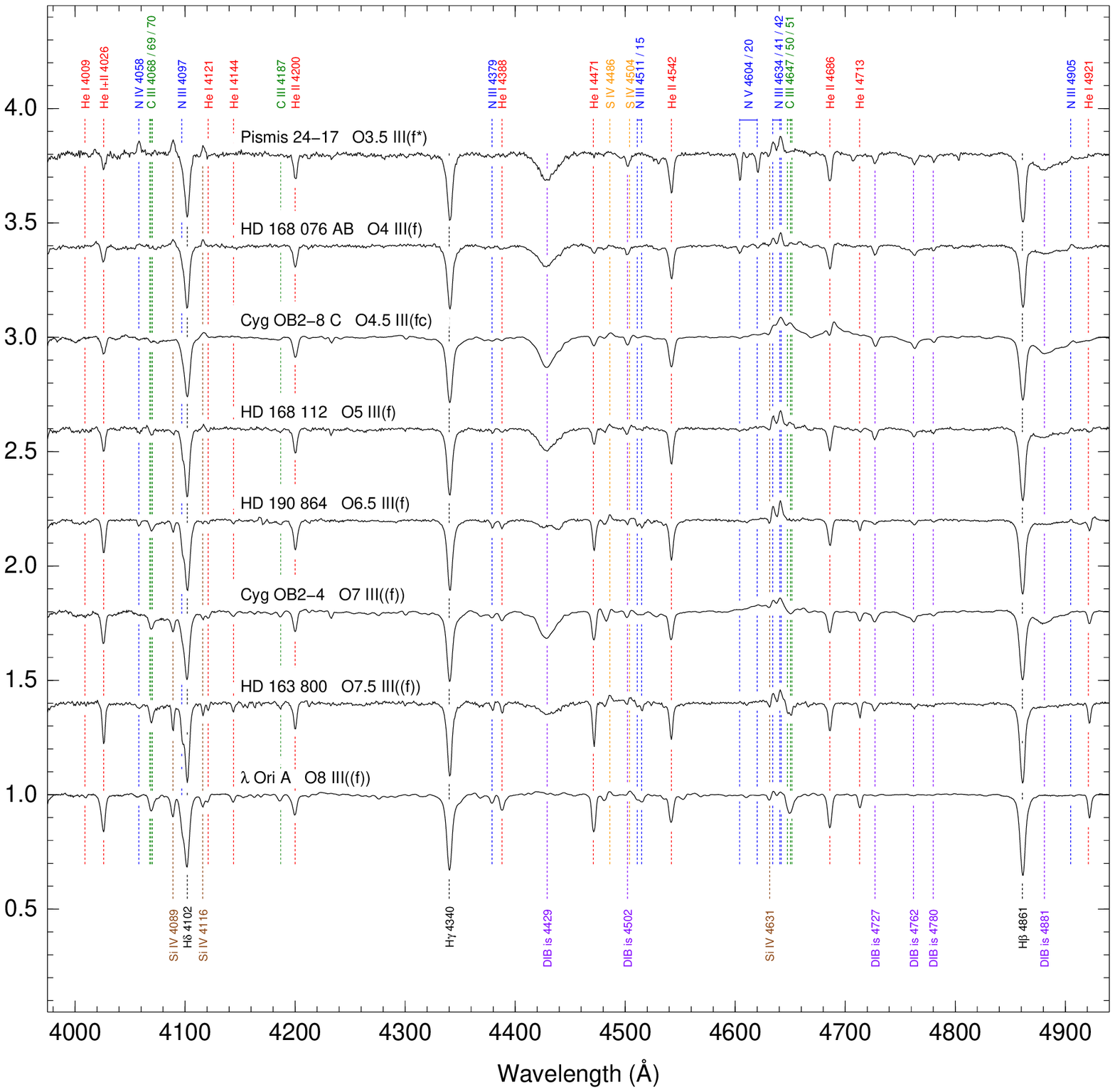}}
\caption{Same as Fig.~\ref{LC_Ia} for luminosity class III. [See the electronic version of the journal for a color version of this figure.]}
\label{LC_IIIa}
\end{figure*}	

\addtocounter{figure}{-1}

\begin{figure*}
%\centerline{\includegraphics*[width=1.01\linewidth]{LC_IIIb.ps}}
\centerline{\includegraphics*[width=1.01\linewidth]{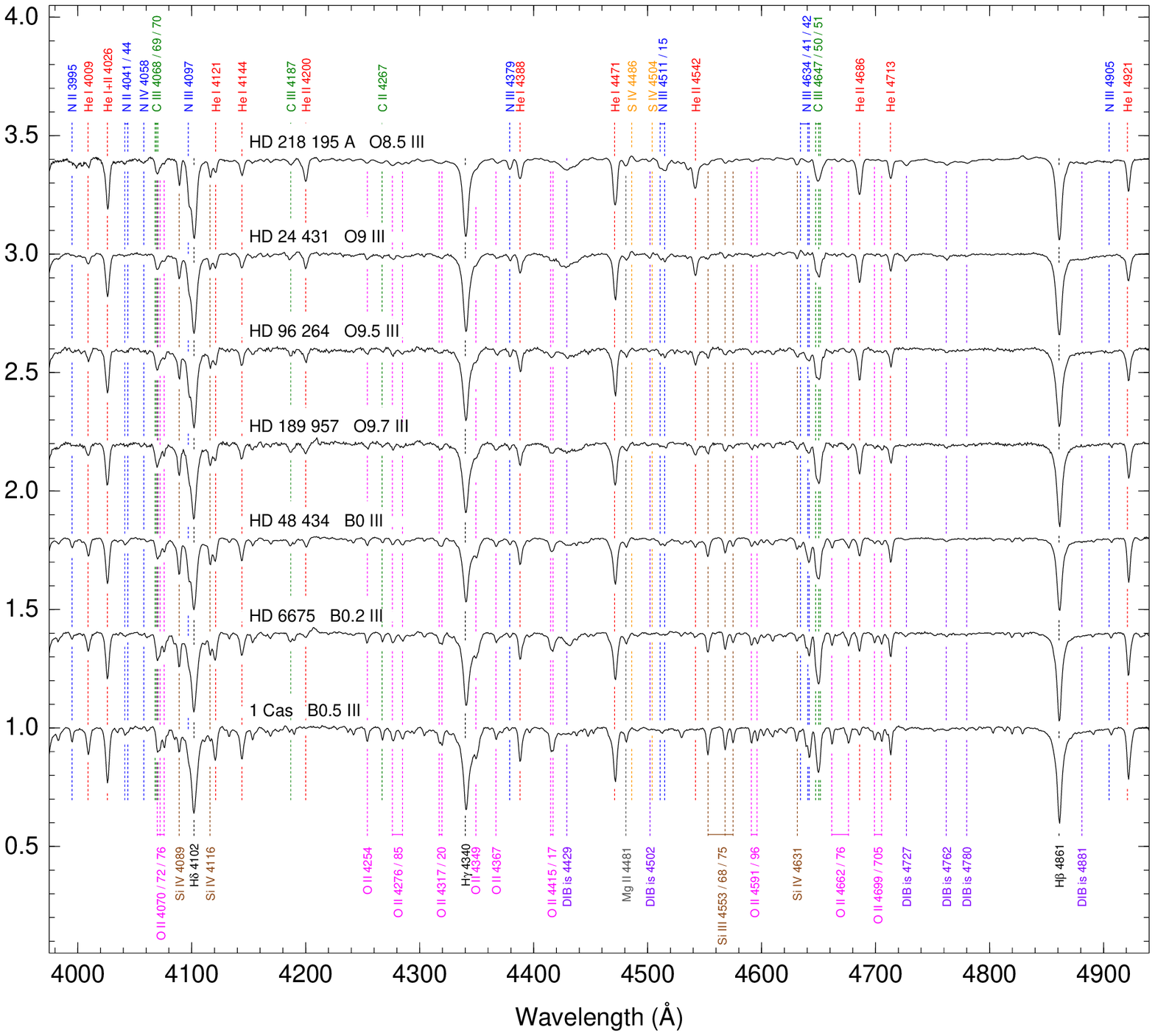}}
\caption{(continued).}
\label{LC_IIIb}
\end{figure*}	

\begin{figure*}
%\centerline{\includegraphics*[width=1.01\linewidth]{LC_Va.ps}}
\centerline{\includegraphics*[width=1.01\linewidth]{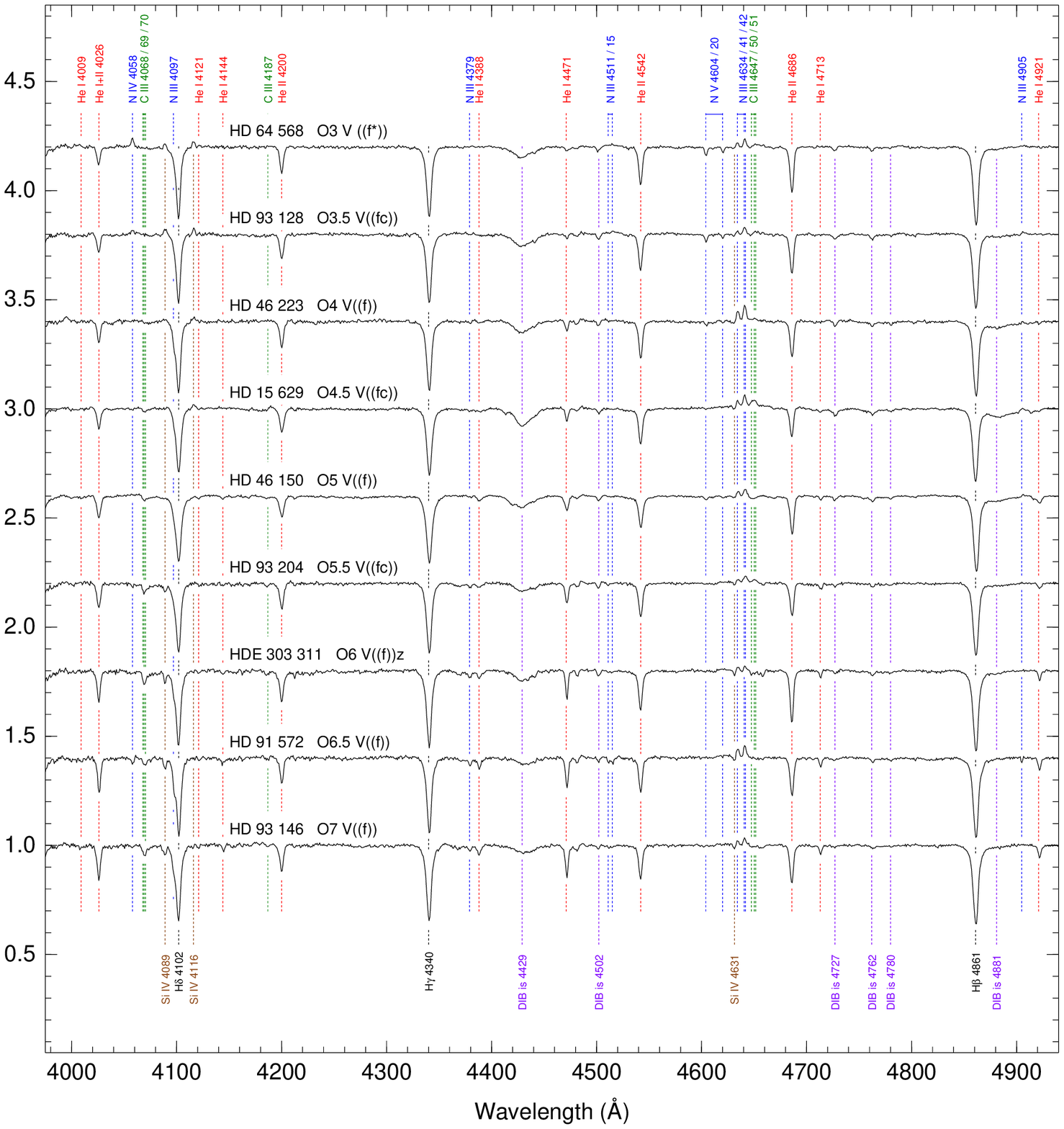}}
\caption{Same as Fig.~\ref{LC_Ia} for luminosity class V. [See the electronic version of the journal for a color version of this figure.]}
\label{LC_Va}
\end{figure*}	

\addtocounter{figure}{-1}

\begin{figure*}
%\centerline{\includegraphics*[width=1.01\linewidth]{LC_Vb.ps}}
\centerline{\includegraphics*[width=1.01\linewidth]{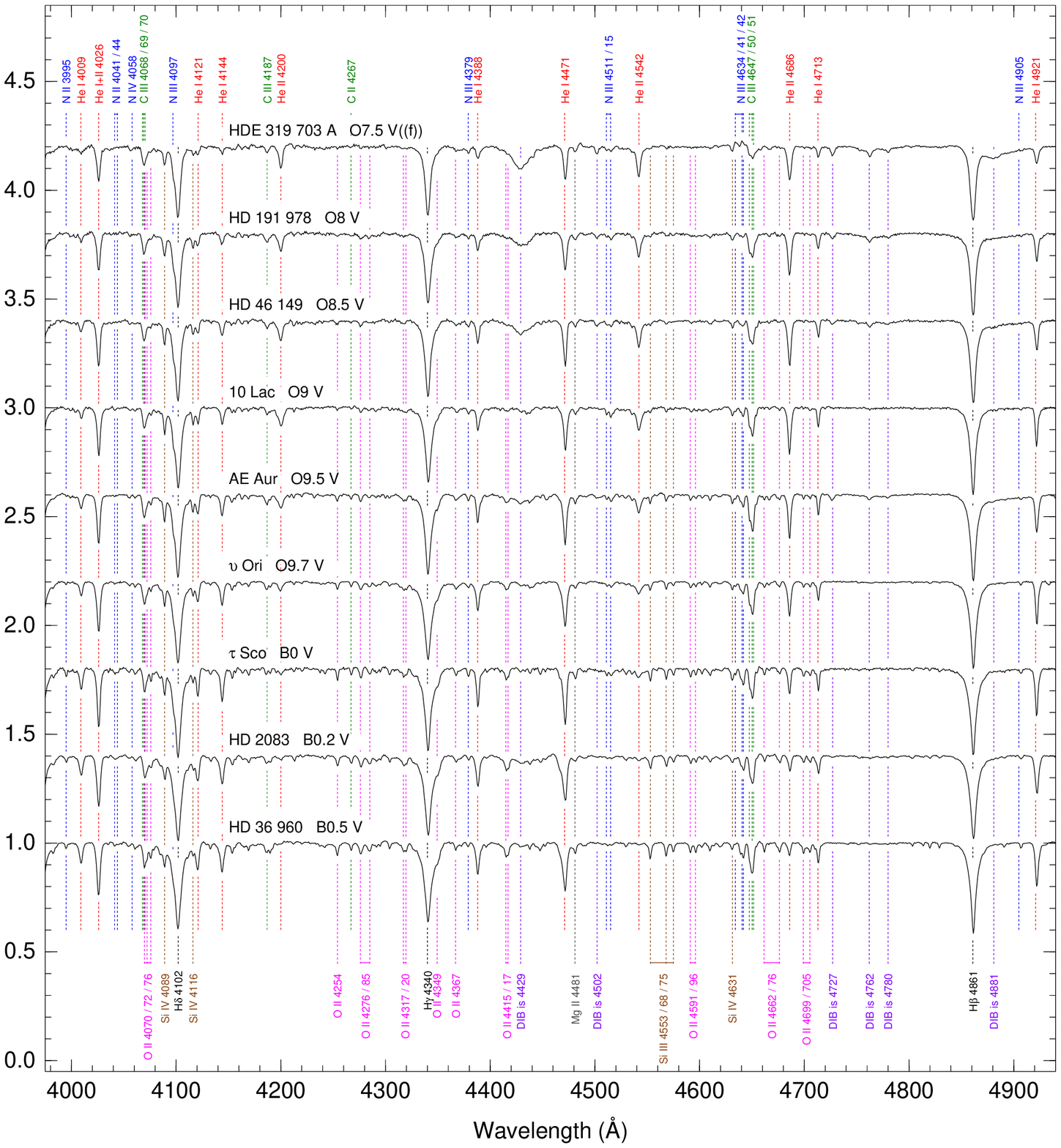}}
\caption{(continued).}
\label{LC_Vb}
\end{figure*}	

\begin{figure*}
%\centerline{\includegraphics*[width=1.01\linewidth]{ST_O4.ps}}
\centerline{\includegraphics*[width=1.01\linewidth]{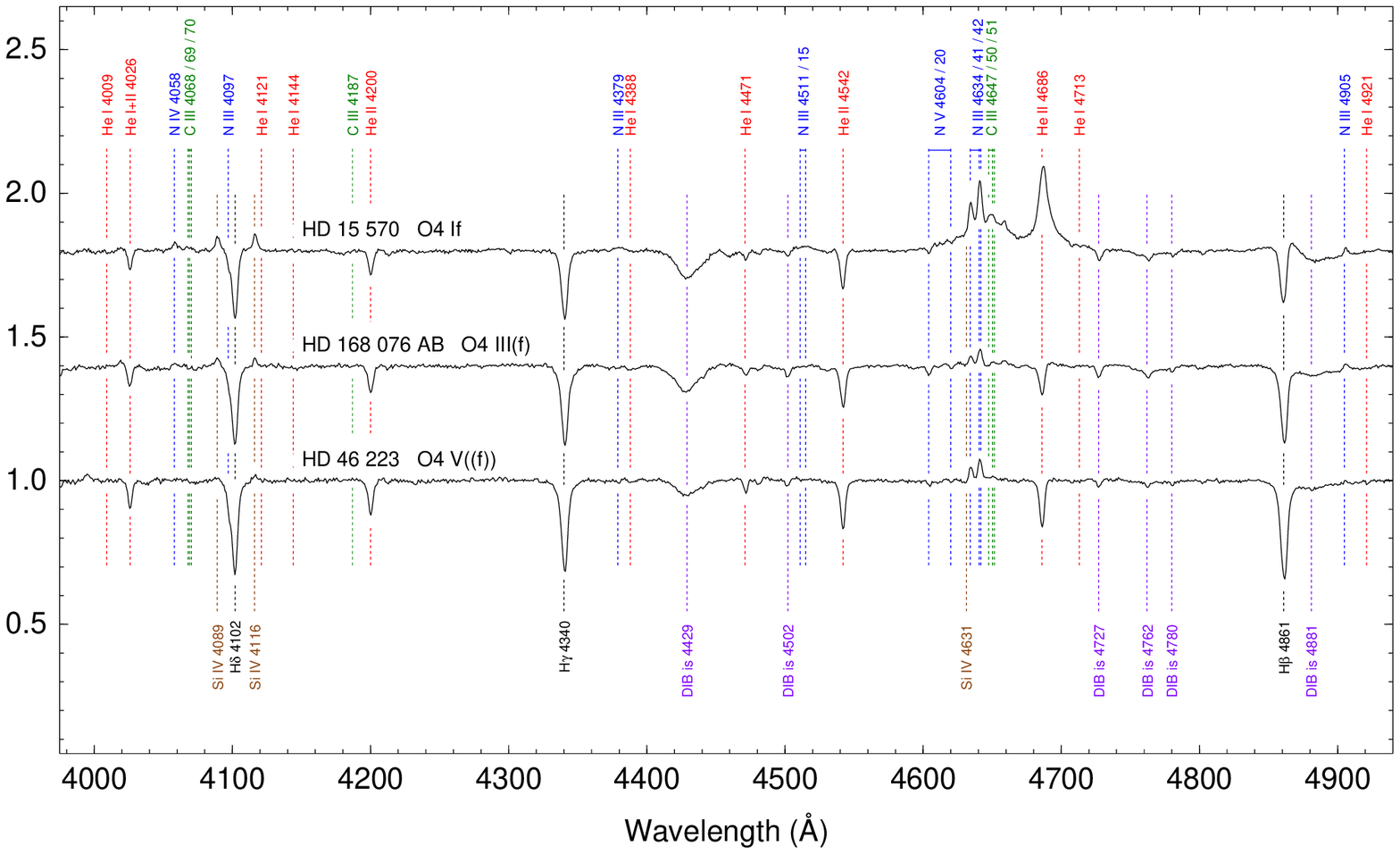}}
\caption{Luminosity effects at spectral type O4. [See the electronic version of the journal for a color version of this figure.]}
\label{ST_O4}
\end{figure*}	

\begin{figure*}
%\centerline{\includegraphics*[width=1.01\linewidth]{ST_O6.5.ps}}
\centerline{\includegraphics*[width=1.01\linewidth]{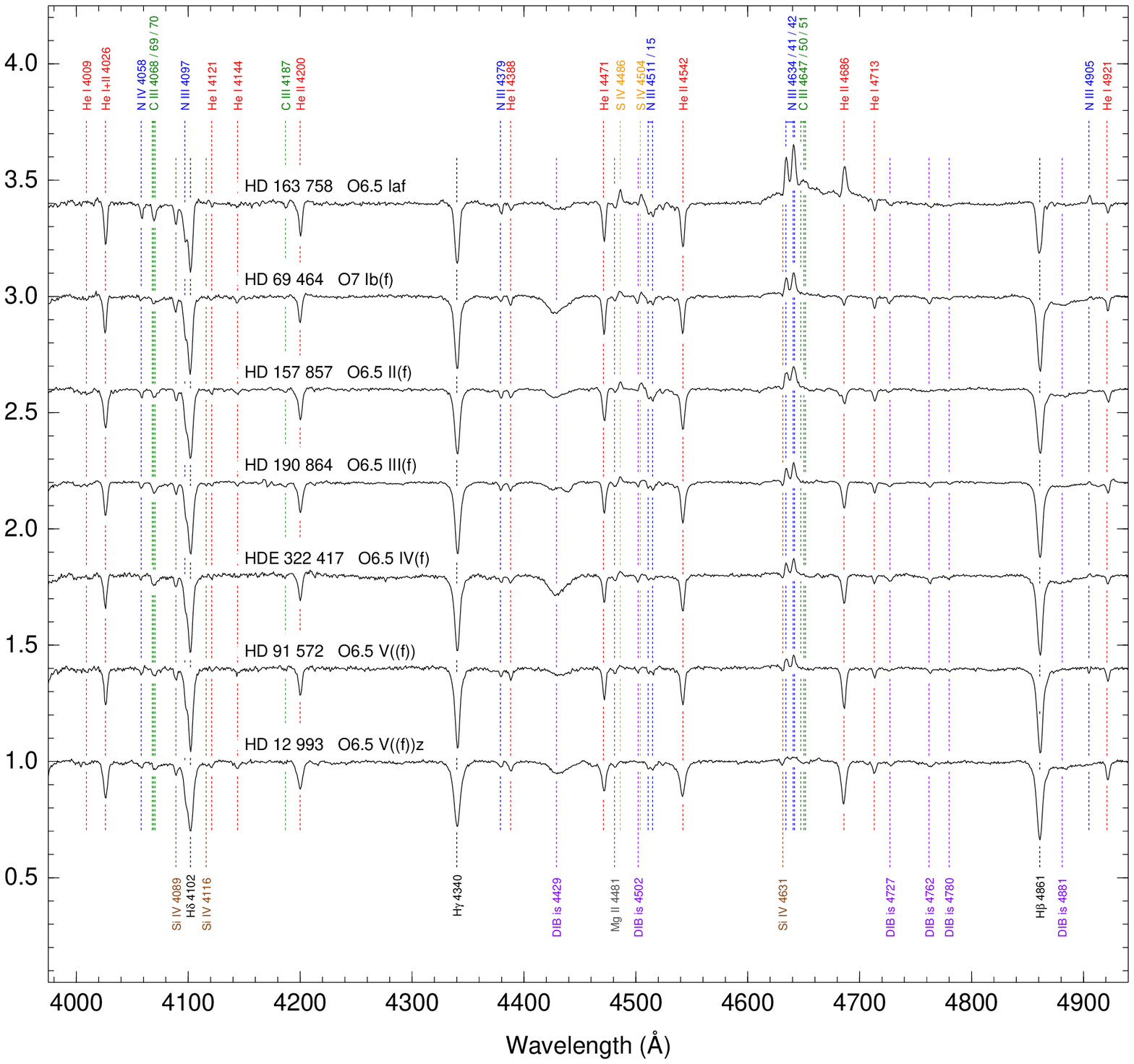}}
\caption{Luminosity effects at spectral type O6.5. [See the electronic version of the journal for a color version of this figure.]}
\label{ST_O6.5}
\end{figure*}	

\begin{figure*}
%\centerline{\includegraphics*[width=1.01\linewidth]{ST_O8.ps}}
\centerline{\includegraphics*[width=1.01\linewidth]{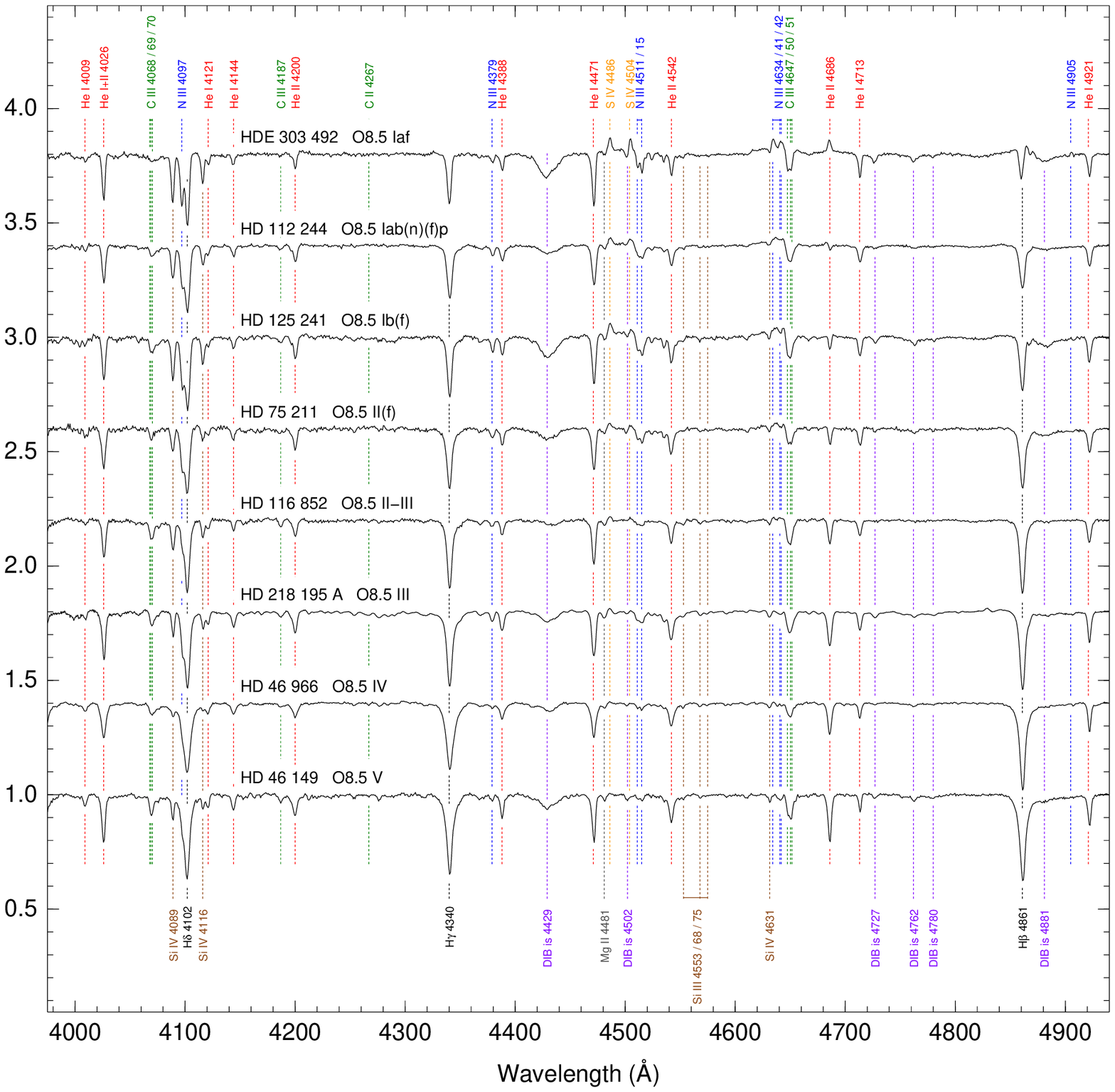}}
\caption{Luminosity effects at spectral type O8. Note that HD 112\,244 is not in our standard list because of its broad lines.
[See the electronic version of the journal for a color version of this figure.]}
\label{ST_O8}
\end{figure*}	

\begin{figure*}
%\centerline{\includegraphics*[width=1.01\linewidth]{ST_O8.5.ps}}
\centerline{\includegraphics*[width=1.01\linewidth]{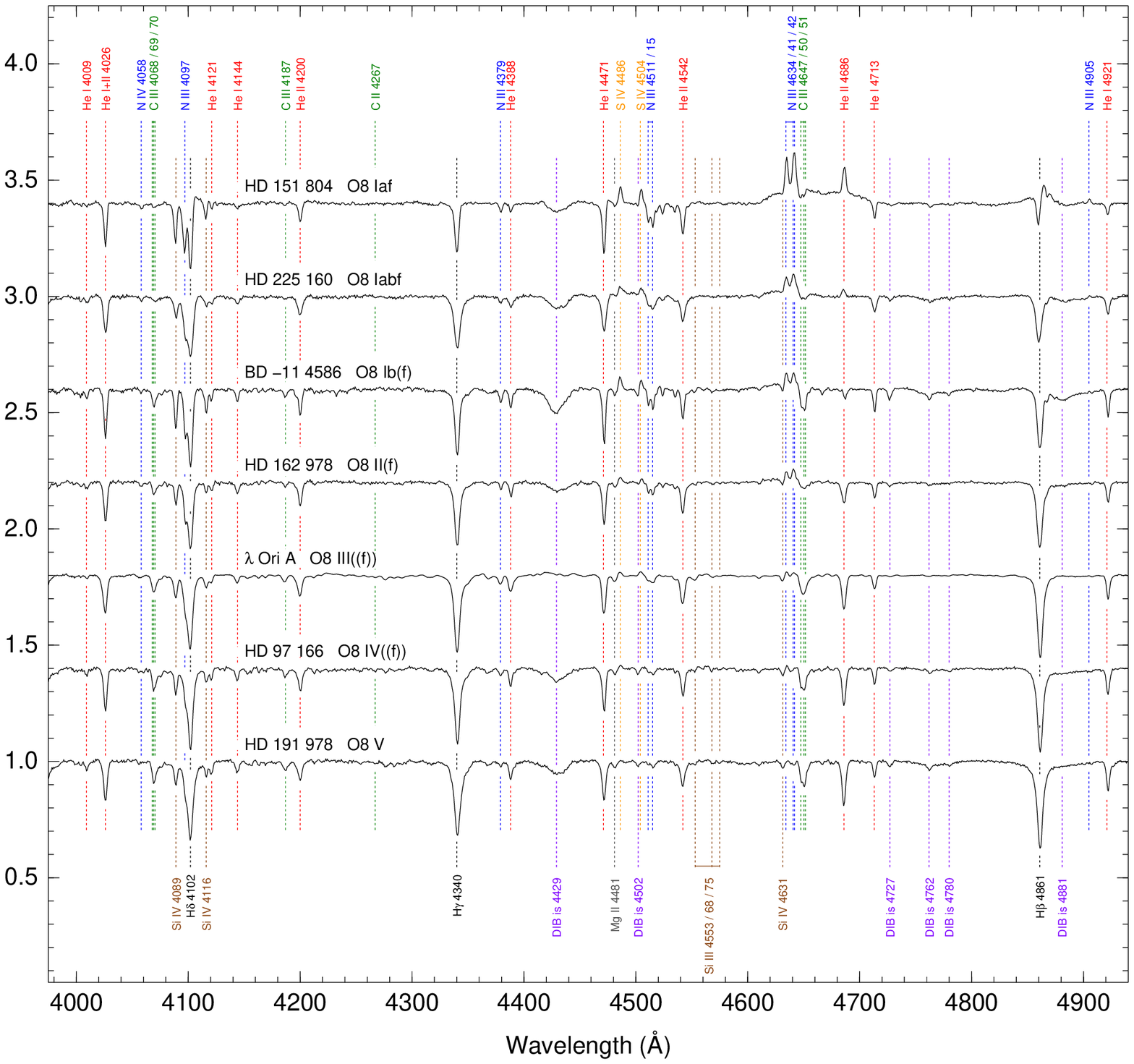}}
\caption{Luminosity effects at spectral type O8.5. [See the electronic version of the journal for a color version of this figure.]}
\label{ST_O8.5}
\end{figure*}	

\begin{figure*}
%\centerline{\includegraphics*[width=1.01\linewidth]{ST_O9.ps}}
\centerline{\includegraphics*[width=1.01\linewidth]{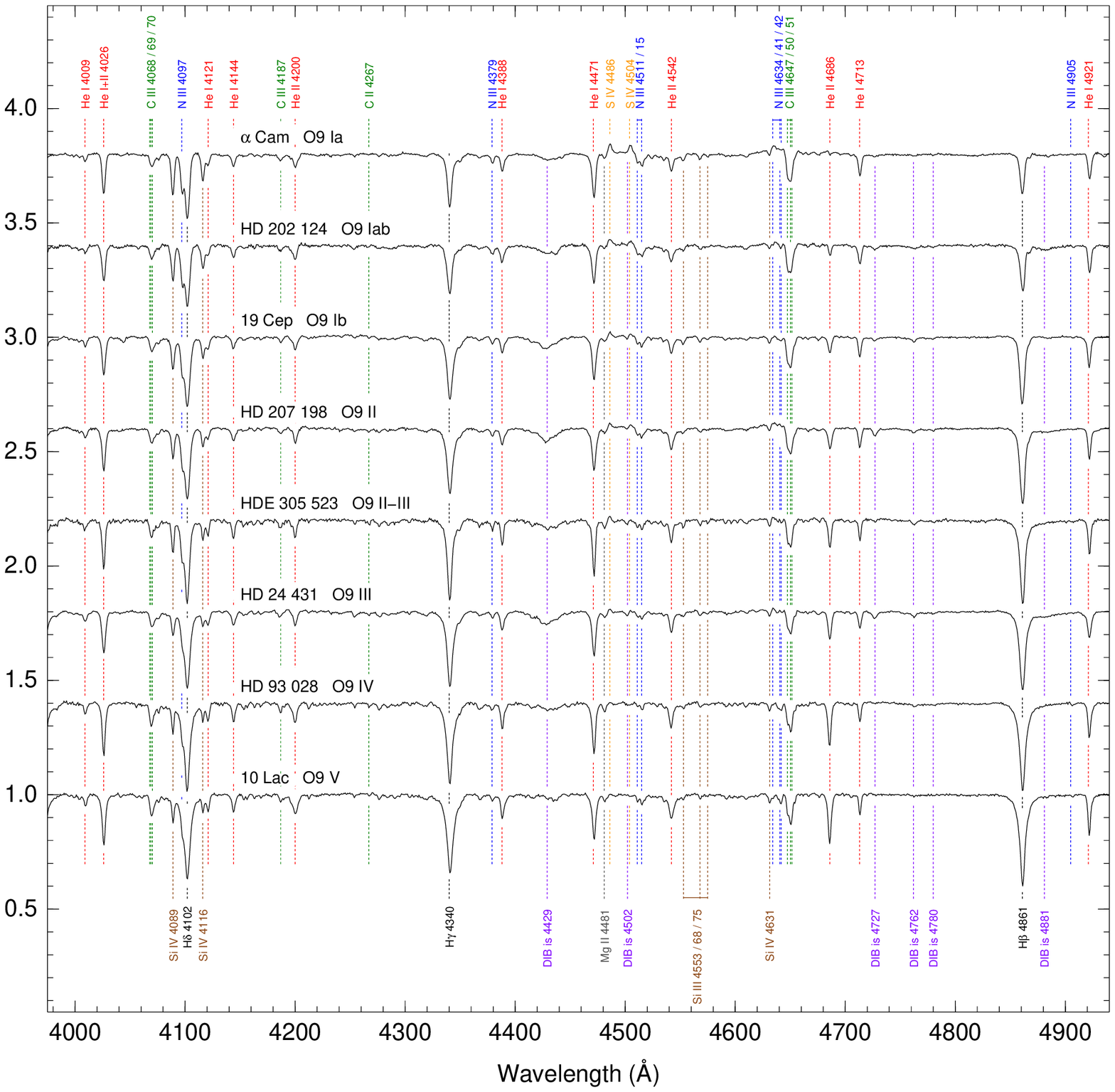}}
\caption{Luminosity effects at spectral type O9. [See the electronic version of the journal for a color version of this figure.]}
\label{ST_O9}
\end{figure*}	

\subsubsection{Spectral-Type Criteria at O8--O9}

\begin{table}
\caption{Spectral-type criteria at types O8.5--B0 (comparisons between absorption-line pairs).}
\small
\begin{tabular}{lcc}
\\
\hline
\multicolumn{1}{c}{Spectral} & \HeII{4542}/\HeI{4388}          & Si\,{\sc iii}~$\lambda 4552$/\HeII{4542} \\
\multicolumn{1}{c}{type}     & and                             &                                          \\
                             & \HeII{4200}/\HeI{4144}          &                                          \\
\hline
O8                           & $>$                             & N/A                                      \\
O8.5                         & $\geq$                          & N/A                                      \\
O9                           & $=$                             & $\ll$                                    \\
O9.5                         & $\leq$                          & $<$                                      \\
O9.7$^a$                     & $<$                             & $\leq$ to $\geq$                         \\ 
B0                           & $\ll$                           & $\gg$                                    \\
\hline
\multicolumn{3}{l}{$^a$ Now used at all luminosity classes.} \\
\end{tabular}
\label{O8.5_B0}
\end{table}

The primary horizontal classification criterion for the O stars has been
the helium ionization ratio \HeII{4542}/\HeI{4471}.  It
has a value of unity at type O7 and is very sensitive toward either side.
However, when the ratio becomes very unequal, its estimation is more
difficult; nevertheless, it has been applied throughout the O-type
sequence.  At the earliest types, there is no comparable absorption-line alternative, 
but at late O types the ratios \HeII{4542}/\HeI{4388} and 
\HeII{4200}/\HeI{4144} are very sensitive.  In previous
work, they have been allowed to increase with luminosity class at a given
spectral type, but here we adopt them as the primary spectral-type
criteria at types O8.5--B0 and define type O9 by values of unity in
both of these ratios. As a result, there may be small systematic
differences between the present and previous classifications (although
\citealt{Walbetal00} had already adopted these procedures), and the   
spectral types of some fundamental standards have been revised, e.g.,
$\alpha$~Cam and 19~Cep from O9.5 to O9. It is believed that these new
definitions will yield more reproducible and consistent classifications
for late O stars. The definition of type O9.7 remains \HeII{4542} equal to 
Si\,{\sc iii} $\lambda$4552, with a range from slightly greater to slightly less than allowed.
This spectral type was formerly used only for luminosity classes higher than
III, but here it has been newly applied at the lower luminosity classes as well, to improve the overall
consistency at late-O types. Thus, three standard stars have been moved: HD 189\,957 from O9.5 III to O9.7 III,
$\upsilon$ Ori from B0 V to O9.7 V, and $\tau$ Sco from B0.2 V to B0 V. It is expected that this
redefinition will increase the number of stars classified as O by moving previous B0 V to III objects
to the O9.7 V to III categories. The criteria at types O8.5--B0 are summarized in 
Table~\ref{O8.5_B0}.

\subsubsection{Luminosity-Class Criteria}

\begin{table}
\caption{O8-O8.5 \HeII{4686} luminosity criterion.}
\begin{tabular}{lcc}
\\
\hline
\multicolumn{1}{c}{Lum.}  & O8                  & O8.5                   \\
\multicolumn{1}{c}{class} &                     &                        \\
\hline
Ia                        & strong emission     & weak emission          \\
Iab                       & weak emission       & neutral                \\
Ib                        & near neutral        & very weak absorption   \\
II                        & \multicolumn{2}{c}{weak absorption}          \\
III                       & \multicolumn{2}{c}{strong absorption}        \\
V                         & \multicolumn{2}{c}{very strong absorption}   \\
\hline
\end{tabular}
\label{O8_O8.5}
\end{table}

\begin{table}
\caption{O9-O9.7 luminosity criteria (comparisons between absorption-line pairs).}
\small
\begin{tabular}{lcc}
\\
\hline
\multicolumn{1}{c}{Lum.}  & \HeII{4686}/\HeI{4713}      & Si\,{\sc iv}~$\lambda 4089$/\HeI{4026} \\
\multicolumn{1}{c}{class} &                             &                                        \\
\hline
Ia                        &  $\sim 0$                   & $>$                                    \\
Iab                       &  $\ll$ to $<$               & $\geq$ to $\leq$                       \\
Ib                        &  $\leq$                     & $\leq$                                 \\
II                        &  $=$                        & $<$                                    \\
III                       &  $>$                        & $<$ to $\ll$                           \\
V                         &  $\gg$                      & $\ll$                                  \\
\hline
\end{tabular}
\label{O9_9.7}
\end{table}

The first luminosity classification for stars earlier than types O8--O9
was introduced by \citet{Walb71a,Walb73a}; it is based upon the selective   
emission \citep{Walb01} effects in \HeII{4686} and  
N\,{\sc iii}~$\lambda\lambda$4634--4640--4642, i.e., the Of effect.
It was in part based on the inference that the negative luminosity effect in the
corresponding absorption lines at late-O types is caused by emission
filling by the same effect. At late-O types, the increasing intensity of
the Si\,{\sc iv} lines at $\lambda\lambda$4089, 4116 relative to nearby He~I
lines provides an independent luminosity criterion (Table~\ref{O8_O8.5}). In some spectra, for
whatever reasons (e.g., companions, metallicity, resolution effects on
lines of different intrinsic widths...), these independent criteria can
be somewhat discrepant; examples can be seen in the present atlas and
sample. In the MK process, the general approach is to examine the entire
spectrum and adopt an ``average'' over all available criteria; if the
discrepancies are judged to be too great, a ``p'' (for peculiar) is
added to the spectral type.  Here, we have preferred to adopt the
behavior of \HeII{4686} as the primary luminosity criterion for 
definiteness, allowing some range in the Si\,{\sc iv} at a given class. The
values of the \HeII{4686}/\HeI{4713} ratio at spectral
types O8--O8.5 and O9--9.7 are given in Tables~\ref{O8_O8.5}~and~\ref{O9_9.7},
respectively; the corresponding morphology at   
earlier types is defined in the atlas. Again, it is believed that this
procedure will yield more reproducible and consistent luminosity classes;
the effects on the calibration remain to be investigated.

On this basis, we can now readily distinguish luminosity class IV at
spectral types O6--O8 in data of the present quality; these types were
previously little used if at all. Here the \HeII{4686} absorption 
is intermediate between those of classes V and III\footnote{Classes IV and II
and supergiant subclasses are not used at spectral types earlier than O6;
thus, there is a range in the appearance of \HeII{4686} at class III
for the earlier types.}. Inversely, many previously known and new examples 
of type O~Vz \citep{Walb09b}, in which \HeII{4686} absorption is   
stronger than any other He\,{\sc ii} or He\,{\sc i} lines, hypothesized to be caused by 
an ``inverse Of effect'' and possibly related to extreme youth, are readily 
seen in the atlas and normal sample.

\subsubsection{Ofc Stars}

As already reported by \citet{Walbetal10a}, this study has revealed a
new category of O-type spectra, denoted as Ofc, in which emission lines
of C\,{\sc iii} $\lambda\lambda$4647--4650--4652 reach intensities similar to
the adjacent ones of N\,{\sc iii} $\lambda\lambda$4634--4640--4642 that are
included in the definition of the Of category.  This phenomenon is
strongly peaked at spectral type O5 at all luminosity classes and, as 
discussed in the earlier paper, likely corresponds to a sharply defined 
sensitivity of the ionic level populations to the atmospheric parameters. 
Figure~\ref{Ofc} here presents the complete violet through green spectral range 
in the current sample of 8 Northern Ofc spectra, and the atlas illustrates the
behavior of these features at adjacent spectral types.  It can be seen
that in some of the hottest Ofc spectra, Si\,{\sc iv} $\lambda$4654 and C\,{\sc iv}
$\lambda$4658 become comparable to the C\,{\sc iii} (see also \citealt{Walbetal02b}).

A related notational point is the elimination of the ``+'' sign following
the ``f'', previously used to denote emission in Si\,{\sc iv} $\lambda\lambda$4089, 
4116.  That notation unfortunately created confusion with superluminosity, 
as used for late-O and early-B supergiants.  It is no longer regarded as
essential, as the Si\,{\sc iv} emission is now well established as a common
feature that responds to temperature and gravity in normal O-type
spectra, and many other selective emission features are being identified
\citep{Walb01,WernRauc01,Cortetal09}. The degrees of the f-parameter itself are left
unchanged and they are still defined in terms of the {\em qualitative} appearance of 
\HeII{4686} and N\,{\sc iii} $\lambda\lambda$4634--4640--4642 combined,
e.g., absorption or emission in the former (see Table~\ref{qualifiers} for details).

\subsection{Peculiar Categories}

	In this subsection we describe the characteristics and membership in the
sample of this paper of the different peculiar categories of O stars. The spectral
classifications of this and the next subsection are shown in Table~\ref{spectralclas}. 
Stars within these two subsections and in Table~\ref{spectralclas} are sorted by their GOS ID 
(see \citealt{Maizetal04b}), whose first numbers correspond to the (rounded) Galactic longitude.

\subsubsection{Ofc Stars}
\label{sec:Ofc}

\begin{figure*}
%\centerline{\includegraphics*[width=\linewidth]{Ofc.ps}}
\centerline{\includegraphics*[width=\linewidth]{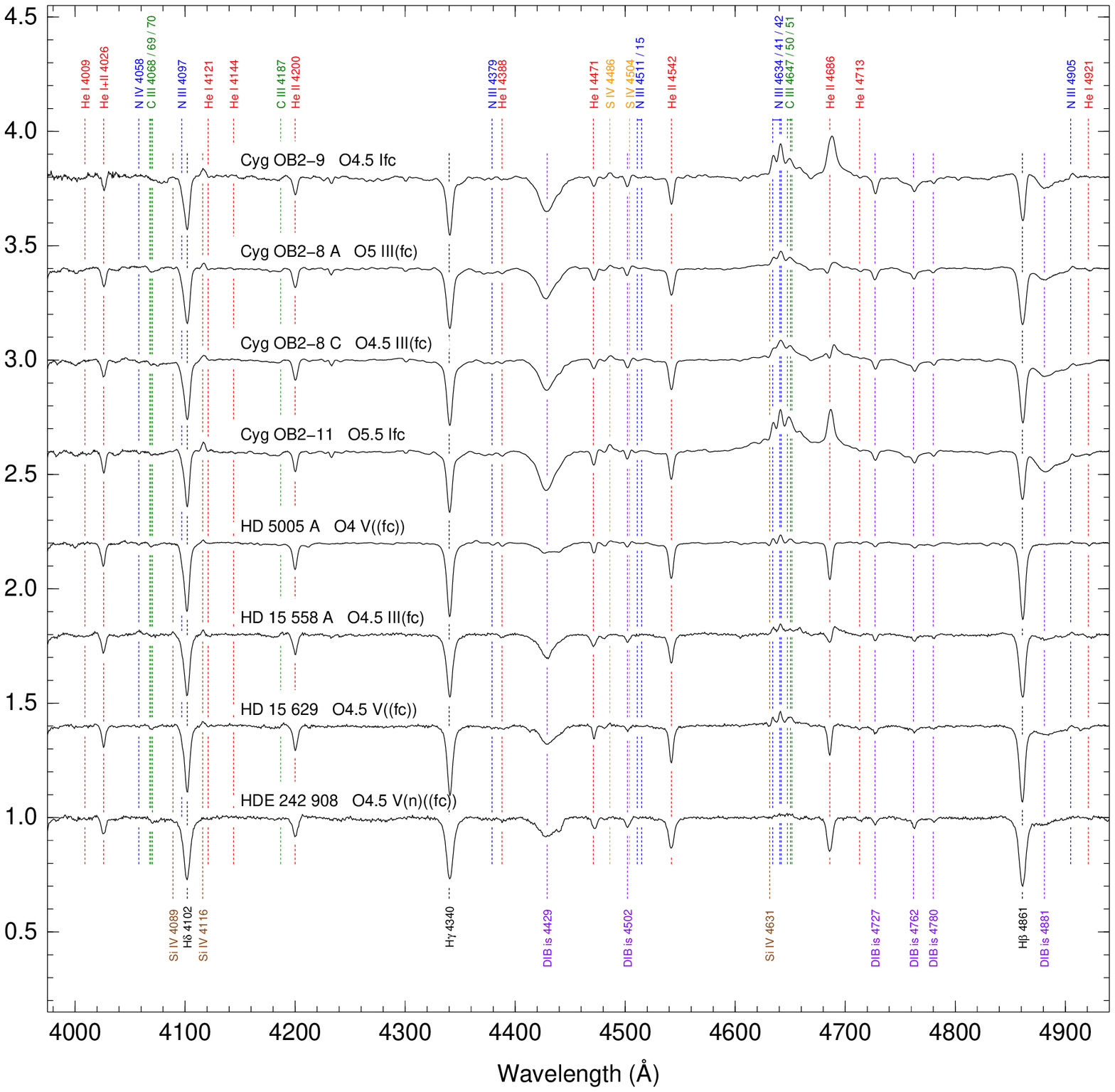}}
\caption{Spectrograms for Ofc stars. [See the electronic version of the journal for a color version of this figure.]}
\label{Ofc}
\end{figure*}

	This category was described in the previous subsection and by \citet{Walbetal10a}. The
spectrograms for the stars here are shown in Fig.~\ref{Ofc}. Note that the Ofc stars that are also SB2s are listed here instead of
in \ref{sec:SB2}.

\paragraph{Cyg OB2-9 = LS III +41 36 = [MT91]~431}

\object[NAME VI CYG 9]{}
This object is a single-lined spectroscopic binary and a non-thermal radio source with a binary
period of 2.35 years deduced from radio data \citep{VanLetal08}. The period was confirmed with 
optical data by \citet{Nazeetal08c} and the first orbital solution was provided by \citet{Nazeetal10}.
See Fig.~\ref{chart2} for a chart. 

\paragraph{Cyg OB2-8 A = BD +40 4227 = [MT91]~465}

\object[NAME VI CYG 8A]{}
\object[NAME VI CYG 8]{}

\citet{DeBeetal04} identified this system as an O6 + O5.5 spectroscopic binary. In our $R$~$\sim$~2500 data we are unable to 
separate the two components but the composite spectrum shows broad lines. Nevertheless, at the original resolution of our CAHA 
data ($R$~$\sim$~3000) we do see double lines and we can assign spectral types to this system of O5.5 III (fc) + O5.5 III (fc).
See Fig.~\ref{chart1} for a chart. 

\paragraph{Cyg OB2-8 C = LS III +41 38 = [MT91]~483}

\object[NAME VI CYG 8C]{}
\object[NAME VI CYG 8]{}
Note that the current version of the WDS catalog has Cyg OB2-8 C and D interchanged with respect to the most common usage.
See Fig.~\ref{chart1} for a chart. 

%\paragraph{Cyg OB2-11 = BD +41 3807 = [MT91]~734}

\object[NAME VI CYG 11]{}

\paragraph{HD 5005 A}

\object[HD 5005 A]{}
\object[HD 5005]{}
We obtained individual spectrograms for the four bright components in this system (A, B, C, and D) and we found all of them to
be O stars. B, C, and D are located at separations from A of 1\farcs529, 3\farcs889, and 8\farcs902, respectively \citep{Maiz10}.
The AB components are blended in all previous observations to our knowledge, resulting in a mid-O spectral type. We have deconvolved 
them spatially, with the remarkable results of an early-Ofc type for A and a late-O for B.
The strong C\,{\sc iii} $\lambda\lambda$4647--4650--4652 absorption in the latter eliminates the former's emission in this feature from
the composite spectrum. This system demonstrates the importance of spatial resolution for the analysis of O stars and provides a caution 
for more distant objects. The A component in this system appears unresolved in \citet{Masoetal09}.
See Fig.~\ref{chart2} for a chart. 
%In IC 1590.

\paragraph{HD 15\,558 A}

\object[HD 15558 A]{}
\object[HD 15558]{}
The B component is located at a separation of 9\farcs883 and a $\Delta m$ of 2.81 magnitudes in the $z$ band and
turned out to have an early-B spectral type. 
\citet{DeBeetal06a} find A to be a double-lined spectroscopic binary with spectral
types O5.5 III(f) + O7 V and they suggest that it could be a triple because the minimum mass is very large. Our spectrograms show no
evidence of multiple velocity components but the observed lines are broad.
See Fig.~\ref{chart3} for a chart. 
%BD +60 513 and BD +60 497 are too distant
%\citet{Turnetal08}.
%RXB: SB2O+O.
%Constant or SB1 according to \citet{Hilletal06}.
%In IC 1805.

\paragraph{HD 15\,629}

\object[HD 15629]{}
The spectrum is nearly identical to that of HD 5005 A.
See Fig.~\ref{chart3} for a chart. 
%BD +60 513 and BD +60 497 are too distant
%Constant radial velocity: \citet{Hilletal06}.
%\citet{DeBeetal06a}.
%In IC 1805.

\paragraph{HDE 242\,908}

\object[HDE 242908]{}
See Fig.~\ref{chart4} for a chart. 
%In NGC 1893.

\subsubsection{ON/OC Stars}
\label{sec:ON/OC}

The relative intensities of the N\,{\sc iii}~$\lambda\lambda$4634, 4640
and C\,{\sc iii}~$\lambda$4650 features are well delineated in the ON
spectra \citep{Walb76,Walb03} at all luminosity classes with the present  
observational parameters, as shown in Figure~\ref{ONOCa}.  Several previously  
marginal cases have become clear here, and some new ones have been added.   
We recall that cases with the N\,{\sc iii}~$\lambda$4640 blend stronger than
C\,{\sc iii}~$\lambda$4650 are classified ON, while those with the former weaker
than the latter, but still much stronger than in normal spectra, are
denoted as ``Nstr'' (for N strong).

The different degrees of line broadening among these spectra are
consistently specified in the classifications.  
%In particular, we note                                                             %Southern sample
%the discovery of two new ONnn stars, HD~102\,415 and HD~117\,490.  They are
%very similar to HD~191\,423, which was extensively analyzed by Howarth \& 
%Smith (2001), who determined that is has the highest rotational velocity 
%known for an O-type star, 400~km~sec$^{-1}$.  
The relationship between 
rotational velocity and surface nitrogen enrichment in massive stars is a 
subject of considerable current interest \citep{MaedMeyn00,Huntetal08b,Huntetal09}.
Two rapidly rotating ON giants are contained in this paper, HD 13\,268 and HD 191\,423;
a number of others have been found in our southern sample and will be discussed subsequently.
%The ONnn stars are extreme in both regards and are thus
%of special interest in that connection. Three ONn/(n) objects in Figure~\ref{ONOCa}
%may be less extreme or higher inclination examples.  In fact, it may be
%significant that all the ON stars of luminosity classes III and II here 
%are rapid rotators.

The OC spectra are perhaps somewhat less striking in these data, because
the resolution is marginal to demonstrate the salient deficiency of 
N\,{\sc iii}~$\lambda$4097 in the blueward wing of H$\delta$.  That nitrogen line has a
comparable depth to Si\,{\sc iv}~$\lambda$4089 or even the Balmer line itself,
in normal and ON supergiant spectra. C\,{\sc iii}~$\lambda$4650 is stronger in
OC than in normal spectra of the same types.  Less extreme cases are 
denoted as ``Nwk'' (for N weak).  

Note that the ON/OC stars that are also SB2s are listed here instead of
in \ref{sec:SB2}.

\paragraph{BD +36 4063}

\object[BD +36 4063]{}
This object is an interacting binary (\citealt{Willetal09b}, see also
{\tt http://www. lowell.edu/workshops/Contifest/abstracts. php?w=Howarth}).
Its ON nature was discovered by \citet{Math89}

\paragraph{HD 201\,345}

\object[HD 201345]{}
The protoype late-ON dwarf has been reassigned luminosity class IV here. It was suggested
to be a SB by \citet{Lest73}.

\paragraph{HD 191\,423}

\object[HD 191423]{}
This object is the most rapid rotator of type O known to date \citep{HowaSmit01}.
It has the prototype ONnn spectrum \citep{Walb03}.

\paragraph{HD 191\,781}

\object[HD 191781]{}
This is the prototype late-ON supergiant.

%\paragraph{HD 12\,323}

\object[HD 12323]{}
%SB1O.

\paragraph{HD 13\,268}

\object[HD 13268]{}
This object was not present in version 1 of GOSC. Its ONn nature was discovered by \citet{Math89}.
%\citet{DeBeetal08}: Variable on short time scales.

%\paragraph{HD 14\,633}

\object[HD 14633]{}
%SB1O.

\paragraph{$\delta$ Ori AaAb = Mintaka AaAb = \\ HD~36\,486~AaAb}

\object[HD 36486 AaAb]{}
\object[HD 36486 A]{}
\object[HD 36486]{}
This object is in the complex $\delta$ Ori system \citep{Harvetal02}. B and C are relatively distant while Ab is at a separation of
0\farcs325 from Aa with a $\Delta m$ of 1.48 in the $z$ band \citep{Maiz10}. Here we are unable to spatially separate the spectra 
of Aa and Ab. Aa is a double-lined spectroscopic binary: \citet{Harvetal02} use tomographic separation to give spectral types of 
O9.5 II and B0.5 III for Aa1 and Aa2, respectively. In our spectra we are unable to detect the double lines. Aa is also an
eclipsing binary with an amplitude of 0.097 magnitudes \citep{Lefeetal09}. The orbital elements of the AaAb orbit are given by 
\citet{Zascetal09}. The new Hipparcos calibration gives a revised distance of $221^{+33}_{-25}$ pc \citep{Maizetal08c}, substantially
less than that of the Orion association. 

.
%\citet{Masoetal98}: B, C components farther away than 30\arcsec\ (C is HD 36\,485).
%\citet{Maizetal04b} gave spectrum for AaAb.

\paragraph{$\zeta$ Ori A = Alnitak A = HD 37\,742}

\object[HD 37742]{}
We were able to extract the individual spectra of A and B (= HD 37\,743), separated by 2\farcs424 and with a $\Delta m$ of 2.424 
magnitudes in the $z$ band.
%\citet{Masoetal98} give B component as B2 III at 2\farcs42 with $\Delta m = 2.2$. Tycho-2 gives 
%$\Delta B$ = 1.7. Note that B is HD 37\,743.
%\citet{Turnetal08}.
The new Hipparcos calibration gives a distance of $239^{+43}_{-32}$ pc \citep{Maizetal08c}, consistent with that of $\delta$~Ori~AaAb,
indicating a small distance between these two objects of the Orion belt\footnote{The results for the third belt star, $\epsilon$~Ori
= Alnilam, place it farther away but with a much larger uncertainty.}. In some observations, the luminosity class of $\zeta$~Ori~A
appears as II.
\citet{Bouretal08a} detected a weak magnetic field and \citet{Lefeetal09} found an intrinsic variability with an amplitude of 
0.029 magnitudes.

\paragraph{HD 48\,279 A}

\object[HD 48279 A]{}
\object[HD 48279]{}
In high-resolution data currently under separate investigation, this spectrum appears as full-fledged ON,
i.e. with N\,{\sc iii}~$\lambda$4640 $>$ C\,{\sc iii}~$\lambda$4650, indicating that it may be variable.
We placed B (6\farcs860 away) on the slit and obtained an F spectral type for that component.
See Fig.~\ref{chart6} for a chart.
%\citet{Masoetal98} give a B component at 6\farcs66 with $\Delta m = 2.7$, CHECK.
%A third component with $\Delta m = 1.1$ at 35\farcs8 is HD 288\,966, which Simbad lists as spectral type A.
%\citet{Turnetal08}.
%\citet{Mahyetal09} detect no RV variations and give a spectral type of ON7.5 V.

\subsubsection{Onfp Stars}
\label{sec:Onfp}  

The Onfp category was defined by \citet{Walb72,Walb73a} to describe Of
spectra displaying \HeII{4686} emission with an absorption reversal.  
Independently of that characteristic, nearly all of them have broadened 
absorption lines indicative of rapid rotation, as denoted by the ``n''.  
\citet{ContLeep74} designated such spectra as Oef, suggesting a 
relationship to the Be stars. \citet{Walbetal10b} have investigated a 
sample of these objects in the Magellanic Clouds, listing only eight known 
Galactic counterparts; several new ones are reported here. The properties of the category are extensively 
discussed in that paper and will not be repeated here.
The spectrograms for the present stars in this category are shown in Fig.~\ref{Onfpa}.
Note that the Onfp stars in this paper that are also SB2s are listed here instead of
in \ref{sec:SB2}.

%Two of the previous Galactic objects, HD~152\,248 and HD~192\,281, do not             %Southern sample
%show Onfp characteristics in the present data and are not included in the
%category here. HD~152\,248 was investigated at high resolution by Sana et
%al. (2001), who showed that it is a double-lined spectroscopic binary with
%both stellar $\lambda$4686 components in absorption and likely
%colliding-wind emission.  
One of the previous Galactic objects, HD~192\,281, does not     
show Onfp characteristics in the present data and is not included in the
category here. HD~192\,281 has a weak P~Cygni profile at
\HeII{4686} here (i.e., no emission blueward of the absorption component),
although variability cannot be entirely ruled out; see also \citet{DeBeRauw04}.
%On the other hand, we have found four additional Onfp spectra                     %Southern sample
%in the present sample, which are included in Figure~12: HD~1337 (AO~Cas;
%Bagnuolo et al. 1999), HD~47\,129 (Plaskett's Star; Linder et al. 2008),
%HD~117\,797, and HD~167\,971 (De Becker et al. 2005). Interestingly,
%three of them are known spectroscopic binaries, indicating that such
%systems may comprise a significant fraction of the Onfp category, and two
%of them are only intermittently Onfp as shown.  Also, exceptionally, two 
%of the new objects do not have broadened absorption lines; while one (AO~Cas) 
%is not formally Of because of its late spectral type.  

AO~Cas and MY~Ser are only intermittently ``Onfp'' as shown, while
\citet{Lindetal08} show complex, variable \HeII{4686} profiles
as a function of phase in HD~47\,129. Exceptionally, HD 47\,129 and MY~Ser do not
have broadened absorption lines, while AO Cas is not formally Of because
of its late spectral type. It is noteworthy that three of the five new Onfp spectra reported here correspond
to well known spectroscopic binaries.

\paragraph{HD 175\,754}

\object[HD 175754]{}
The weak Onfp \HeII{4686} profile noted here was first discovered in a high-resolution study in progress.

\paragraph{MY Ser = HD 167\,971}

\object[HD 167971]{}
This system was classified as O8 I + O5-8 V + O5-8 V by \citet{Leitetal87}. \citet{DeBeetal05} analyzed XMM observations and
suggested that the X-ray emission originates in the interaction between the winds of the two main-sequence stars, with the
supergiant located further away. \citet{Lefeetal09} found eclipses with an amplitude of 0.237 magnitudes. In our data we
detect that the system is at least a SB2, with a main O8 Iafp component and a secondary O4/5 spectrum. The radio emission was studied
by \citet{Blometal07}. This system could not be resolved with HST/FGS at $\sim$10 mas (Ed Nelan, private communication). 
See Fig.~\ref{chart1} for a chart. 
%In NGC 6604.

\paragraph{V442 Sct = HD 172\,175}

\object[HD 172175]{}
This star was suggested to be Onfp by \citet{Walb82a} and is clearly confirmed here.
%RXB: Variable spectrum or SB1?

\paragraph{$\lambda$ Cep = HD 210\,839}

\object[HD 210839]{}
This star is one of the original, prototype Onfp objects.
The new Hipparcos calibration gives a revised distance of $649^{+112}_{-83}$ pc \citep{Maizetal08c}.
%RXB: Variable spectrum.
%\citet{Lefeetal09}: Intrinsic variability amplitude of 0.036 magnitudes.

\paragraph{BD +60 2522}

\object[BD +60 2522]{}
This star is one of the original, prototype Onfp objects.

\paragraph{AO Cas = HD 1337}

\object[HD 1337]{}
This object is an eclipsing binary with an amplitude of 0.198 magnitudes \citep{Lefeetal09}.
We were able to detect its SB2 character, as well as weak emission wings at \HeII{4686} in one observation,
leading to its association with the Onfp category.

\paragraph{HD 14\,442}

\object[HD 14442]{}
This object was studied by \citet{DeBeRauw04}.

\paragraph{HD 14\,434}

\object[HD 14434]{}
This object was studied by \citet{DeBeRauw04}.

\paragraph{HD 47\,129 = Plaskett's star}

\object[HD 47129]{}
This object is a well known SB2; \citet{Lindetal08} give spectral types of O8 III/I + O7.5 III and suggest it is a binary system in a post 
RLOF stage. We do not clearly detect double lines in our spectra.
%Previously undetected companion in AstraLux data (uncertain). Preliminary values: PA of 252.3, separation of
%1\farcs23, $\Delta z = 4.5$, maybe one in  \citet{Turnetal08}.
%SB2O. 1 OSN, SB2 visible only in H beta and in line asymmetries.

\subsubsection{Of?p Stars}
\label{sec:Of?p}  

This class of objects was recently discussed by \citet{Walbetal10a}. 
Magnetic fields have been detected on three of the five known Galactic members of the class.
The spectrograms for the stars from the present sample in this category are shown in Fig.~\ref{Ofp}.

\paragraph{HD 191\,612}
 
\object[HD 191612]{}
This object is a magnetic oblique rotator with a 538 day period \citep{Wadeetal10} and a SB2 with a period of 1542 days \citep{Howaetal07}.
The latter reference estimates the spectral type of the secondary as B1. 
In our 2007 data the star appears in the O8 ``minimum'' state of its rotational cycle and in our 2009 data at the O6 ``maximum''. 
%Of?p, also a SB2.
%SB2O. 3 OSN (OSN Sep 09 included), clearly variable, SB2?
%RXB: variable spectrum.
%\citet{Walbetal03}.
%\citet{Walbetal04a}.
%\citet{Nazeetal07}.
%\citet{Howaetal08}.

\paragraph{HD 108}

\object[HD 108]{}
The object appears as O8 in both our 2007 and 2009 observations. That is not surprising since it is currently at the mimimum of its $\sim$50 year
magnetic/rotational cycle \citep{Martetal10}.
%\citet{Nazeetal04b}: Large optical variations but with a stable X-ray spectrum.
%RXB: SB1.

\paragraph{NGC 1624-2 = MFJ Sh 2-212 2 = \\ 2MASS~04403728+5027410}

\object[NGC 1624-2]{}
This object was not present in version 1 of GOSC.
Its Of?p character was detected by \citet{Walbetal10a}. Previous classifications were given by \citet{Moffetal79} and
\citet{ChinWink84}.
We placed NGC 1624-9 on the slit and obtained an F spectral type for that component.
See Fig.~\ref{chart5} for a chart.
%In Sh 2-212.

\subsubsection{Oe Stars}
\label{sec:Oe}

The properties of Oe stars are discussed by \citet{Neguetal04}.
The spectrograms for our stars in this category are shown in Fig.~\ref{Oe}.

\paragraph{HD 17\,520 B}

\object[HD 17520 B]{}
\object[HD 17520]{}
This object was not present in version 1 of GOSC. 
It is separated by only 0\farcs316 from the A component with a $\Delta m$ of 0.67 mag in the $z$ band \citep{Maiz10}.
We were able to separate the spectra of A and B and we detected that the emission lines that make the integrated spectrum
have an Oe type \citep{Walt92,Hilletal06} originate in B. There is also evidently strong He\,{\sc i} emission that gives 
those line a double appearance. 
See Fig.~\ref{chart3} for charts.
%In IC 1848.
%Is this the Oe component?

\paragraph{X Per = HD 24\,534}

\object[HD 24534]{}
We draw attention to the surprisingly different He\,{\sc i} profiles within the individual Oe spectra, e.g. the double emission 
only in \HeI{4713} here.
%SB1O.
%\citet{Neguetal04}: B0 Ve.
%RXB: Variable spectrum.
%\citet{Lefeetal09}: Intrinsic variability amplitude of 0.044 magnitudes.

\paragraph{V1382 Ori = HD 39\,680}

\object[HD 39680]{}
Note the double He\,{\sc i}~$\lambda\lambda$4713,5016 lines in this spectrum and the absence of such profiles in other He\,{\sc i} lines.
%\citet{Masoetal98} give a component at 45\farcs7. It is HD 39\,700, which Simbad gives as an A0 star, CHECK.
%\citet{Neguetal04}: O8.5 Ve.
%\citet{Lefeetal09}: Intrinsic variability amplitude of 0.131 magnitudes.

\paragraph{HD 45\,314}

\object[HD 45314]{}
\citet{Masoetal98} indicate the existence of a B component at a separation of 0\farcs05. However, the $\Delta m$ is not given, so 
its presence is not included in the object name.
%RXB: Variable spectrum.
%\citet{Neguetal04}: B0 IVe.
%\citet{Lefeetal09}: Intrinsic variability amplitude of 0.043 magnitudes.
%\citet{Vinketal09}: Polarization in H$\alpha$.

\paragraph{HD 60\,848}

\object[HD 60848]{}
Note the double He\,{\sc i}~$\lambda\lambda$4713,5016 lines in this spectrum and the absence of such profiles in other He\,{\sc i} lines.
%RXB: Variable spectrum.
%\citet{Neguetal04}: O9.5 IVe.
%\citet{Lefeetal09}: Intrinsic variability amplitude of 0.054 magnitudes.
%\citet{Vinketal09}: Polarization in H$\alpha$.

\subsubsection{Double- and triple-lined spectroscopic binaries}
\label{sec:SB2}

	In the last part of this subsection we include the double- (SB2) and triple- (SB3) lined spectroscopic 
binaries that do not belong to any of the other peculiar categories. More than for any other peculiar categories,
membership here is determined by spectral resolution and time coverage, given the large ranges of velocity differences 
and periods existent among massive spectrocopic binaries. Therefore, we have included in this category examples that
have been identified as SB2s or SB3s by other authors (in most cases using higher-resolution spectroscopy) but that are 
single lined in our spectra. In those cases, we point to the relevant reference. Our classifications 
were obtained with MGB varying seven input parameters: the spectral types, luminosity classes, and velocities of both the
primary and secondary; and the flux fraction of the secondary (see Fig.~\ref{SB2} for sample final fits).

\paragraph{HD 167\,771} 

\object[HD 167771]{}
We were able to detect the SB2 character of this object \citep{MorrCont78,Sticetal97}.

\paragraph{HD 168\,075}

\object[HD 168075]{}
This binary has a 43.6 day period 
and has been classifed as O6.5 V ((f)) + B0-1 V \citep{Gameetal08a,Barbetal10,Sanaetal09}. 
We do not detect double lines in our spectra.
See Fig.~\ref{chart1} for a chart. 
%\citet{Boscetal99b}, \citet{Evanetal05}: SB1 with O6-7 V ((f)) + B0: from the composite spectrum.
%SB2O (previous to \citet{Sanaetal09}, SB1?).
%In NGC 6611.

\paragraph{HD 166\,734}

\object[HD 166734]{}
This system is a SB2 with a spectral classification of O7 Ib(f) + O8-9 I given by \citet{Walb73a}. The orbit is analyzed by \citet{Contetal80} and 
the eclipses are described by \citet{OterWils05}. We do not see double lines in our spectra though for one epoch the line profiles are clearly
asymmetric.

\paragraph{HD 191\,201 A}

\object[HD 191201 A]{}
\object[HD 191201]{}
This object has another component (B) at a separation of 0\farcs97 with $\Delta m = 1.8$ \citep{Masoetal09}. We were able to
spatially separate the A and B spectra. The latter is of early-B type while the former shows double lines with spectral types 
O9.5 III and B0 IV.
%\citet{Masoetal09}: B component at 0\farcs97 with $\Delta m = 1.8$, CHECK. Dim component at 5\farcs4.
%In NGC 6871.
%\citet{Burketal97}: O9.5 V-III + O9.5 V-III

\paragraph{HDE 228\,766}

\object[HD 228766]{}
We were able to detect the SB2 character of this object \citep{Walb73e,MassCont77,Rauwetal02}.
%Also an OIafpe.
%SB2O. 3 OSN (OSN Sep 09 included), possibly SB2 but confusion between focus and S/N effects.
%\citet{Rauwetal02}.

\paragraph{Y Cyg = HD 198\,846}

\object[HD 198846]{}
We were able to detect the SB2 character of this object \citep{Burketal97}.
%SB2OE. 3 OSN (OSN Sep 09 included), excellent SB2 in last one.
%\citet{Burketal97}: O9 V + O9.5 V

\paragraph{Cyg OB2-5 A = V279 Cyg A = \\ BD~+40~4220~A}

\object[NAME VI CYG 5A]{}
\object[NAME VI CYG 5]{}
This object has a B component at a separation of 0\farcs934 A with a $\Delta m$ of 3.02 magnitudes in the $z$ band \citep{Maiz10},
which we are able to spatially separate in our data (the B component appears to be a mid-O star but is not 
included in this paper because of the low S/N of its spectrogram). Cyg~OB2-5~A is a peculiar and likely contact binary with well-marked 
eclipses between the two O supergiants \citep{Lindetal09}. \citet{Kennetal10} suggest the existence of a fourth 
component\footnote{Note that their D component is B here.} besides B (which is an O giant, see below) and the 
two unresolved O supergiants in A. Our data show variability between epochs but the spectra are too peculiar to give two accurate
spectral types for the A components. This spectrum was classified as O7 Ianfpe by \citet{Walb73a}, which has led to confusion with
the Onfp category, but such membership was not intended. Moreover, our data show that both components have narrow lines, which
are essentially equal.

\paragraph{HD 195\,592}

\object[HD 195592]{}
\citet{DeBeetal10}: indicate that this is a SB2 with O9.7 I + B spectral types and a 5.063 day period and that the system could 
harbor a third component with a $\sim$20 day period. We do not see double lines in our spectra.
%\citet{Lefeetal09}: Intrinsic variability amplitude of 0.062 magnitudes.

\paragraph{HD 199\,579}

\object[HD 199579]{}
\citet{Willetal01} detect this object as a SB2 with spectral types O6 V((f)) + B1-2 V. We do not see double lines in our
spectra.

\paragraph{HD 206\,267 AaAb}

\object[HD 206267 AaAb]{}
\object[HD 206267 A]{}
\object[HD 206267]{}
This system has two components, Aa and Ab, unresolved here with a separation of 0\farcs118 and a $\Delta m$ of 1.1
\citep{Masoetal09}. 
We were able to detect its SB2 character.
See Fig.~\ref{chart2} for a chart. 
%\citet{Masoetal98}: B component too dim to have an effect on spectrum.
%\citet{Lefeetal09}: Intrinsic variability amplitude of 0.049 magnitudes.
%\citet{Burketal97}: O6.5 V((f)) + O9.5: V
%In IC 1396 (Trumpler 37).

\paragraph{LZ Cep = HD 209\,481}

\object[HD 209481]{}
This object is a SB2 with a period of 3.07 days
\citep{Howaetal91} that experiences eclipses with an amplitude of 0.099 magnitudes
\citep{Lefeetal09}.
The new Hipparcos calibration gives a revised distance of $1027^{+244}_{-165}$ pc \citep{Maizetal08c}.

\paragraph{DH Cep = HD 215\,835}

\object[HD 215835]{}
We were able to detect the SB2 character of this object.
%\citet{Lindetal07}.
%In NGC 7380.
%\citet{Burketal97}: O5.5 III(f) + O6 III(f).

\paragraph{BD +60 497}

\object[BD +60 497]{}
We were able to detect the SB2 character of this object.
%O6.5 V ((f)) + O8.5-O9.5 V ((f)): \cite{RauwDeBe04}.
%SB2 character: \citet{Hilletal06}.
%In IC 1805.

\paragraph{HD 16\,429 A}

\object[HD 16429 A]{}
\object[HD 16429]{}
Our spectra include light from both the Aa and Ab components, separated by 0\farcs295 \citep{Maiz10}, but $\Delta m$ is larger
than 2.0, so the presence of the secondary spectrum is not included in the name. A third component, B, is located farther away 
(6\farcs777) and its light was easily separated from that of the primary. Ab is a SB2 \citep{McSw03}. That author gives spectral 
types of O9.5 II for Aa and O8 III-IV + B0 V? for Ab. We do not see double lines in our spectra.
We placed HD 16\,429 B on the slit and obtained an F spectral type for that component, in agreement with the HD catalog.
See Fig.~\ref{chart4} for a chart. 
%\citet{Turnetal08}.
%SB3O. 2 OSN (1 OSN Sep 09 included), SB2 not visible.
%\citet{Lefeetal09}: Intrinsic variability amplitude of 0.059 magnitudes.

\paragraph{HD 17\,505 A}

\object[HD 17505 A]{}
\object[HD 17505]{}
We were able to separate the spectrum from that of B, located at a separation of 2\farcs153 with a $\Delta m$ of 1.75 in the $z$
band \citep{Maiz10}. 
%\citet{Masoetal98}: B component at 2\farcs15 with $\Delta m = 1.7$, CHECK.
%\citet{Masoetal98}: Other faint components at distances beyond 15\arcsec. Note that G is really HD 17\,520 A.
%\citet{Turnetal08}.
\citet{Hilletal06} find this system to be SB3O, with spectral types O6.5 III ((f)) + O7.5 V ((f)) + O7.5 V ((f)). We were unable
to detect double lines in our spectra.
See Fig.~\ref{chart3} for charts.
%In IC 1848.

\paragraph{HD 18\,326}

\object[HD 18326]{}
We were able to detect the SB2 character of this object.

\paragraph{CC Cas = HD 19\,820}

\object[HD 19820]{}
\citet{Hilletal94} give spectral types of O8.5 III + B0 V for this double-lined spectroscopic binary. \citet{Lefeetal09} detect
eclipses with an amplitude of 0.108 magnitudes.
The SB2 nature of the object manifests itself in asymmetries in some of the lines in our spectra but we were unable to clearly
separate the effects of the two components to produce two spectral types.

\paragraph{LY Aur A = HD 35\,921 A}

\object[HD 35921 A]{}
\object[HD 35921]{}
We were able to separate the spectrum from that of B, located at a separation of 0\farcs598 with a $\Delta m$ of 1.87 in the $z$
band \citep{Maiz10}. This system is an eclipsing binary with an amplitude of 0.722 magnitudes \citep{Lefeetal09}
We were able to detect the SB2 character of this object.
%\citet{Maizetal04b} gave spectrum for AB.
%\citet{Turnetal08}.
%SB2OE. 1 OSN (+1 OSN Oct 09), SB2 not visible and poor focus.

\paragraph{HD 37\,366 A}

\object[HD 37366 A]{}
\object[HD 37366]{}
\citet{Boyaetal07a} determine that the this system is a SB2 with spectral types O9.5 V + B0-1 V. 
We were unable to detect double lines in our spectra.
We placed C on the slit and obtained an A spectral type for that component.

\paragraph{HD 48\,099}

\object[HD 48099]{}
\citet{Mahyetal10} determine that this SB2 has spectral types O5.5 V ((f)) + O9 V. 
We were unable to detect double lines in our spectra.

\paragraph{HD 46\,149}

\object[HD 46149]{}
\citet{Mahyetal09} estimate that this SB2 has spectral types O8 V + B0-1 V. 
We were unable to detect double lines in our spectra.
%In NGC 2244.

\paragraph{$\iota$ Ori = HD 37\,043}

\object[HD 37043]{}
An Ab component is located at a separation of 0\farcs13 \citep{Masoetal09} but it is too dim to have a significant effect in the
observed spectrum. This object may have been ejected from the Trapezium cluster \citep{Hoogetal00}. \citet{Sticetal87} give
spectral types of O9 III + B1 III for this SB2. 
We were unable to detect double lines in our spectra. However, the spectral type varies between O9 and O8,5, so we no longer use this
classical MK standard as such.
The new Hipparcos calibration gives a revised distance of $785^{+182}_{-124}$ pc \citep{Maizetal08c}.

\paragraph{HD 54\,662}

\object[HD 54662]{}
\citet{Boyaetal07a} attempted a tomographic reconstruction of the spectra of this system and could only determine it to within
O6.5~V + O7-9.5~V.
We were unable to detect double lines in our spectra.
%\citet{Vinketal09}: Polarization in H$\alpha$.

\paragraph{HD 53\,975}

\object[HD 53975]{}
\citet{Giesetal94} give spectral types of O7.5 V + B2-3 V for this SB2.
We were unable to detect double lines in our spectra.

\begin{figure*}
%\centerline{\includegraphics*[width=\linewidth]{ONOCa.ps}}
\centerline{\includegraphics*[width=\linewidth]{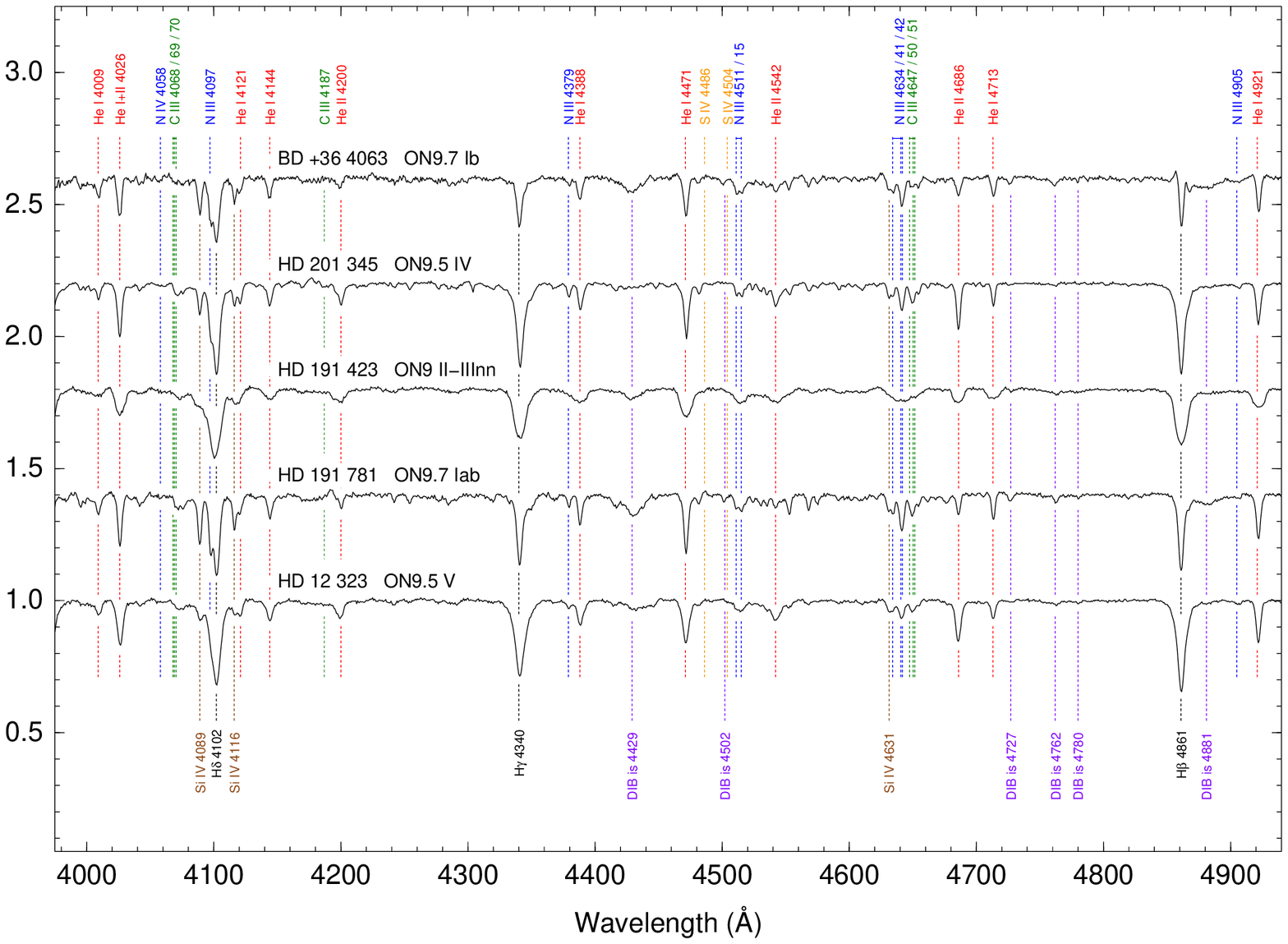}}
\caption{Spectrograms for ON/OC stars. [See the electronic version of the journal for a color version of this figure.]}
\label{ONOCa}
\end{figure*}	

\addtocounter{figure}{-1}

\begin{figure*}
%\centerline{\includegraphics*[width=\linewidth]{ONOCb.ps}}
\centerline{\includegraphics*[width=\linewidth]{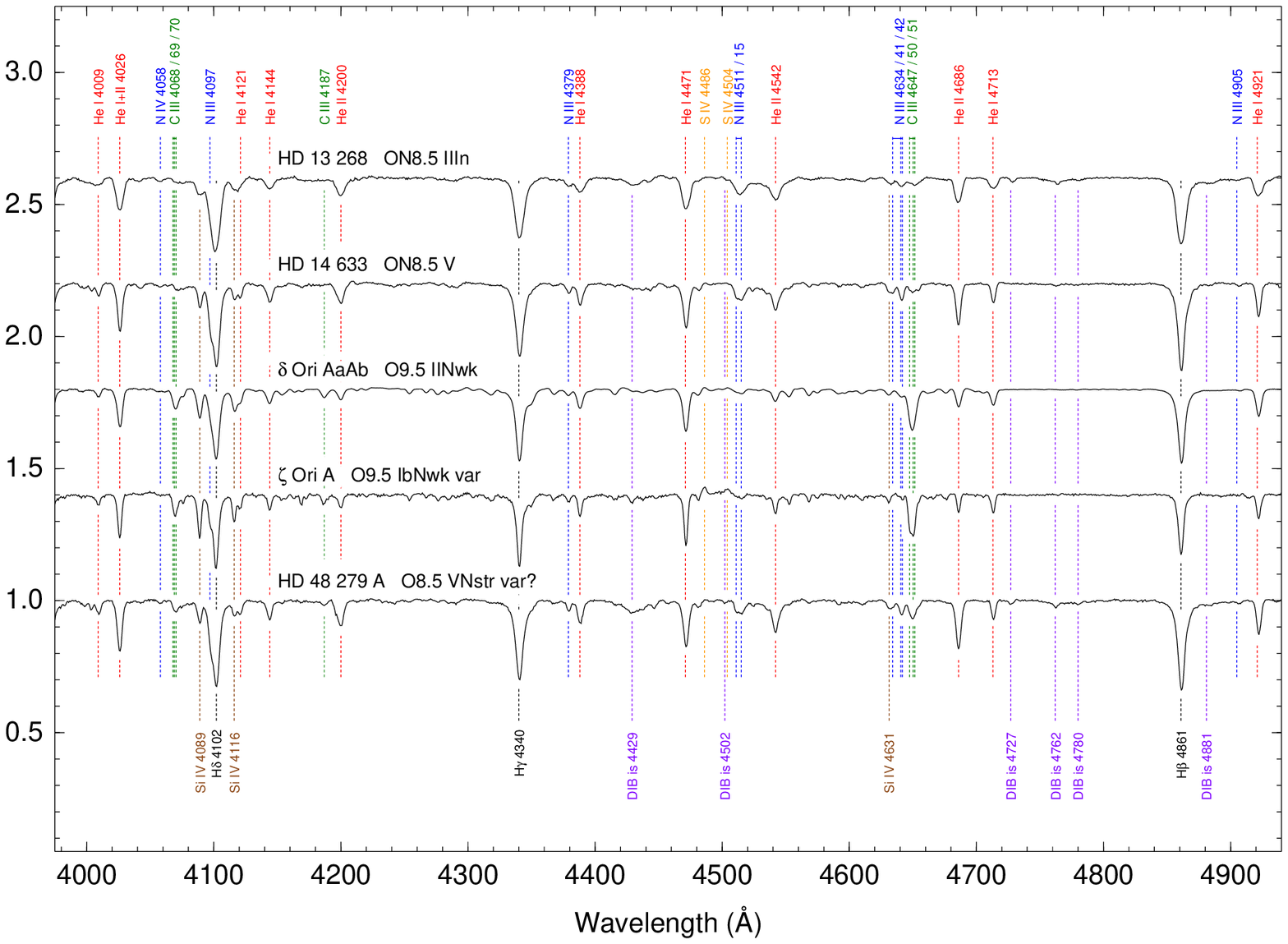}}
\caption{(continued).}
\label{ONOCb}
\end{figure*}	

\begin{figure*}
%\centerline{\includegraphics*[width=\linewidth]{Onfpa.ps}}
\centerline{\includegraphics*[width=\linewidth]{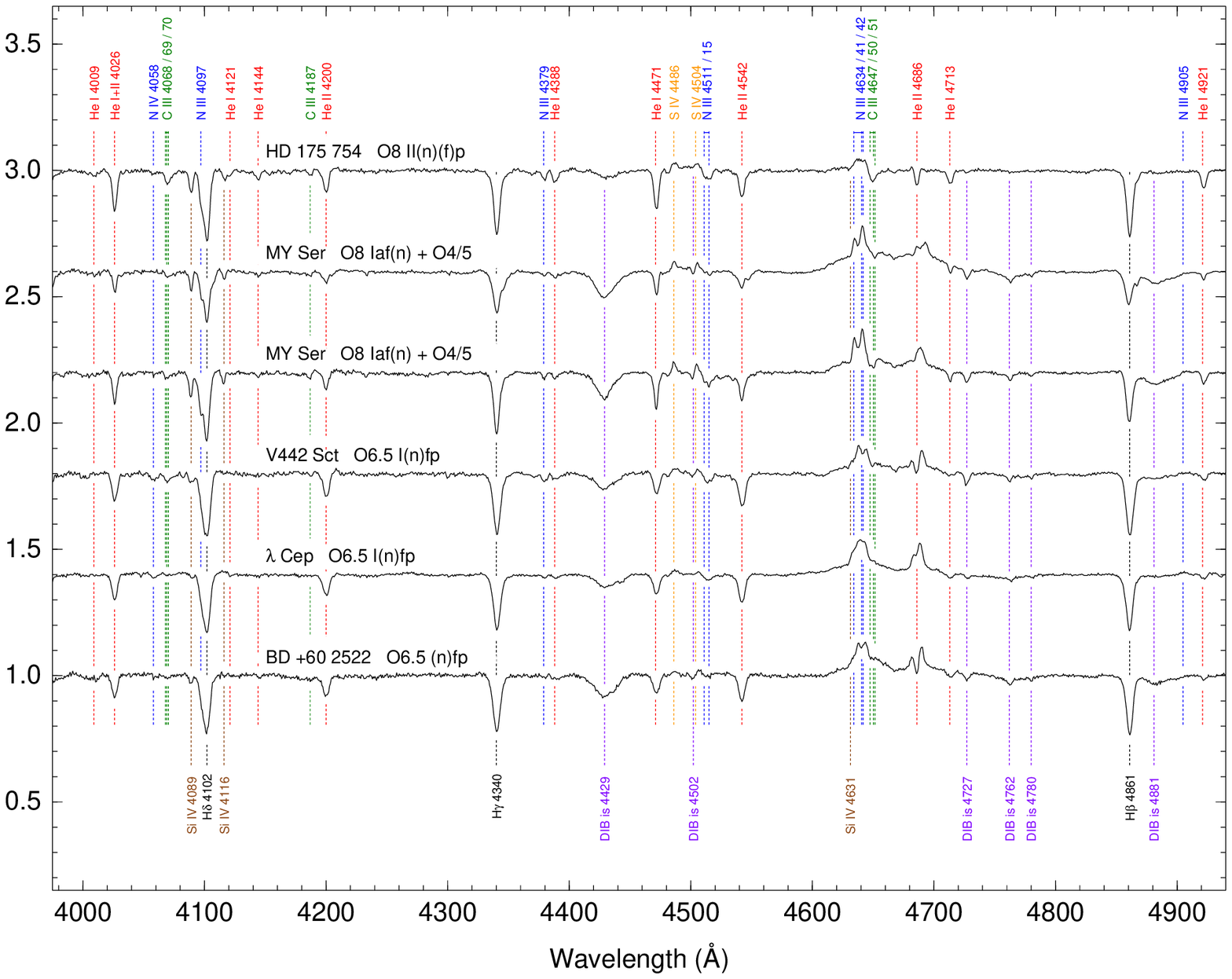}}
\caption{Spectrograms for Onfp stars. MY Ser and AO Cas are each shown in two different phases
(near quadrature and near conjunction). [See the electronic version of the journal for a color version of this figure.]}
\label{Onfpa}
\end{figure*}	

\addtocounter{figure}{-1}

\begin{figure*}
%\centerline{\includegraphics*[width=\linewidth]{Onfpb.ps}}
\centerline{\includegraphics*[width=\linewidth]{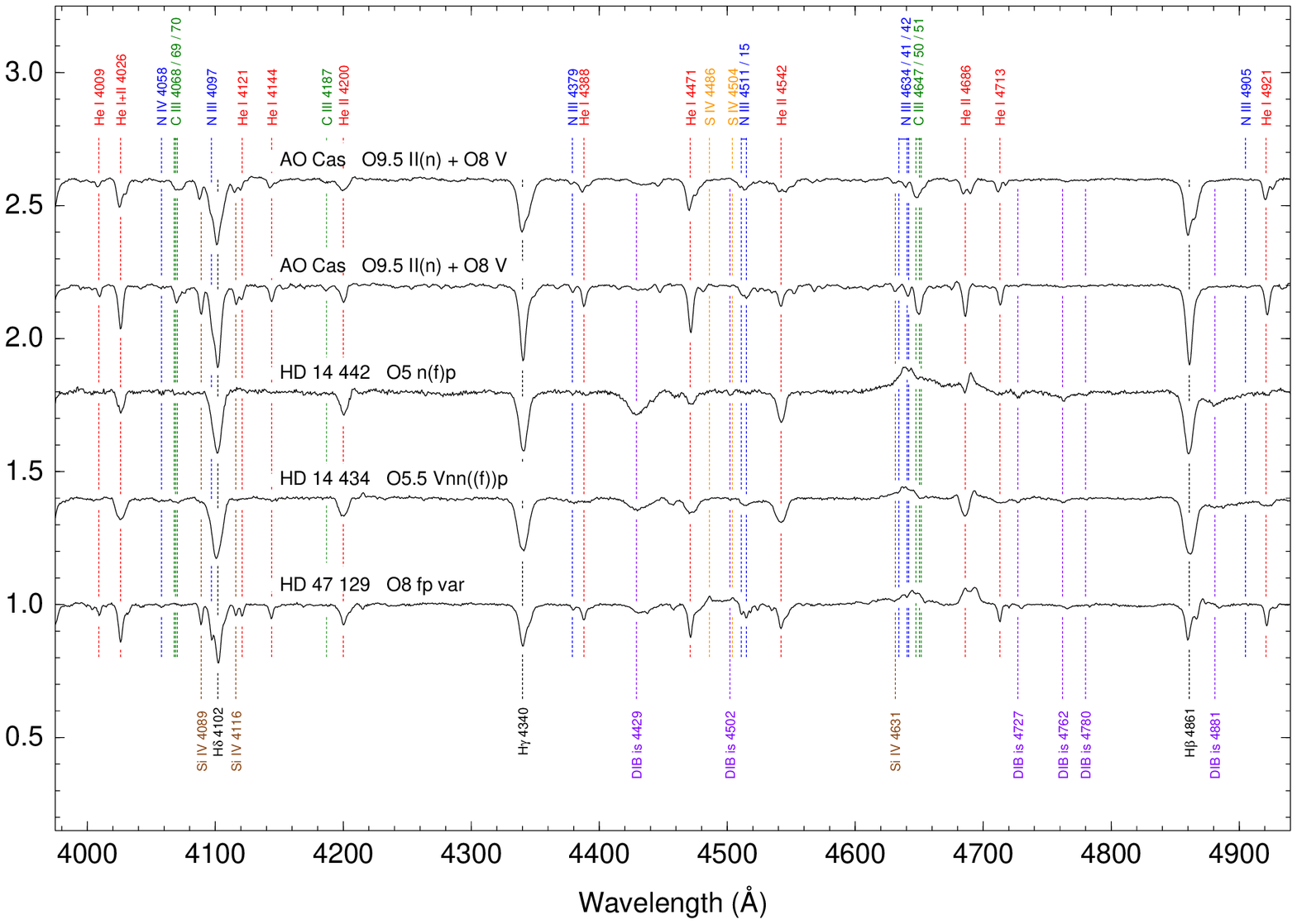}}
\caption{(continued).}
\label{Onfpb}   
\end{figure*}	

\begin{figure*}
%\centerline{\includegraphics*[width=\linewidth]{Ofp.ps}}
\centerline{\includegraphics*[width=\linewidth]{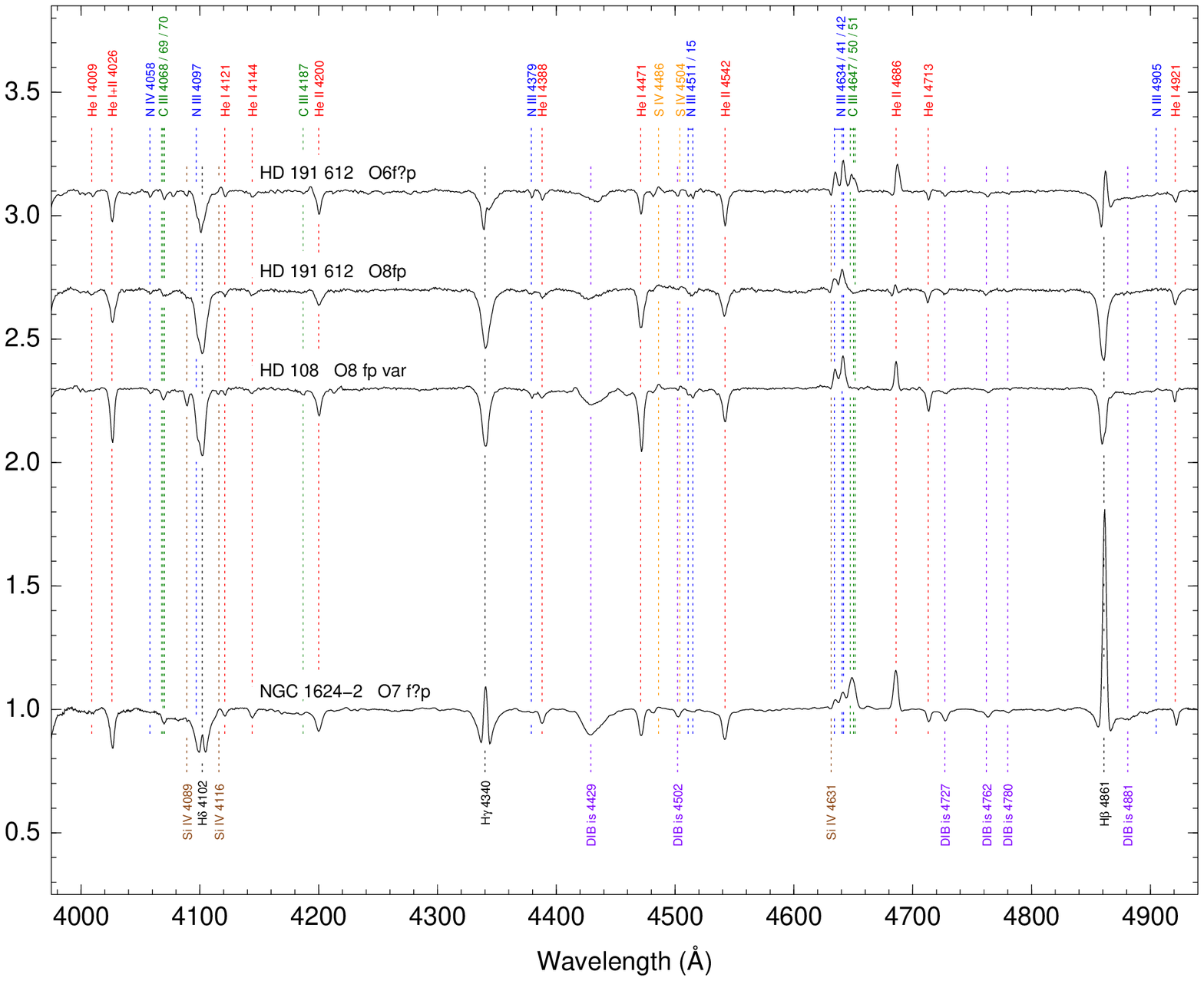}}
\caption{Spectrograms for Of?p stars. HD 191\,612 is shown in its two states. [See the electronic version of the journal for a color version of this figure.]}
\label{Ofp}
\end{figure*}	

\begin{figure*}
%\centerline{\includegraphics*[width=\linewidth]{Oe.ps}}
\centerline{\includegraphics*[width=\linewidth]{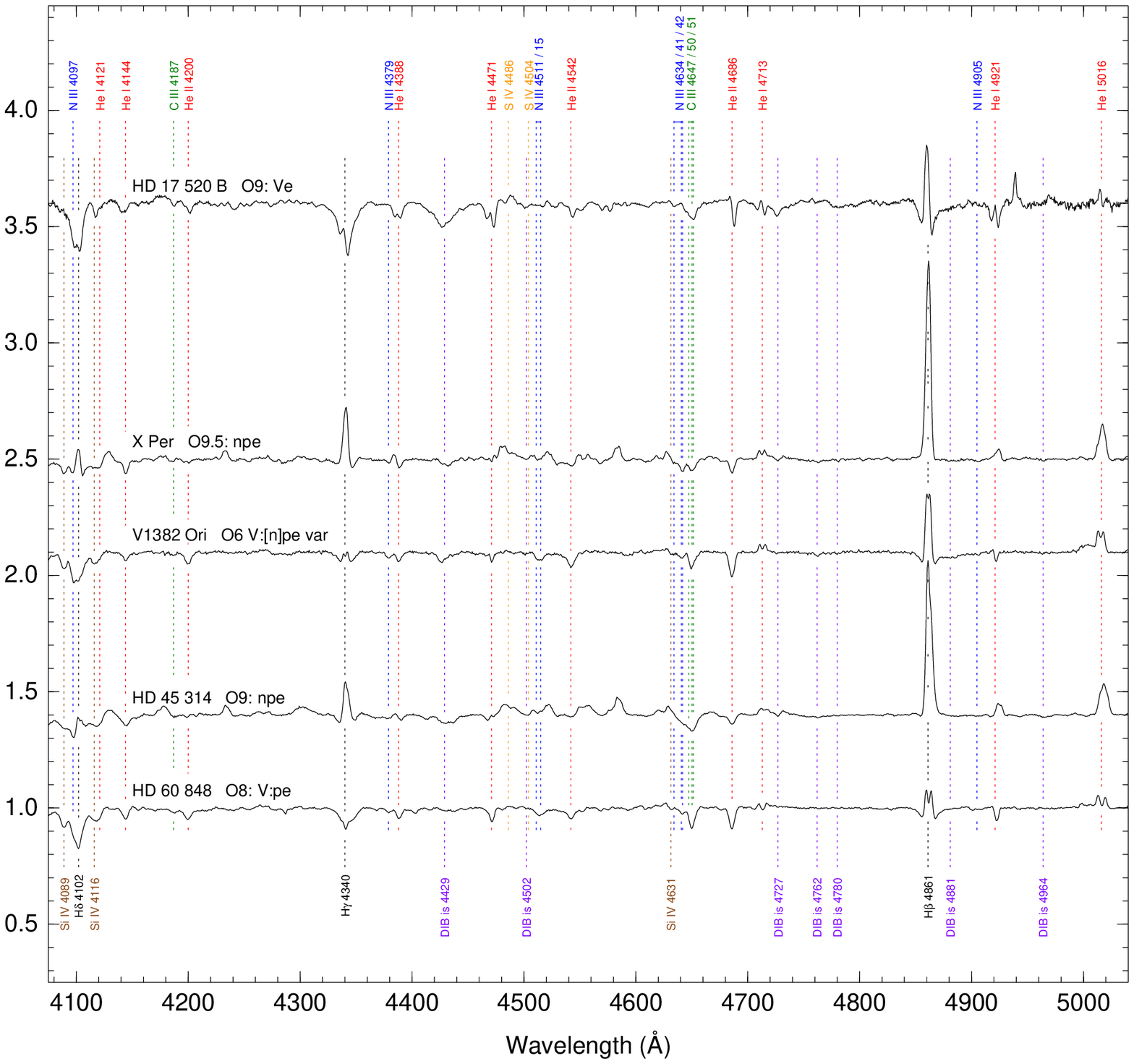}}
\caption{Spectrograms for Oe stars. The wavelength range is slightly different to that of other plots to show the \HeI{5016} line.
[See the electronic version of the journal for a color version of this figure.]}
\label{Oe}
\end{figure*}	

\begin{figure*}
%\centerline{\includegraphics*[width=\linewidth]{SB2a.ps}}
\centerline{\includegraphics*[width=\linewidth]{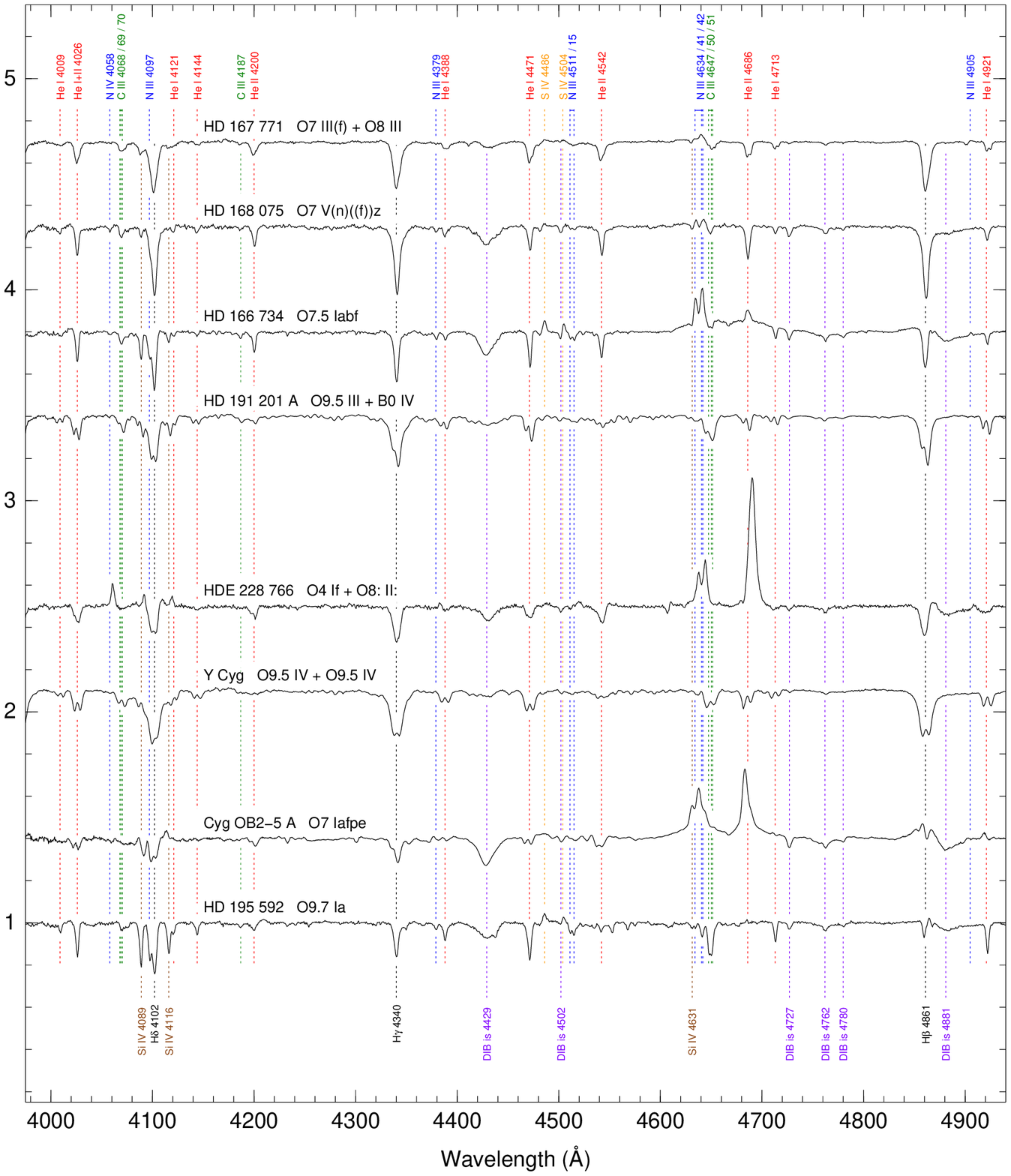}}
\caption{Spectrograms for double- and triple-lined spectroscopic binaries. [See the electronic version of the journal for a color version of this figure.]}
\label{SBa}   
\end{figure*}	

\addtocounter{figure}{-1}

\begin{figure*}
%\centerline{\includegraphics*[width=\linewidth]{SB2b.ps}}
\centerline{\includegraphics*[width=\linewidth]{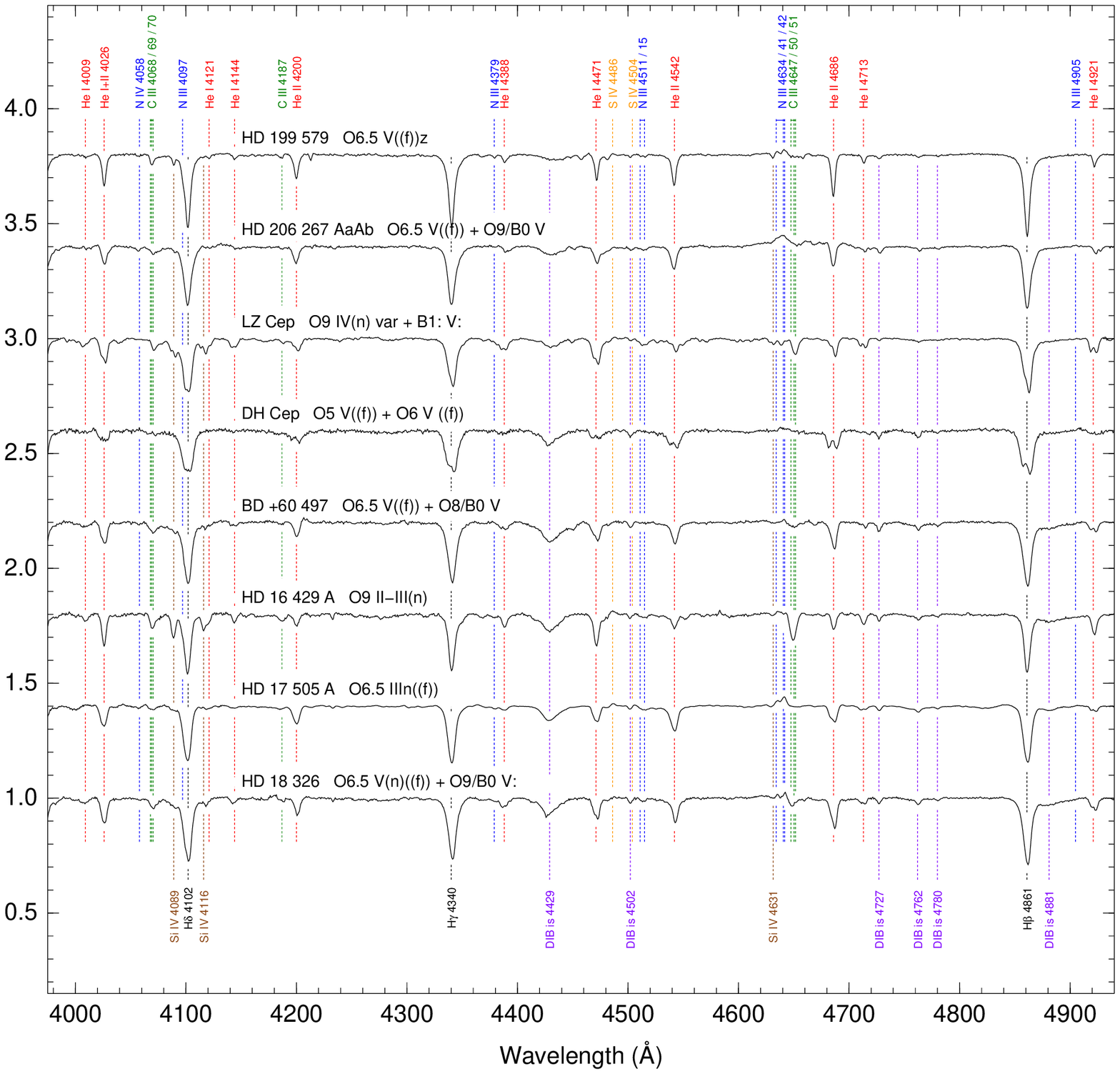}}
\caption{(continued).}
\label{SBb}   
\end{figure*}	

\addtocounter{figure}{-1}

\begin{figure*}
%\centerline{\includegraphics*[width=\linewidth]{SB2c.ps}}
\centerline{\includegraphics*[width=\linewidth]{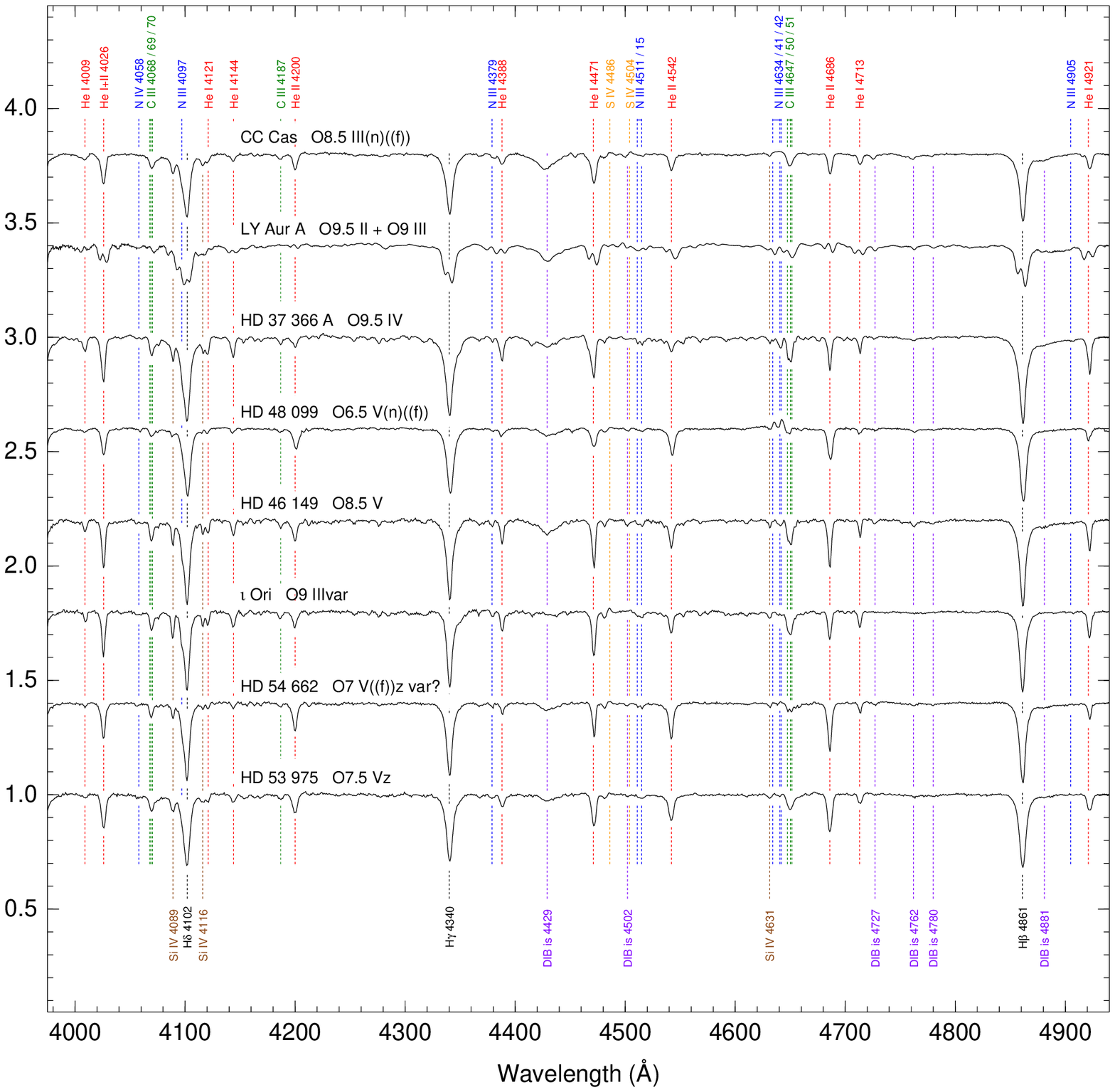}}
\caption{(continued).}
\label{SBc}   
\end{figure*}	

\subsection{Normal Sample}
\label{sec:Nor}

	In this subsection we briefly describe the O stars in our
sample that do not belong to any of the categories of the previous subsection.
	
\paragraph{$\zeta$ Oph = HD 149\,757} 

\object[HD 149757]{}

This object appears to have been ejected by a supernova explosion from Upper Scorpius \citep{Hoogetal00}. It is a very fast
rotator. Its revised Hipparcos distance with the new calibration is $112^{+3}_{-3}$ pc \citep{Maizetal08c}, making it the nearest
O star. 
%\citet{VillHerr05}
%\citet{Neguetal04}: O9.5 IV.
%\citet{Lefeetal09}: Intrinsic variability amplitude of 0.027 magnitudes.
%\citet{Vinketal09}: Polarization in H$\alpha$.

%\paragraph{HD 164\,438} 

\object[HD 164438]{}

\paragraph{HD 167\,659} 

\object[HD 167659]{}
\citet{Masoetal98} give a companion at 0\farcs08 but do not provide a $\Delta m$, so its effect is not included in the
name. This star has been recently found out to be a SB1 \citep{Gameetal08a}.

%\paragraph{HD 157\,857}

\object[HD 157857]{}
%RXB: Variable spectrum.

%\paragraph{HD 167\,633}

\object[HD 167633]{}

\paragraph{HD 165\,319}

\object[HD 165319]{}
This object was not present in version 1 of GOSC.
In \citet{Maiz04b} it appeared as B0 Ia \citep{Morgetal55}.

\paragraph{HD 168\,076 AB}

\object[HD 168076 AB]{}
\object[HD 168076]{}

This system has a separation of 0\farcs148 and a $\Delta m$ of 1.7 magnitudes in the $H$ band. \citet{Sanaetal09} give a
composite spectrum of O3.5 V ((f+)) + O7.5 V. See Fig.~\ref{chart1} for a chart. 
%Star appears elongated in WFPC2 images.
%\citet{Evanetal05}: O4 V ((f+)).
%In NGC 6611.

\paragraph{BD $-$13 4927}

\object[BD -13 4927]{}
See Fig.~\ref{chart1} for a chart. 
%\citet{Evanetal05}: O7 II (f).
%\citet{Sanaetal09}: O7 II (f).
%In NGC 6611.

\paragraph{BD $-$12 4979}

\object[BD 12 4979]{}
This object was not present in version 1 of GOSC. 
See Fig.~\ref{chart1} for a chart.

%\paragraph{HD 168\,112}

\object[HD 168112]{}
%\citet{Rauwetal05a}: No radial velocity variations detected.
%RXB: Double lines seen in echelle spectrum.

%\paragraph{HD 171\,589}

\object[HD 171589]{}

%\paragraph{BD $-$11 4586}

\object[BD -11 4586]{}

%\paragraph{HD 169\,582}

\object[HD 169582]{}
%RXB: Variable spectrum?

\paragraph{HD 173\,010}
 
\object[HD 173010]{}
This is an extremely luminous star. It might be classifed Ia+ but for the absence of \HeII{4686} emission as in the LMC counterparts \citep{Cortetal09}.
%RXB: Variable spectrum?

\paragraph{HD 173\,783}

\object[HD 173783]{}
The N\,{\sc iii} spectrum is exceedingly strong but C\,{\sc iii} is not abnormally weak.

%\paragraph{9 Sge = QZ Sge = HD 188\,001}

\object[HD 188001]{}
%RXB: SB1?, variable spectrum.
%\citet{Lefeetal09}: Intrinsic variability amplitude of 0.047 magnitudes.

\paragraph{HDE 344\,783}

\object[HDE 344783]{}
This object was not present in version 1 of GOSC. 
See Fig.~\ref{chart1} for a chart. 
%In NGC 6823 with HDE 344\,782+3+4.

\paragraph{HDE 344\,782}

\object[HDE 344782]{}
This object was not present in version 1 of GOSC. 
The spectrum includes two components (B and C) that are too dim to affect the spectral type.
See Fig.~\ref{chart1} for a chart. 
%In NGC 6823 with HDE 344\,782+3+4.

\paragraph{HDE 344\,784 A}

\object[HDE 344784 A]{}
\object[HDE 344784]{}
Note that in \citet{Maiz04b} this object appears as BD +22 3782.
See Fig.~\ref{chart1} for a chart. 
%In NGC 6823 with HDE 344\,782+3+4.

%\paragraph{HD 186\,980}

\object[HD 186890]{}

\paragraph{Cyg X-1 = V1357 Cyg = HDE 226\,868}

\object[HD 226868]{}
This system is the prototypical high-mass, black-hole X-ray binary (\citealt{CabNetal09} and references therein).
%\citet{Yanetal08}

%\paragraph{HD 190\,864}

\object[HD 190864]{}
%In NGC 6871.

\paragraph{HD 190\,429 B}

\object[HD 190429 B]{}
\object[HD 190429]{}
This object was not present in version 1 of GOSC. 
We were able to clearly separate the spectrum from that of A, located at a
separation of 1\farcs959 with a $\Delta m$ of 0.61 in the $z$ band \citep{Maiz10}. 

\paragraph{HD 190\,429 A}

\object[HD 190429 A]{}
\object[HD 190429]{}
As already mentioned, we were able to separate the A and B spectra. Note that \citet{Masoetal98} give an Ab component with a 
separation of 0\farcs09 but with a large $\Delta m$, so its existence is not mentioned in the object name (i.e. it is A instead of
AaAb).
%\citet{Turnetal08}.
%\citet{Contetal95a}: NIR spectral morphology similar to WN stars.

\paragraph{HD 191\,201 B}

\object[HD 191201 B]{}
\object[HD 191201]{}
This object was not present in version 1 of GOSC. 
%In NGC 6871.

%\paragraph{HD 192\,639}

\object[HD 192639]{}
%\citet{Lefeetal09}: Intrinsic variability amplitude of 0.048 magnitudes.

\paragraph{HD 193\,443 AB}

\object[HD 193443 AB]{}
\object[HD 193443]{}

\citet{Masoetal09} give a separation between the A and B components of 0\farcs13 with $\Delta m = 0.3$. We were unable to resolve
the system. The C component is too dim to have an effect on spectrum  \citep{Masoetal98}. 

%\paragraph{HDE 228\,841}

\object[HDE 228841]{}
%RXB: Variable spectrum.

%\paragraph{HD 193\,514}

\object[HD 193514]{}

%\paragraph{V2011 Cyg = HD 192\,281}

\object[HD 192281]{}
%\citet{DeBeRauw04}
%RXB: SB1?, horrible orbit, \citet{Bara93}.

%\paragraph{HDE 229\,232}

\object[HDE 229232]{}
%RXB: Curious, variable spectrum.

\paragraph{HD 189\,957}

\object[HD 189957]{}
This is the standard for the newly introduced O9.7 III category.

%\paragraph{HD 191\,978}

\object[HD 191978]{}

\paragraph{HD 193\,322 AaAb}

\object[HD 193322 AaAb]{}
\object[HD 193322 A]{}
\object[HD 193322]{}

Aa and Ab are separated by 0\farcs055 with a small $\Delta m$ \citep{Maiz10,Masoetal09} and are unresolved in our data. The B
component is at a separation of 2\farcs719 and we were able to separate its spectrum from that of AaAb: it is an early B star. This
complex system was studied by \citet{McKietal98}, who found out that Aa is a SB1 with Ab in a 31-year period orbit around it. See
\citet{Robeetal10} for a recent study of this system and the surrounding cluster, Collinder 419.
The new Hipparcos calibration gives a revised distance of $708^{+255}_{-145}$ pc \citep{Maizetal08c}.
%\citet{Turnetal08}.

%\paragraph{HD 192\,001}

\object[HD 192001]{}

%\paragraph{HDE 229\,196}

\object[HDE 229196]{}
%RXB: Variable spectrum?

\paragraph{V2185 Cyg = Schulte 50 = [MT91]~421}

\object[V2185 Cyg]{}
This object was not present in version 1 of GOSC.
See Fig.~\ref{chart2} for a chart.

\paragraph{Cyg OB2-22 A = Schulte 22 A = [MT91]~417~A}

\object[Schulte 22 A]{}
\object[Schulte 22]{}
Cyg OB2-22 A and B are separated by 1\farcs521 with a $\Delta m$ of 0.59 in the $z$ band \citep{Maiz10}. 
We were able to extract the individual spectra of A and B (see below for B).
See also \citet{Walbetal02b}. This star has an extremely high spectroscopic/evolutionary mass and has not been resolved at high resolution,
including HST/ACS \citep{Maiz10}, HST/FGS, and Gemini AO (Ed Nelan, Doug Gies, private communications), although a tighter structure cannot be ruled out.
See Fig.~\ref{chart2} for charts. 

\paragraph{Cyg OB2-22 B = Schulte 22 B = [MT91]~417~B}

\object[Schulte 22 B]{}
\object[Schulte 22]{}
Cyg OB2-22 B itself is a double system composed of Ba and Bb. Their separation is 0\farcs216 and their $\Delta m$ is of 2.34
magnitudes in the $z$ band \citep{Maiz10}, which is too large to have the effect of Bb included in the name of the object.
See Fig.~\ref{chart2} for charts. 

\paragraph{NSV 13\,148 = Schulte 24 = [MT91]~480}

\object[NSV 13148]{}
This object was not present in version 1 of GOSC. 
See Fig.~\ref{chart1} for a chart.

\paragraph{Cyg OB2-8 B = Schulte 8 B = [MT91]~462}

\object[NAME VI CYG 8B]{}
\object[NAME VI CYG 8]{}
This object was not present in version 1 of GOSC.
See Fig.~\ref{chart1} for a chart. 

%\paragraph{Cyg OB2-4 = LS III +41 29 = [MT91]~217}

\object[NAME VI CYG 4]{}

\paragraph{Cyg OB2-8 D = Schulte 8 D = [MT91]~473}

\object[NAME VI CYG 8D]{}
\object[NAME VI CYG 8]{}
This object was not present in version 1 of GOSC.
Note that the current version of the WDS catalog has Cyg OB2-8 C and D interchanged with respect to the most common usage.
See Fig.~\ref{chart1} for a chart. 

\paragraph{Cyg OB2-7 = Schulte 7 [MT91]~457}

\object[NAME VI CYG 7]{}
See Fig.~\ref{chart1} for a chart. 

%\paragraph{HD 188\,209}

\object[HD 188209]{}
%\citet{Lefeetal09}: Intrinsic variability amplitude of 0.051 magnitudes.

%\paragraph{HD 202\,124}

\object[HD 202124]{}
%RXB: variable spectrum?
%\citet{Lefeetal09}: Intrinsic variability amplitude of 0.060 magnitudes.

%\paragraph{68 Cyg = V1809 Cyg = HD 203\,064}

\object[HD 203064]{}
%\citet{Lefeetal09}: Intrinsic variability amplitude of 0.031 magnitudes.

\paragraph{10 Lac = HD 214\,680}

\object[HD 214680]{}
The revised Hipparcos distance to this object with the new calibration is $542^{+77}_{-59}$ pc \citep{Maizetal08c}.

\paragraph{HD 206\,183}

\object[HD 206183]{}
This object was not present in version 1 of GOSC.
In \citet{Maizetal04b} a classification of B0 V was given.

\paragraph{HD 204\,827 AaAb}

\object[HD 204827 AaAb]{}
\object[HD 204827]{}
This object was not present in version 1 of GOSC.
\citet{Masoetal98} give a separation of 0\farcs12 and a $\Delta m = 1.2$ for the Aa + Ab system. B is more
than three magnitudes fainter.
In \citet{Maizetal04b} a classification of B0.2 V was given.

%\paragraph{HD 210\,809}

\object[HD 210809]{}

\paragraph{HD 207\,538}

\object[HD 207538]{}
This object was not present in version 1 of GOSC.
In \citet{Maizetal04b} a classification of B0.2 V was given but now is the standard for the newly introduced O9.7 IV category.

\paragraph{HD 207\,198}

\object[HD 207198]{}
\citet{Masoetal98} give a companion with $\Delta m = 3.5$ at a separation of 18\farcs3, so its effect does
not modify the name. 

%\paragraph{19 Cep = HD 209\,975}

\object[HD 209975]{}
%\citet{Lefeetal09}: Intrinsic variability amplitude of 0.028 magnitudes.

%\paragraph{HD 218\,915}

\object[HD 218915]{}
%RXB: SB2?. 2 OSN (+1 OSN Oct 09), SB2 not visible.

\paragraph{HD 218\,195 A}

\object[HD 218195 A]{}
\object[HD 218195]{}

A B component at a separation of 0\farcs919 with a $\Delta m$ of 2.56 in the $z$ band was detected by \citet{Maiz10}. We were able
to extract its spectrum independently of A and we obtained an early-B spectral type. 

%\paragraph{HD 216\,532}

\object[HD 216532]{}
%RXB: Variable spectrum.

%\paragraph{HD 216\,898}

\object[HD 216898]{}

%\paragraph{HD 217\,086}

\object[HD 217086]{}
%A weak companion ($\Delta m$ of 3.58 in the $z$ band) exists at a separation of 2\farcs865 \citep{Maiz10}.
%Previously undetected companion in AstraLux data. Preliminary values: PA of 166.4, separation of 3\farcs40, 
%$\Delta z = 6.6$. Marked as YSO in Simbad, see ApJS 163, 306 (2006).
%\citet{Turnetal08}.

%\paragraph{HD 225\,146}

\object[HD 225146]{}

%%\paragraph{HD 225\,160}
%\citet{NeguMarc03}

\object[HD 225160]{}

\paragraph{HD 5005 C}

\object[HD 5005 C]{}
\object[HD 5005]{}
We obtained individual spectra for the four bright components in this system (A, B, C, and D) and we found all of them to be O stars.
See Fig.~\ref{chart2} for a chart. 
%In IC 1590.

\paragraph{HD 5005 B}

\object[HD 5005 B]{}
\object[HD 5005]{}
We obtained individual spectra for the four bright components in this system (A, B, C, and D) and we found all of them to be O stars.
This object was not present in version 1 of GOSC.
The luminosity class from \HeII{4686}/\HeI{4713} conflicts with a very weak Si\,{\sc iv} and probable extreme youth.
See Fig.~\ref{chart2} for a chart. 
%In IC 1590.
 
\paragraph{HD 5005 D}

\object[HD 5005 D]{}
\object[HD 5005]{}
We obtained individual spectra for the four bright components in this system (A, B, C, and D) and we found all of them to be O stars.
This object was not present in version 1 of GOSC.
See Fig.~\ref{chart2} for a chart. 
%In IC 1590.

%\paragraph{BD +60 261}

\object[BD +60 261]{}

%\paragraph{HD 10\,125}

\object[HD 10125]{}
%Previously undetected companion in AstraLux data. Preliminary values: PA of 230.2, separation of
%0\farcs78, $\Delta z = 3.2$.

%\paragraph{HD 13\,022}

\object[HD 13022]{}

%\paragraph{HD 12\,993}

\object[HD 12993]{}

%\paragraph{BD +62 424}

\object[BD +62 424]{}
%Constant radial velocity: \citet{Hilletal06}.

%\paragraph{V354 Per = HD 13\,745}

\object[HD 13745]{}

\paragraph{BD +60 498}

\object[BD +60 498]{}
%BD +60 513 and BD +60 497 are too distant
%In IC 1805.
This object was not present in version 1 of GOSC.
The luminosity class from \HeII{4686}/\HeI{4713} conflicts with a very weak Si\,{\sc iv} and probable extreme youth.
See Fig.~\ref{chart3} for a chart. 

\paragraph{BD +60 499}

\object[BD +60 499]{}
%BD +60 513 and BD +60 497 are too distant for the chart
%Previously undetected companion in AstraLux data. Preliminary values: PA of 225.8, separation of
%4\farcs88, $\Delta z = 5.0$.
See Fig.~\ref{chart3} for a chart. 
%In IC 1805.

\paragraph{BD +60 501}

\object[BD +60 501]{}
%BD +60 513 and BD +60 497 are too distant
%O7 V ((f)), constant radial velocity: \cite{RauwDeBe04}.
%Constant radial velocity: \citet{Hilletal06}.
See Fig.~\ref{chart3} for a chart. 
%In IC 1805.

\paragraph{HD 15\,570}

\object[HD 15570]{}
%BD +60 513 and BD +60 497 are too distant
%Constant radial velocity: \citet{Hilletal06}.
%\citet{DeBeetal06a}.
%\citet{DeBeetal09}: Study of the spectroscopic variability.
See Fig.~\ref{chart3} for a chart. 
%In IC 1805.

%\paragraph{BD +60 513}

\object[BD +60 513]{}
%O7.5 V ((f)), constant radial velocity: \cite{RauwDeBe04}.
%Constant radial velocity: \citet{Hilletal06}.
%In IC 1805.

%\paragraph{HD 14\,947}

\object[HD 14947]{}
%\citet{DeBeetal09}: Study of the spectroscopic variability.

%\paragraph{HD 15\,642}

\object[HD 15642]{}

%\paragraph{HD 18\,409}

\object[HD 18409]{}

\paragraph{HD 17\,505 B}

\object[HD 17505 B]{}
\object[HD 17505]{}
This object was not present in version 1 of GOSC.
We were able to separate the spectrum from that of A, located at a separation of 2\farcs153 with a $\Delta m$ of 1.75 in the $z$
band \citep{Maiz10}. 
See Fig.~\ref{chart3} for charts.
%In IC 1848.
%\cite{Hilletal06}: O8.5 V.

\paragraph{HD 17\,520 A}

\object[HD 17520 A]{}
\object[HD 17520]{}
The A component is separated by only 0\farcs316 from the B component with a $\Delta m$ of 0.67 mag in the $z$ band \citep{Maiz10}
but we were able to separate the two spectra.
%\citet{Masoetal09}: B component at 0\farcs31 with $\Delta m = 0.5$, CHECK.
%Not in \citet{Masoetal98}.
%SB2?. 1 OSN, SB2 seen only in H beta.
\citet{Hilletal06} indicate that in the integrated A + B spectrum, the A component appears to be a SB1. In our spectra we detect
distinct velocity changes between the emission lines (which originate in B, see subsection~\ref{sec:Oe}) and the absorption profile at
dofferent epochs, which supports the SB1 character for A.
See Fig.~\ref{chart3} for charts.
%In IC 1848.

\paragraph{BD +60 586 A}

\object[BD +60 586 A]{}
\object[BD +60 586]{}
%B, A, and C are aligned with a PA of 233 and separations of B and C of 16\arcsec and 7\arcsec, respectively.
%Constant radial velocity: \citet{Hilletal06}.
See Fig.~\ref{chart4} for a chart. 

\paragraph{HD 15\,137}

\object[HD 15137]{}
%\citet{DeBeetal08}: Variable on short time scales.
\citet{McSwetal10} measure the SB1 orbit of this object and suggest that the unseen companion may be a neutron star or black hole.

%\paragraph{HD 16\,691}

\object[HD 16691]{}
%\citet{Contetal95a}: NIR spectral morphology similar to WN stars.
%\citet{DeBeetal09}: Study of the spectroscopic variability.

%\paragraph{HD 16\,832}

\object[HD 16832]{}

%\paragraph{HD 17\,603}

\object[HD 17603]{}

%\paragraph{$\alpha$ Cam = HD 30\,614}

\object[HD 30614]{}
%\citet{Lefeetal09}: Intrinsic variability amplitude of 0.040 magnitudes.

%\paragraph{HDE 237\,211}

\object[HDE 237211]{}
%\citet{NeguMarc03}

\paragraph{HD 24\,431}

\object[HD 24431]{}
A B component is present at a separation of 0\farcs720 but its $\Delta m$ of 2.9 in the $z$ band \citep{Maiz10} 
indicates that it is too weak to have a significant effect in the optical spectrum.
%\citet{Turnetal08}.

\paragraph{$\xi$ Per = Menkhib = HD 24\,912}

\object[HD 24912]{}
The revised Hipparcos distance to this object with the new calibration is $416^{+116}_{-74}$ pc \citep{Maizetal08c}.

%\paragraph{HD 41\,161}

\object[HD 41161]{}

\paragraph{BD +39 1328}

\object[BD +39 1328]{}
The spectrum of this star is rather anomalous: the neutralized \HeII{4686} indicates a high luminosity (as classified), but the
Si\,{\sc iv} and C\,{\sc iii} absorptions are very weak, perhaps indicating a different origin of the \HeII{4686} behavior.
%\citet{NeguMarc03}

%\paragraph{HD 34\,656}

\object[HD 34656]{}

\paragraph{AE Aur = HD 34\,078}

\object[HD 34078]{}
This object appears to have been ejected from the Trapezium cluster (\citealt{Hoogetal00} and references therein).
%\citet{Lefeetal09}: Intrinsic variability amplitude of 0.056 magnitudes.

%\paragraph{HD 36\,483}

\object[HD 36483]{}

\paragraph{HD 35\,619}

\object[HD 35619]{}
A B component is present at a separation of 2\farcs772 but its $\Delta m$ of 2.88 in the $z$ band \citep{Maiz10} 
indicates that it is too weak to have a significant effect in the optical spectrum.

%\paragraph{HD 37\,737}

\object[HD 37737]{}

\paragraph{BD +33 1025}

\object[BD +33 1025]{}
This object was not present in version 1 of GOSC.
See Fig.~\ref{chart4} for a chart.
%In NGC 1893.

\paragraph{HDE 242\,935 A}

\object[HDE 242935 A]{}
\object[HDE 242935]{}
The A component is separated by 1\farcs081 from the B component with a $\Delta m$ of 0.80 mag in the $z$ band \citep{Maiz10}.
We were able to separate the two spectra and obtain an early-B type for the B component.
See Fig.~\ref{chart4} for charts.
%In NGC 1893.

\paragraph{HDE 242\,926}

\object[HD 242926]{}
%Previously undetected companion in AstraLux data. Preliminary values: PA of 171.3, separation of
%1\farcs82, $\Delta z = 3.9$.
See Fig.~\ref{chart4} for a chart.
%In NGC 1893.

\paragraph{HD 93\,521}

\object[HD 93521]{}
This object is a non-radial pulsator \citep{Rauwetal08}.
%RXB: Variable spectrum.

%\paragraph{HD 36\,879}

\object[HD 36879]{}

%\paragraph{HD 42\,088}

\object[HD 42088]{}
%\citet{Vinketal09}: Polarization in H$\alpha$.

\paragraph{HD 44\,811}

\object[HD 44811]{}
%Secondary component at 6\farcs0 but too dim to have an effect ($\Delta m$ = 3.5).
See Fig.~\ref{chart5} for a chart. 

%\paragraph{HD 41\,997}

\object[HD 41997]{}

\paragraph{$\lambda$ Ori A = HD 36\,861}

\object[HD 36861]{}
The A component is separated by 4\farcs342 from the B component with a $\Delta m$ of 1.91 mag in the $z$ band \citep{Maiz10}.
We were able to separate the two spectra and obtain an early-B type for the B component, which has a separate HD number 
(36\,862). The revised Hipparcos distance to $\lambda$ Ori A with new calibration is $361^{+89}_{-60}$ pc \citep{Maizetal08c}.
%\citet{Turnetal08}.

\paragraph{15 Mon AaAb = S Mon AaAb = \\ HD~47\,839~AaAb}

\object[HD 47839 AaAb]{}
\object[HD 47839 A]{}
\object[HD 47839]{}
The three brightest components of this system are Aa, Ab, and B. 15 Mon B is at a separation of 2\farcs976 with respect to Aa and
their $\Delta m$ is of 3.23 magnitudes in the $z$ band \citep{Maiz10}. We were able to spatially separate the B spectra in our
data and confirm that it is of early-B type. Aa and Ab are much closer, with a separation of 0\farcs128 and a $\Delta m$ of 1.43
in the $z$ band, and we were unable to separate them. The Aa-Ab orbit is being followed with somewhat conflicting preliminary
orbits at the present time: see \citet{Maiz10} and references therein. The revised Hipparcos distance to the system with new 
calibration is $309^{+60}_{-43}$ pc \citep{Maizetal08c}. The spectral type of this fundamental MK standard has been found to be
apparently variable between O7 and O7.5 on an undetermined timescale, which is under investigation. Thus, it should be used with 
caution or not at all as a standard now.  
See Fig.~\ref{chart6} for a chart.
%\citet{Turnetal08}.
%\citet{Cvetetal09}: 74 year orbit. 
%\citet{Lefeetal09}: Intrinsic variability amplitude of 0.040 magnitudes.
%In NGC 2264.

%\paragraph{HD 46\,966}

\object[HD 46966]{}
%\citet{Mahyetal09} detect no RV variations and give a spectral type of O8.5 V.

\paragraph{HD 46\,106}

\object[HD 46106]{}
This object was not present in version 1 of GOSC. 
The luminosity class from \HeII{4686}/\HeI{4713} conflicts with a very weak Si\,{\sc iv} and probable extreme youth.
%In NGC 2244.

%\paragraph{HD 46\,202}

\object[HD 46202]{}
%Previously undetected companion in AstraLux data. Preliminary values: PA of 263.0, separation of
%3\farcs88, $\Delta z = 3.6$ (2MASS photometry available).
%\citet{Mahyetal09} detect no radial velocity variations and give a spectral type of O9 V.
%In NGC 2244.

%\paragraph{HD 46\,150}

\object[HD 46150]{}
%\citet{Masoetal98}: Many dim components.
%RXB: Variable spectrum. 
%\citet{Mahyetal09} give a spectral type of 05.5 V ((f)) and detect radial velocity variations.
%In NGC 2244.

\paragraph{HD 46\,056 A}

\object[HD 46056 A]{}
\object[HD 46056]{}
A B component at a separation of 10\farcs419 and a $\Delta m$ of 2.78 magnitudes in the $z$ band \citep{Maiz10} was 
determined to be of early-B spectral type.
\citet{Mahyetal09} suggest that the broad lines of the A component are caused by rapid rotation.
%SB2?. 1 OSN, SB2 not visible (but wide lines). \citet{Mahyetal09} say it is not a SB but a rapid rotator and
%give a spectral type of O8 V n.
%In NGC 2244.

%\paragraph{HD 46\,223}

\object[HD 46223]{}
%\citet{Mahyetal09} detect no RV variations and give a spectral type of O4 V ((f+)).
%In NGC 2244.

\paragraph{$\zeta$ Ori B = HD 37\,743}

\object[HD 37743]{}
This object was not present in version 1 of GOSC. 
We were able to separate the B spectrum from that of A (= HD 37\,742), 
located at a separation of 2\farcs424 and with a $\Delta m$ of 2.26 magnitudes in the $z$ band. In \citet{Maizetal04b} it was 
given as an early B giant but \citet{Garmetal82} listed it as O9.5 IV.

\paragraph{$\sigma$ Ori AB = HD 37\,468 AB}

\object[HD 37468 AB]{}
\object[HD 37468]{}
This system lies at the core of the well-studied $\sigma$ Ori cluster \citep{Caba07,Sheretal08}. The current separation between 
A and B is 0\farcs260; its orbit is followed by \citet{Turnetal08}. We were unable to spatially separate in our spectra
the relatively low-$\Delta m$ (1.57 in the $z$ band) AB pair. See Fig.~\ref{chart5} for charts.
%\citet{Masoetal98}: C, D components are at $\approx 12\arcsec$, E component is farther away than 30\arcsec\
%and is HD 37\,479. 
%\citet{Maizetal04b} gave spectrum for ABC.
%\citet{Caba07}, \citet{Sheretal08}: Description of the bright stars in the $\sigma$ Ori cluster.

%\paragraph{HD 46\,485}

\object[HD 46485]{}
%\citet{Mahyetal09} say that it is a rapid rotator with spectral type O8 V n.
%\citet{Vinketal09}: Polarization in H$\alpha$.

%\paragraph{HD 46\,573}

\object[HD 46573]{}
%\citet{Mahyetal09} derive a 10.67 days period and a spectral type of O7.5 V ((f)) with an unseen low-mass companion.

\paragraph{$\theta^1$ Ori CaCb = HD 37\,022 AB}

\object[HD 37022 AB]{}
\object[HD 37022]{}
\object[NAME TRAPEZIUM]{}
This well-known object is the brightest star in the Trapezium and the main source of ionizing photons in the Orion nebula. In
recent years a bright companion ($\Delta m = 1.3$) has been detected at a small separation (tens of mas) and its orbit is currently
being followed \citep{Krauetal07,Patietal08,Krauetal09b}. Ca is a magnetic oblique rotator with a period of 15.424 $\pm$ 0.001 
days \citep{Nazeetal08d}. The pair Ca-Cb is obviously unresolved in our spectra. See Fig.~\ref{chart6} for a chart.
%\citet{SimDetal06}
%\citet{Vinketal09}: Polarization in H$\alpha$.

\paragraph{$\theta^2$ Ori A = HD 37\,041}

\object[HD 37041]{}
$\theta^2$ Ori A is the other O-type system in the Orion nebula. An Ab companion is spatially unresolved in our data (separation
of 0\farcs396) but its $\Delta m$ is too large (2.62 magnitudes at the $z$ band, \citealt{Maiz10}) to have  asignificant effect in
the onserved spectra. The B and C components are farther away than 30\arcsec\ and they have their own HD numbers (B is HD 37\,042, 
C is HD 37\,062, \citealt{Masoetal98}). The revised Hipparcos distance with the new calibration is
$520^{+201}_{-103}$~pc \citep{Maizetal08c}. The luminosity class IV derived here from \HeII{4686}/\HeI{4713} is unlikely to represent a 
real luminosity effect in this probable ZAMS star.
%\citet{SimDetal06}
%X-ray variable: \citet{Schuetal06}.
%\citet{Turnetal08}.
%\citet{Vinketal09}: Polarization in H$\alpha$.

%\paragraph{V689 Mon = HD 47\,432}

\object[HD 47432]{}
%\citet{Lefeetal09}: Intrinsic variability amplitude of 0.053 magnitudes.

\paragraph{$\upsilon$ Ori = HD 36\,512}

\object[HD 36512]{}
This object was not present in version 1 of GOSC. Previously, it was a B0 V standard but now is the O9.7 V standard.

\paragraph{HD 52\,533 A = BD -02 1885}

\object[HD 52533 A]{}
\object[HD 52533]{}
There are several dim companions in the vicinity of HD 52\,533 A but the brightest one is C, with a $\Delta m$ of = 1.1 and a
separation of 22\farcs6 \citep{Masoetal98}. We placed C on the slit and obtained a G spectral type for that component.
See Fig.~\ref{chart6} for a chart.

%\paragraph{HD 52\,266}

\object[HD 52266]{}

\paragraph{HD 57\,682}

\object[HD 57682]{}

A magnetic field has been discovered in this star \citep{Grunetal09}. The spectral lines are extremely sharp and there are Balmer
profile variations at high resolution very similar to those in the Of?p stars at earlier types.

%\paragraph{HD 55\,879}

\object[HD 55879]{}

%\paragraph{HD 54\,879}

\object[HD 54879]{}

\begin{figure*}
%\centerline{\includegraphics*[width=\linewidth, bb=50 225 545 725]{charts1.ps}}
%\centerline{\includegraphics*[width=\linewidth, bb=50 225 545 725]{fig19a.ps}}
\centerline{\includegraphics*[width=\linewidth]{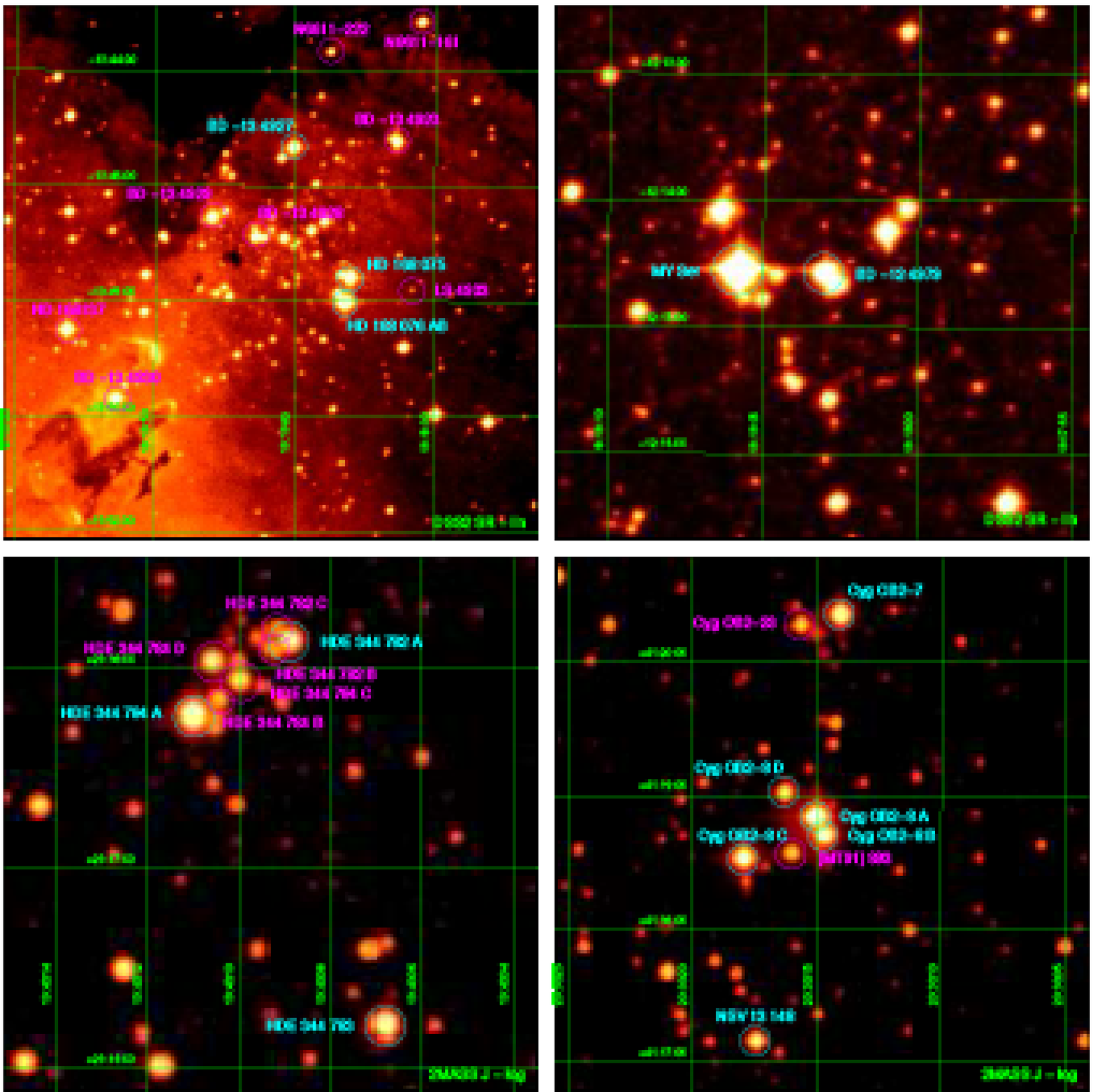}}
%\centerline{Figure unavailable in astro-ph version.}
\caption{Twenty-four fields that include O stars for which we have obtained good-quality spectrograms (marked in cyan or black). 
Objects in magenta or grey [a] have no or only low-S/N spectrogram, [b] are not O stars, and/or [c] are seen as individual 
sources in the image but cannot be separated from a bright spectroscopic companion. Subfields delimited with a magenta or black
square are shown at higher spatial resolution in another panel. The image 
source and intensity scale type (linear or logarithmic) are shown at the lower right corner of each panel.
[See the electronic version of the journal for a color version of this figure.]}
\label{chart1}
\end{figure*}	

\addtocounter{figure}{-1}

\begin{figure*}
%\centerline{\includegraphics*[width=\linewidth, bb=50 225 545 725]{charts2.ps}}
%\centerline{\includegraphics*[width=\linewidth, bb=50 225 545 725]{fig19b.ps}}
\centerline{\includegraphics*[width=\linewidth]{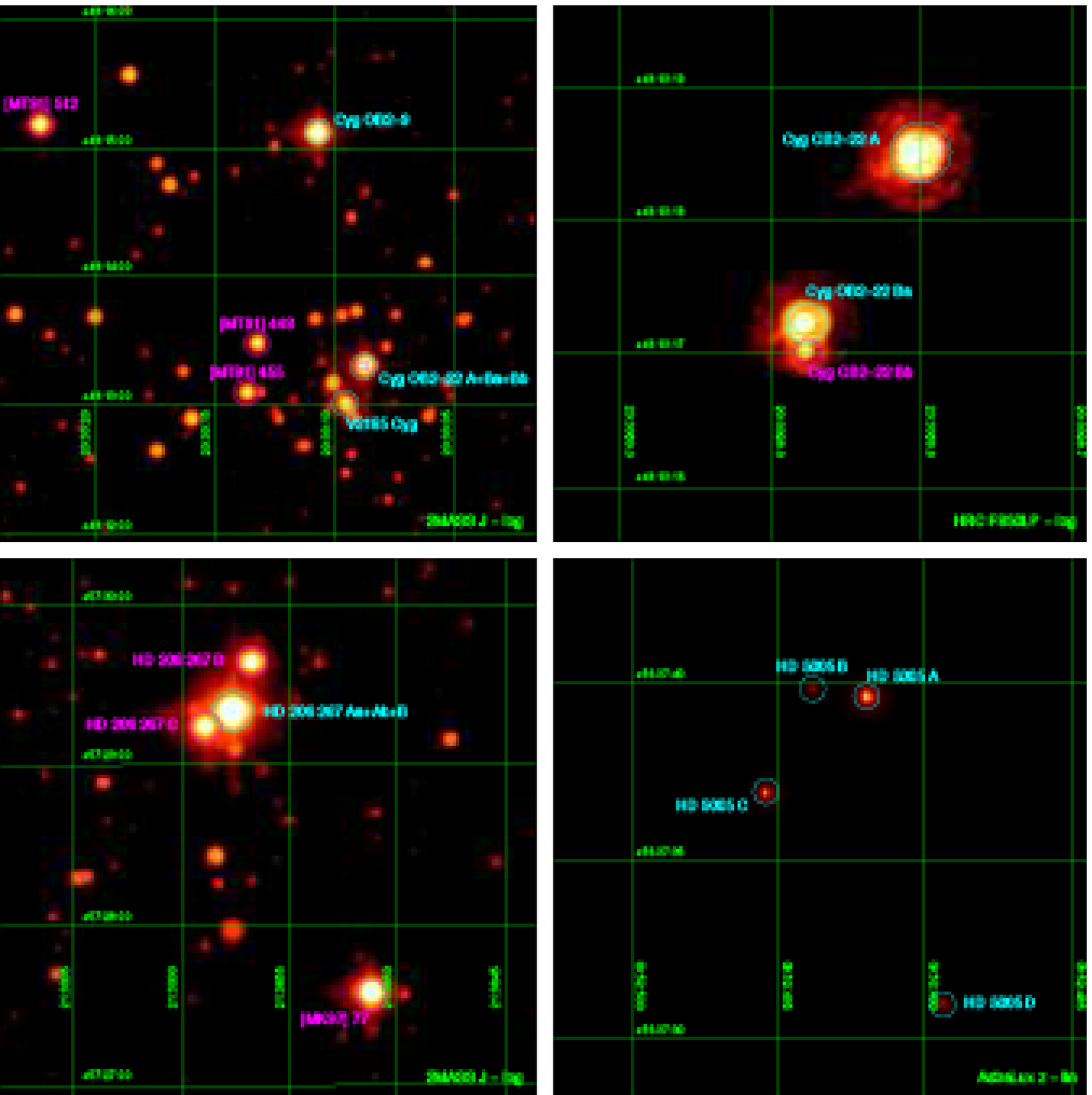}}
%\centerline{Figure unavailable in astro-ph version.}
\caption{(continued).}
\label{chart2}
\end{figure*}	

\addtocounter{figure}{-1}

\begin{figure*}
%\centerline{\includegraphics*[width=\linewidth, bb=50 225 545 725]{charts3.ps}}
%\centerline{\includegraphics*[width=\linewidth, bb=50 225 545 725]{fig19c.ps}}
\centerline{\includegraphics*[width=\linewidth]{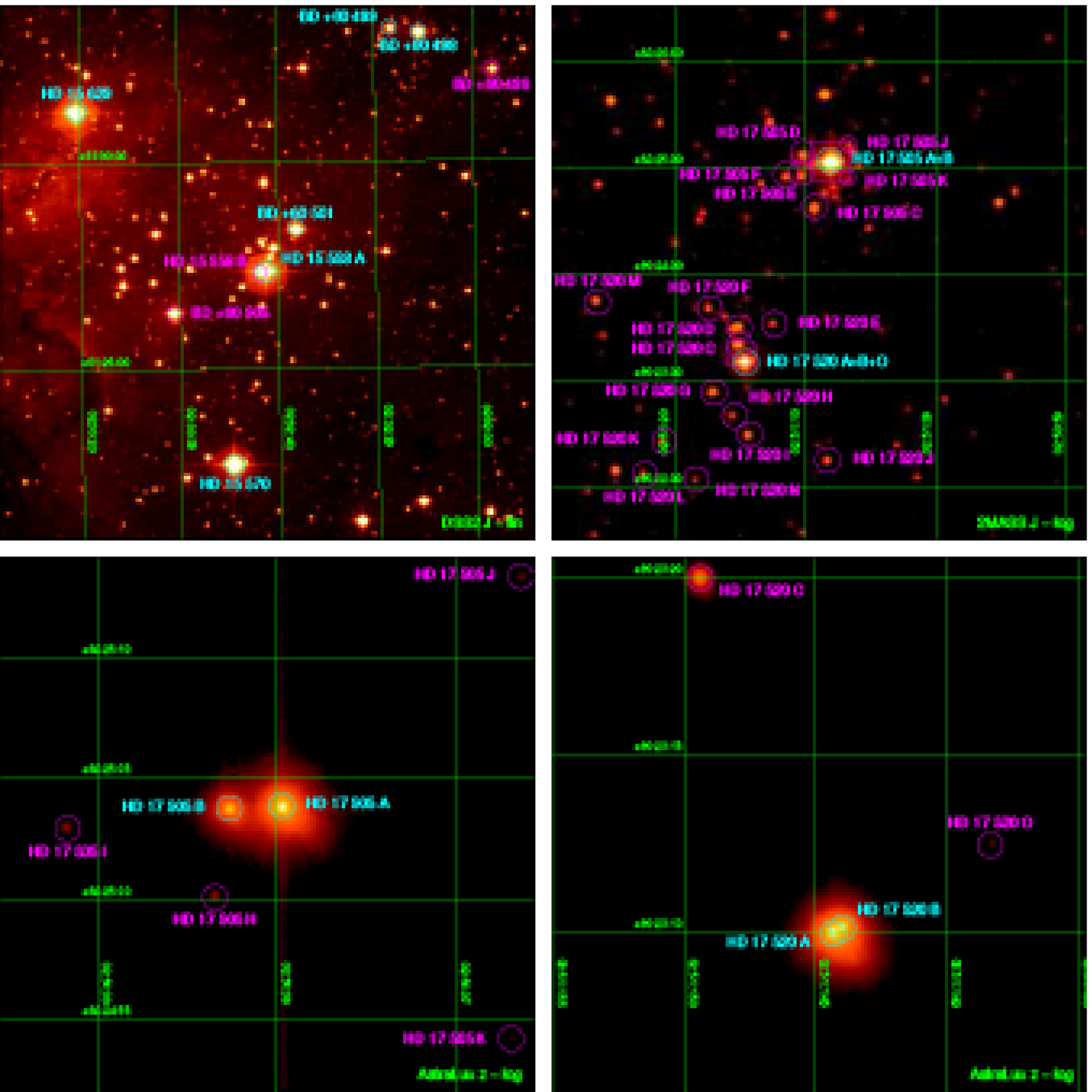}}
%\centerline{Figure unavailable in astro-ph version.}
\caption{(continued).}
\label{chart3}
\end{figure*}	

\addtocounter{figure}{-1}

\begin{figure*}
%\centerline{\includegraphics*[width=\linewidth, bb=50 225 545 725]{charts4.ps}}
%\centerline{\includegraphics*[width=\linewidth, bb=50 225 545 725]{fig19d.ps}}
\centerline{\includegraphics*[width=\linewidth]{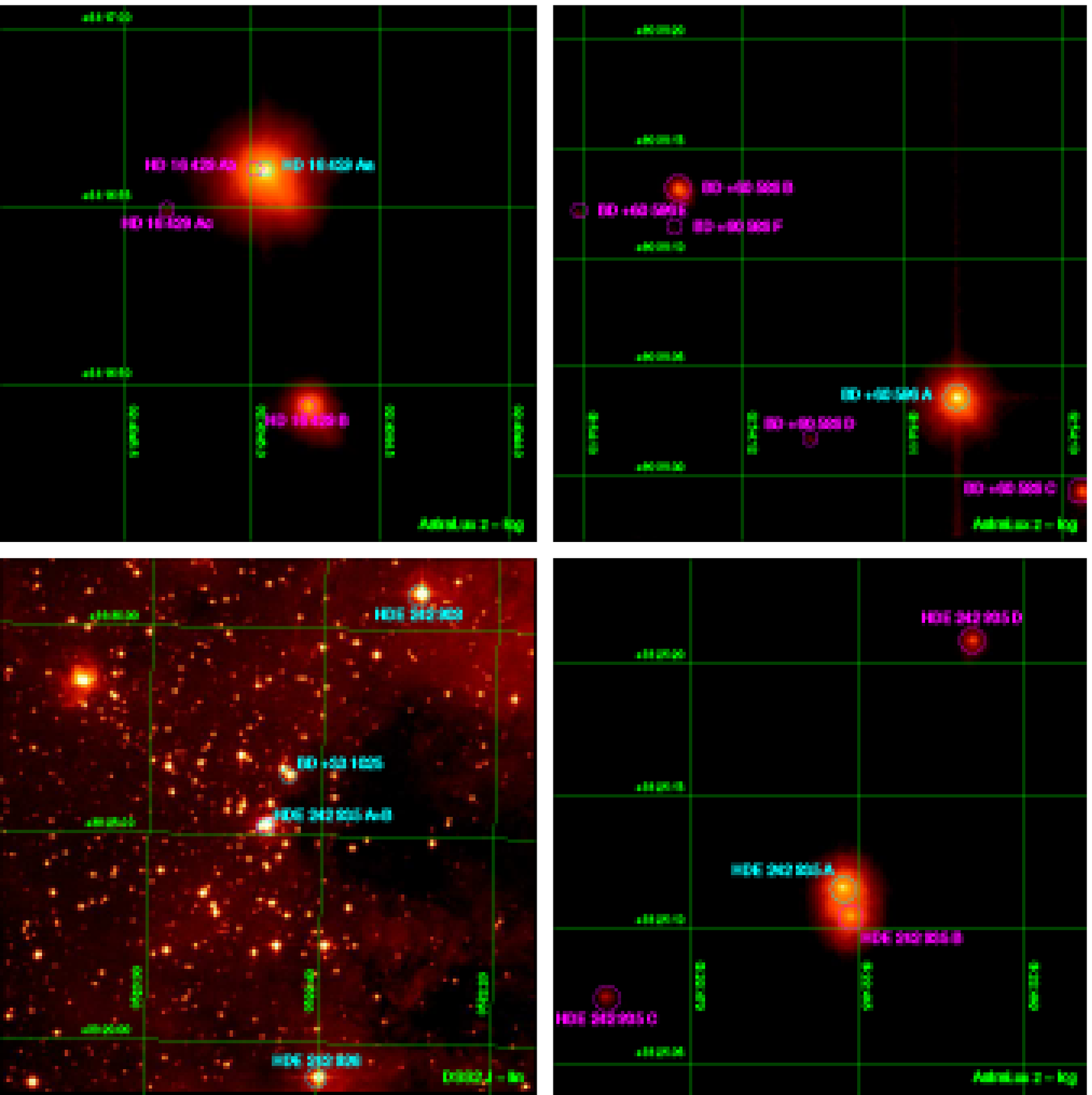}}
%\centerline{Figure unavailable in astro-ph version.}
\caption{(continued).}
\label{chart4}
\end{figure*}	

\addtocounter{figure}{-1}

\begin{figure*}
%\centerline{\includegraphics*[width=\linewidth, bb=50 225 545 725]{charts5.ps}}
%\centerline{\includegraphics*[width=\linewidth, bb=50 225 545 725]{fig19e.ps}}
\centerline{\includegraphics*[width=\linewidth]{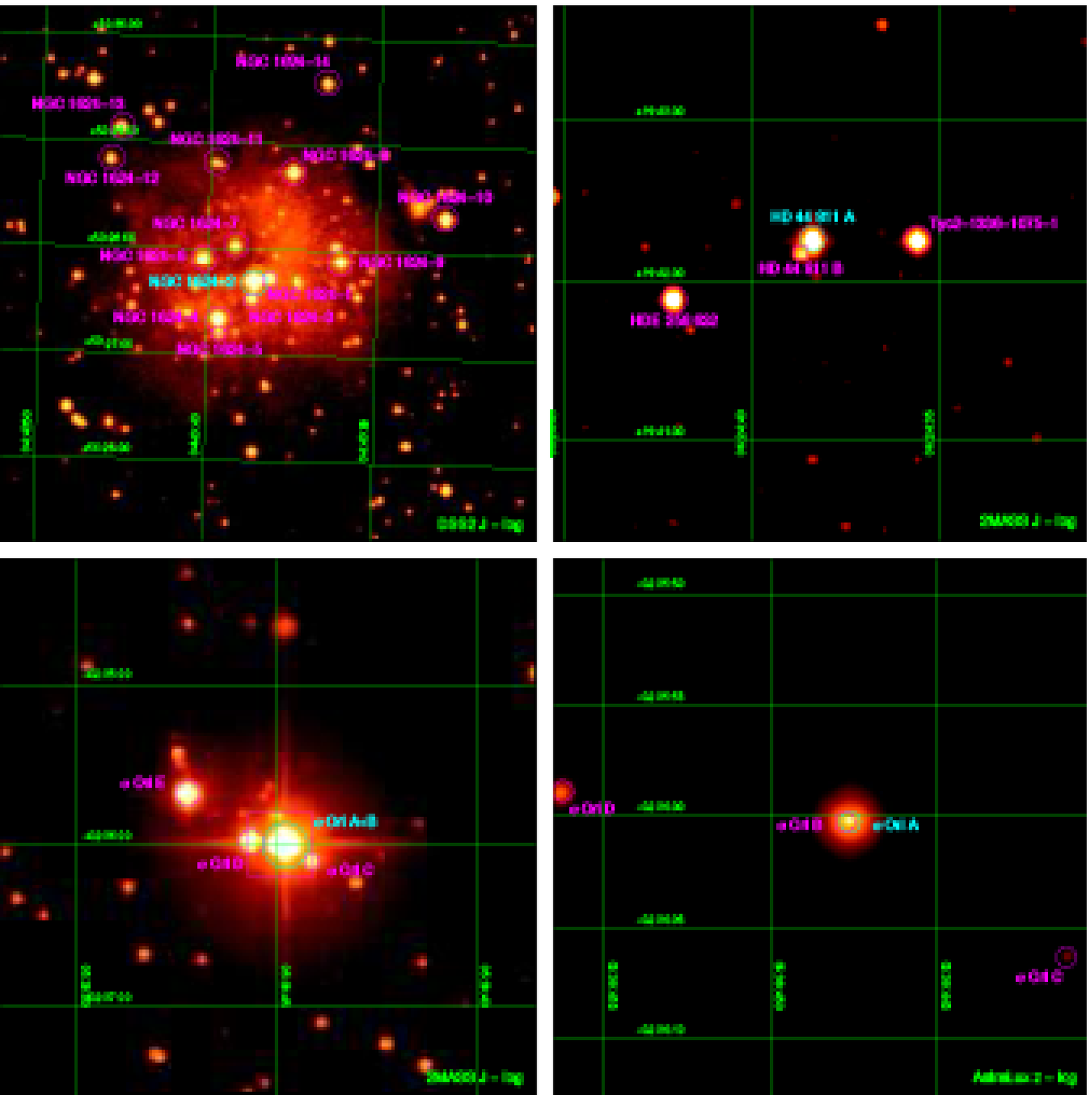}}
%\centerline{Figure unavailable in astro-ph version.}
\caption{(continued).}
\label{chart5}
\end{figure*}	

\addtocounter{figure}{-1}

\begin{figure*}
%\centerline{\includegraphics*[width=\linewidth, bb=50 225 545 725]{charts6.ps}}
%\centerline{\includegraphics*[width=\linewidth, bb=50 225 545 725]{fig19f.ps}}
\centerline{\includegraphics*[width=\linewidth]{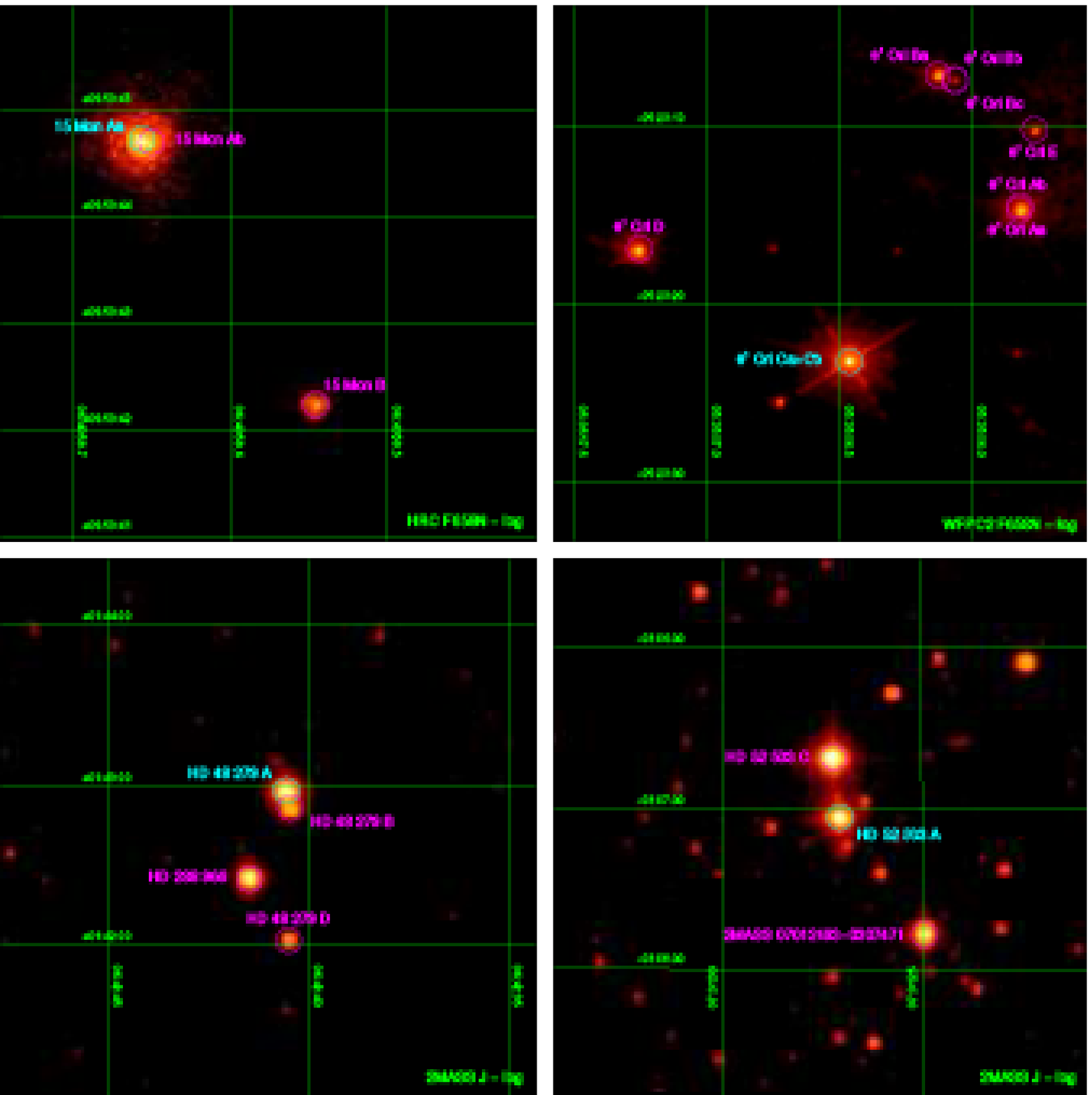}}
%\centerline{Figure unavailable in astro-ph version.}
\caption{(continued).}
\label{chart6}
\end{figure*}	

\section{Summary}    

We have presented the first installment of a massive new survey of Galactic O-type spectra.
On the basis of our extensive sample of high-quality, digital data in hand, we have reviewed the 
classification system and introduced several refinements designed to improve the accuracy and 
consistency of the spectral types. These include the routine use of luminosity class IV at 
spectral types O6-O8, and most importantly, a redefinition of the spectral-type criteria at late-O 
types so that they are uniform at all luminosity classes for a given subtype. As a consequence, 
some objects previously classified as B0 have moved into the newly defined O9.7 type for classes V 
through III, expanding the definition of the O spectral category. The list of standard spectra 
that define the system has been revised and expanded, including representatives of the new 
subcategories, although a few gaps in the two-dimensional grid remain to be filled from future 
observations. A new O-type classification atlas has likewise been provided. These developments, 
as well as enhanced convenience and accuracy of the digital classification in general, have been 
supported by a powerful new classification software tool that superimposes unknown with any 
standard spectrograms sequentially, along with capabilities to match the line widths and even 
double lines in spectroscopic binaries, iteratively with assumed parameters for the components. 
Attention to spatial resolution of close visual multiple systems has provided significantly improved 
information about their spectra, notably for HD~5005, HD~17\,520, and HDE~242\,935, in which the 
previous composite spectral types were misleading. 

As expected from the substantial increases in the quantity, quality, and homogeneity of our sample, 
new members or characteristics of special categories, and even a new category (Ofc, 
\citealt{Walbetal10a} and above) of O-type spectra have been found. These also include the previously 
defined ON/OC, Onfp, Of?p, Oe, and SB categories, all of which have been discussed. Extensive notes 
and references have been given for many individual stars, both normal and peculiar; plots of all the 
spectrograms are provided, as well as charts for crowded regions. Analogous developments will be 
forthcoming in future installments of our program, particularly as we incorporate large numbers of 
fainter stars that have in general been less well observed previously than those presented here.  
Further astronomical and astrophysical discussion and applications of our results will be undertaken
when the full sample is complete.      

\begin{acknowledgements}

%We would like to thank Brian Mason and Doug Gies for their help with the historically confusing issue of the
%assignment of the 25\,638 and 25\,639 HD numbers.
Support for this work was provided by [a] the Spanish Government Ministerio de Ciencia e Innovaci\'on through 
grant AYA2007-64052, the Ram\'on y Cajal Fellowship program, and FEDER funds; [b] the Junta de Andaluc\'{\i}a
grant P08-TIC-4075; [c] NASA through grants GO-10205, GO-10602, and GO-10898 from the Space Telescope 
Science Institute, which is operated by the Association of Universities for Research in Astronomy Inc., under 
NASA contract NAS~5-26555; [d] the Direcci\'on de Investigaci\'on de la Universidad de La Serena (DIULS 
PR09101); and [e] the ESO-Government of Chile Joint Committee Postdoctoral Grant.
This research has made extensive use of [a] Aladin \citep{Bonnetal00};
[b] the SIMBAD database, operated at CDS, Strasbourg, France; and [c] the Washington Double Star Catalog,
maintained at the U.S. Naval Observatory \citep{Masoetal01}.

\end{acknowledgements}

\bibliographystyle{apj}
\bibliography{general}

\eject

\begin{figure*}
%\centerline{\includegraphics*[width=\linewidth]{Nora.ps}}
\centerline{\includegraphics*[width=\linewidth]{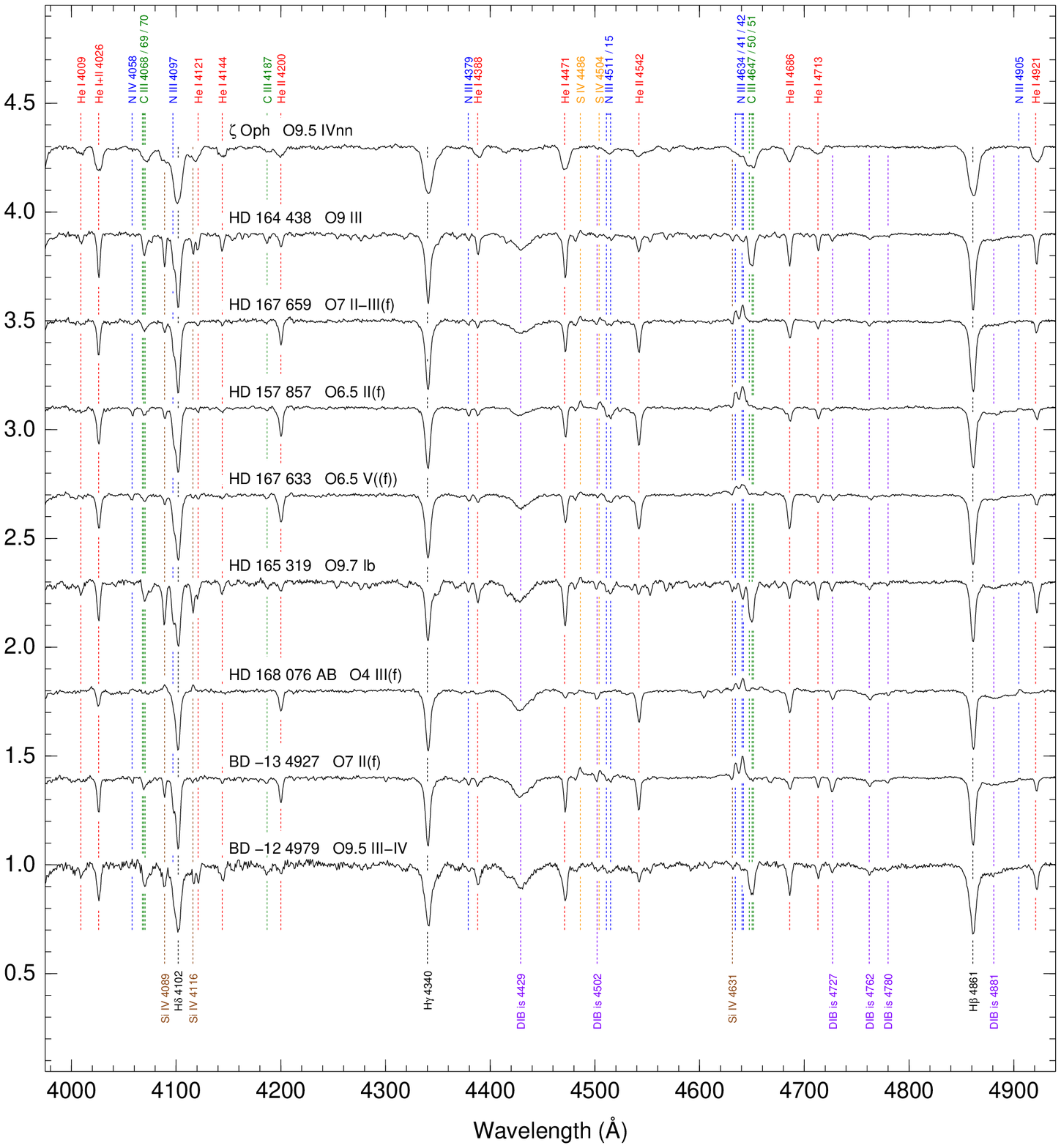}}
\caption{Spectrograms for normal stars. [See the electronic version of the journal for a color version of this figure.]}
\label{Nora}   
\end{figure*}	

\addtocounter{figure}{-1}

\begin{figure*}
%\centerline{\includegraphics*[width=\linewidth]{Norb.ps}}
\centerline{\includegraphics*[width=\linewidth]{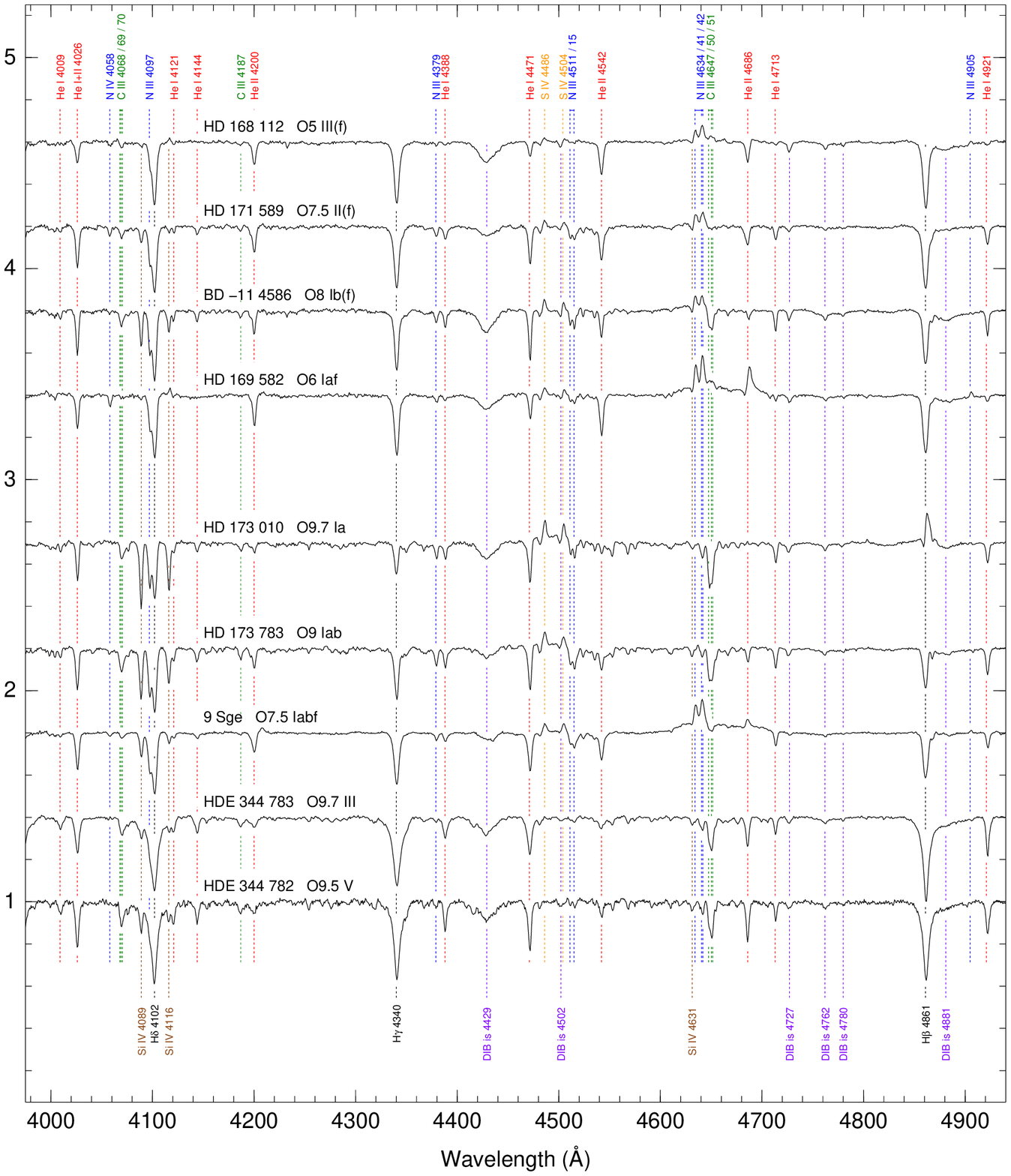}}
\caption{(continued).}
\label{Norb}   
\end{figure*}

\addtocounter{figure}{-1}

\begin{figure*}
%\centerline{\includegraphics*[width=\linewidth]{Norc.ps}}
\centerline{\includegraphics*[width=\linewidth]{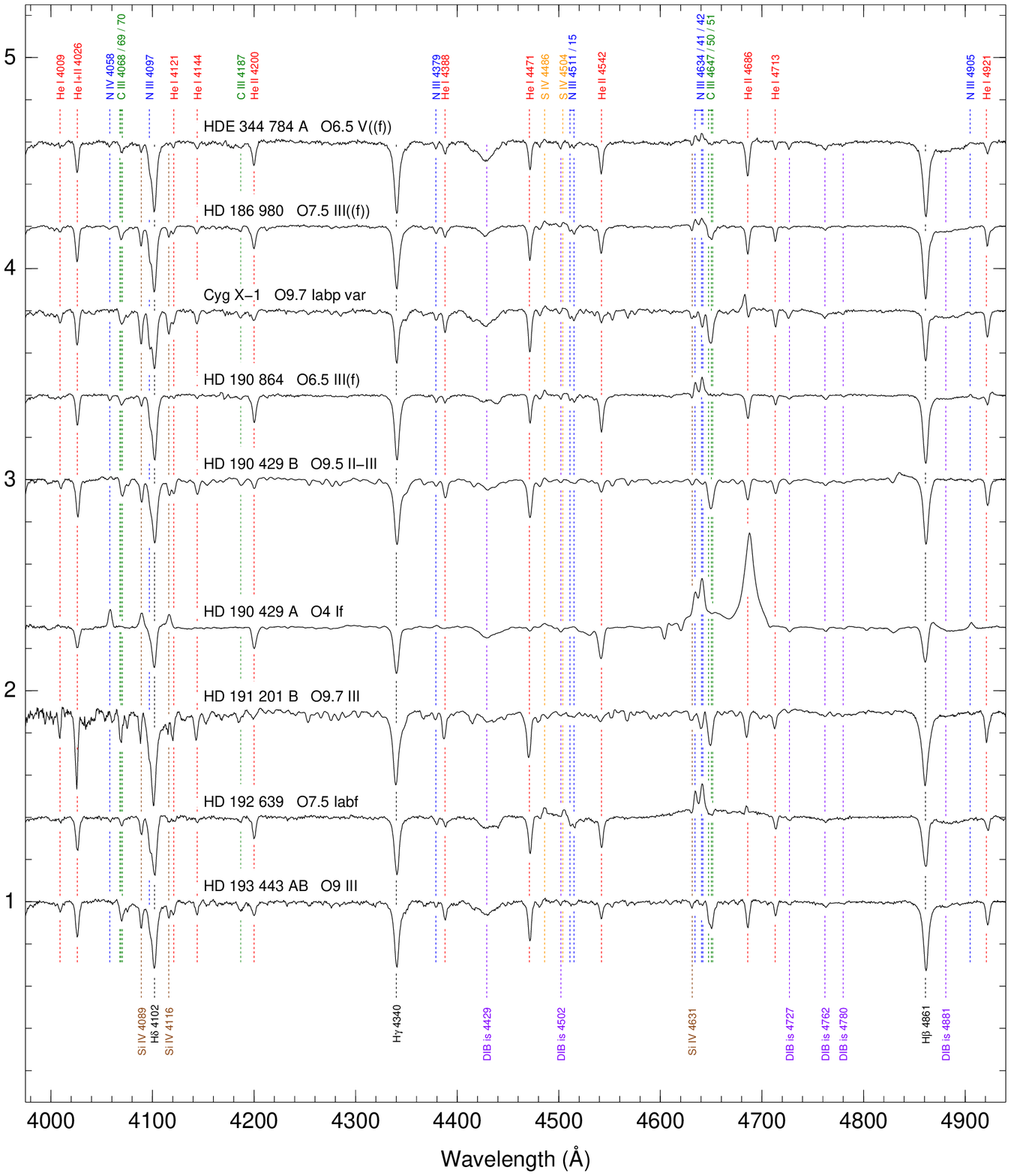}}
\caption{(continued).}
\label{Norc}   
\end{figure*}

\addtocounter{figure}{-1}

\begin{figure*}
%\centerline{\includegraphics*[width=\linewidth]{Nord.ps}}
\centerline{\includegraphics*[width=\linewidth]{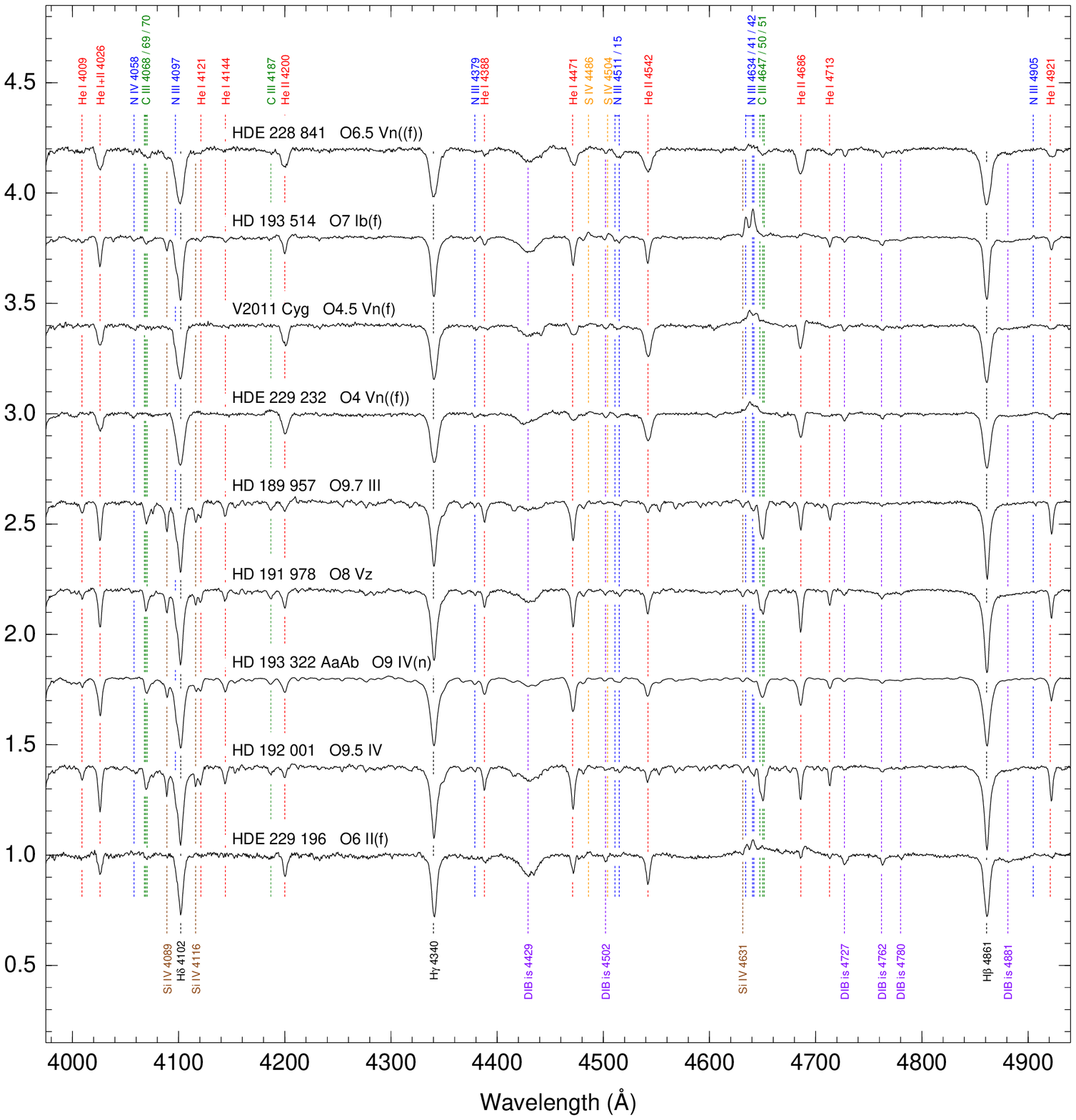}}
\caption{(continued).}
\label{Nord}   
\end{figure*}

\addtocounter{figure}{-1}

\begin{figure*}
%\centerline{\includegraphics*[width=\linewidth]{Nore.ps}}
\centerline{\includegraphics*[width=\linewidth]{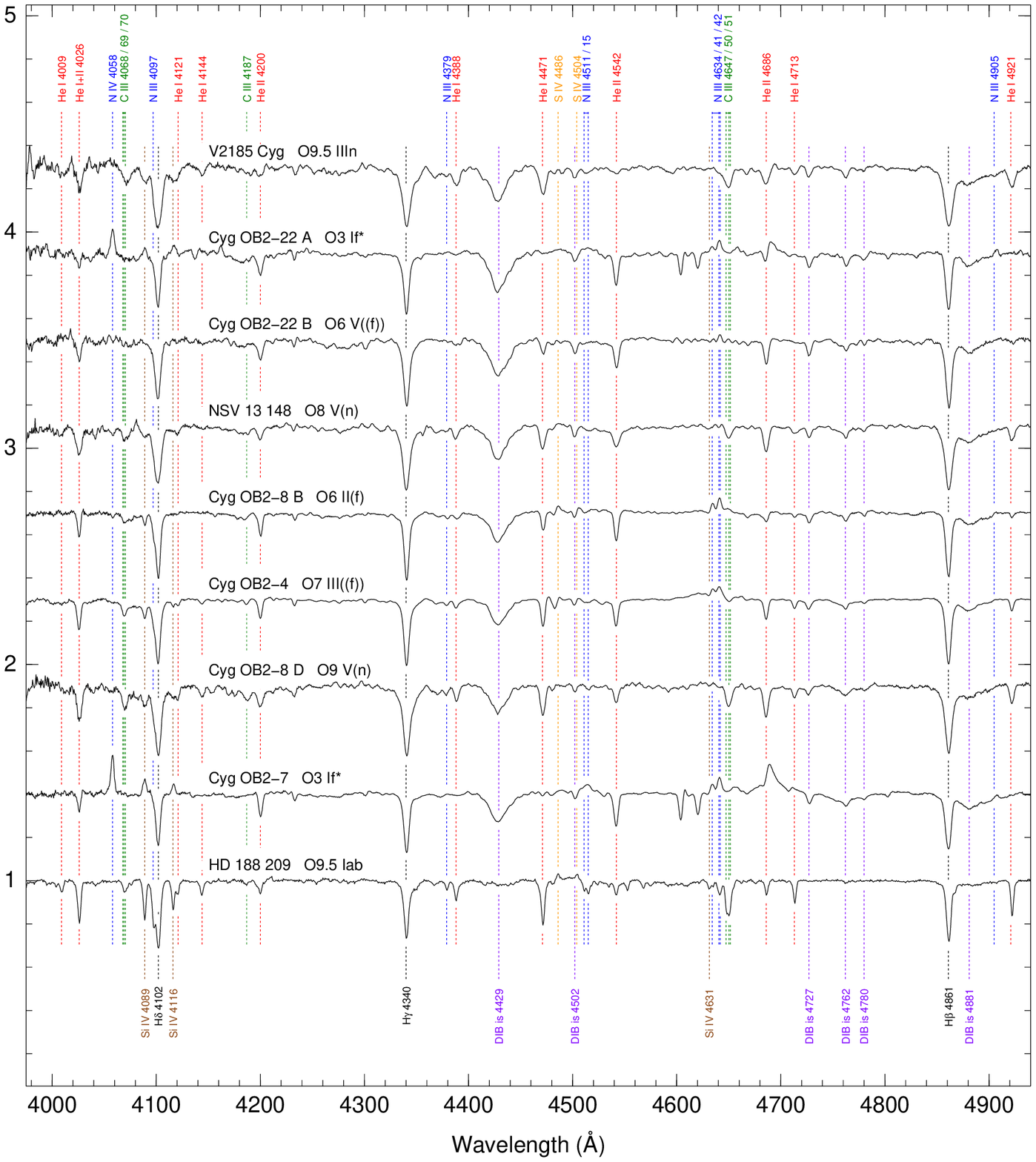}}
\caption{(continued).}
\label{Nore}   
\end{figure*}

\addtocounter{figure}{-1}

\begin{figure*}
%\centerline{\includegraphics*[width=\linewidth]{Norf.ps}}
\centerline{\includegraphics*[width=\linewidth]{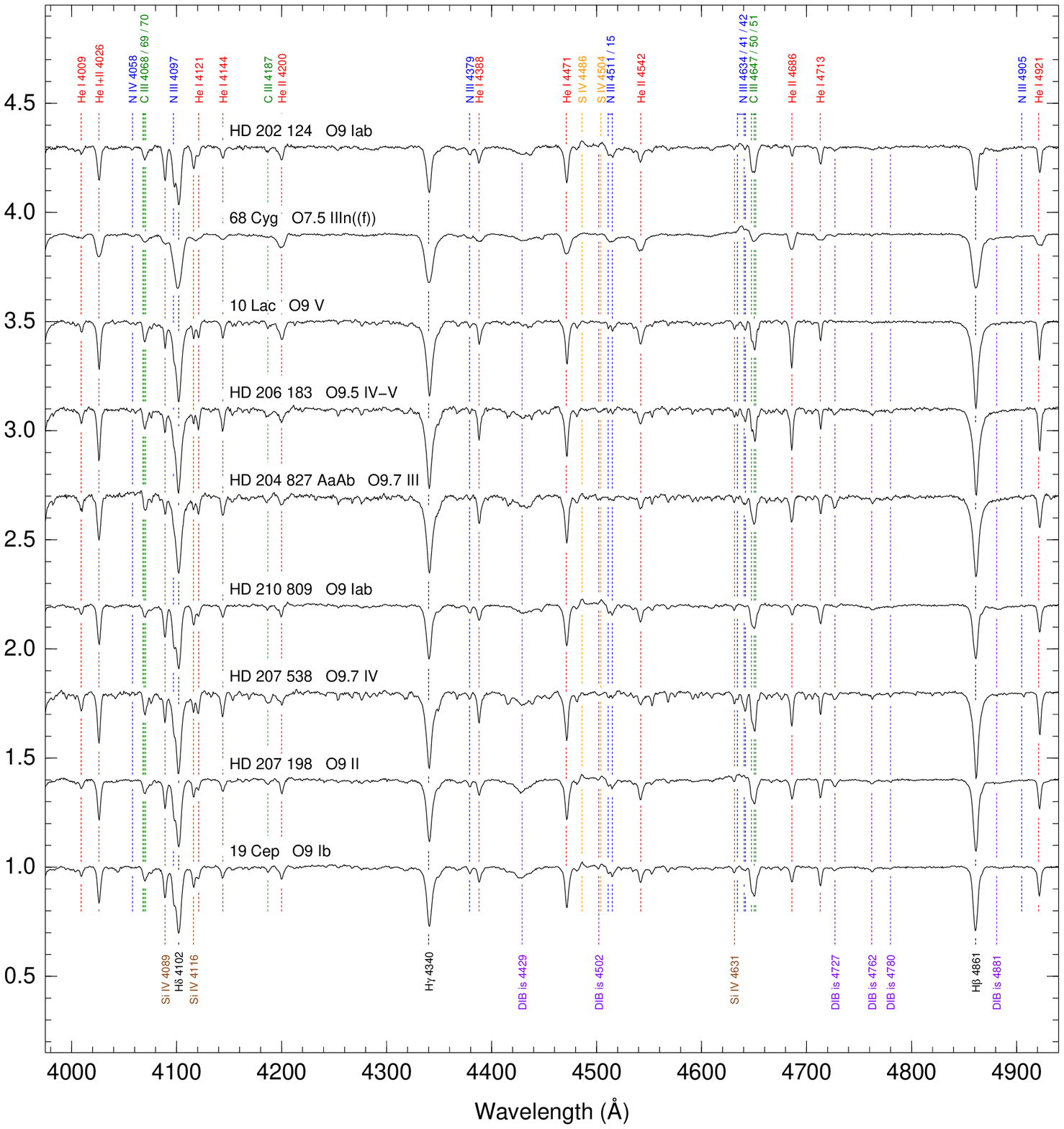}}
\caption{(continued).}
\label{Norf}   
\end{figure*}

\addtocounter{figure}{-1}

\begin{figure*}
%\centerline{\includegraphics*[width=\linewidth]{Norg.ps}}
\centerline{\includegraphics*[width=\linewidth]{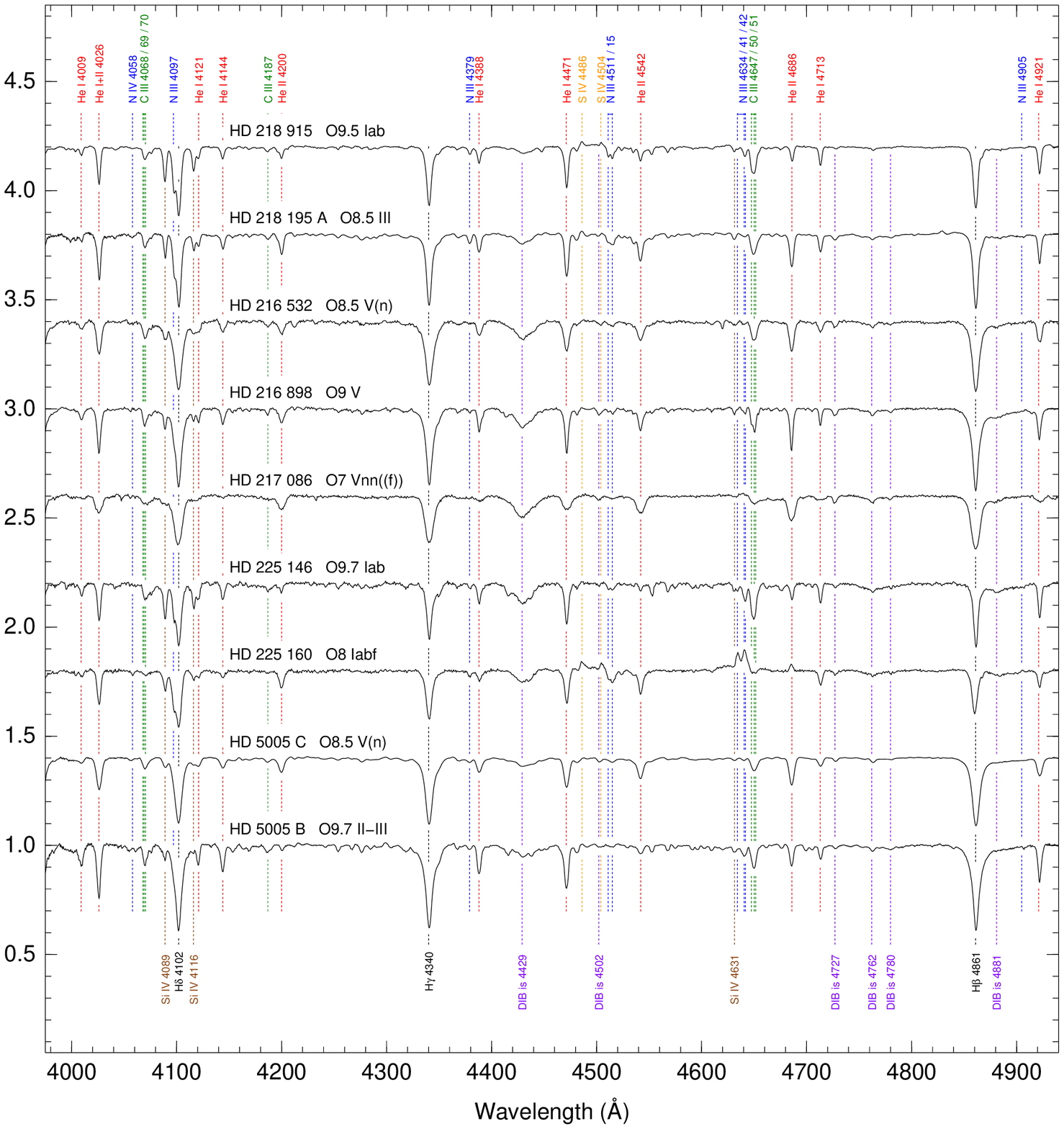}}
\caption{(continued).}
\label{Norg}   
\end{figure*}

\addtocounter{figure}{-1}

\begin{figure*}
%\centerline{\includegraphics*[width=\linewidth]{Norh.ps}}
\centerline{\includegraphics*[width=\linewidth]{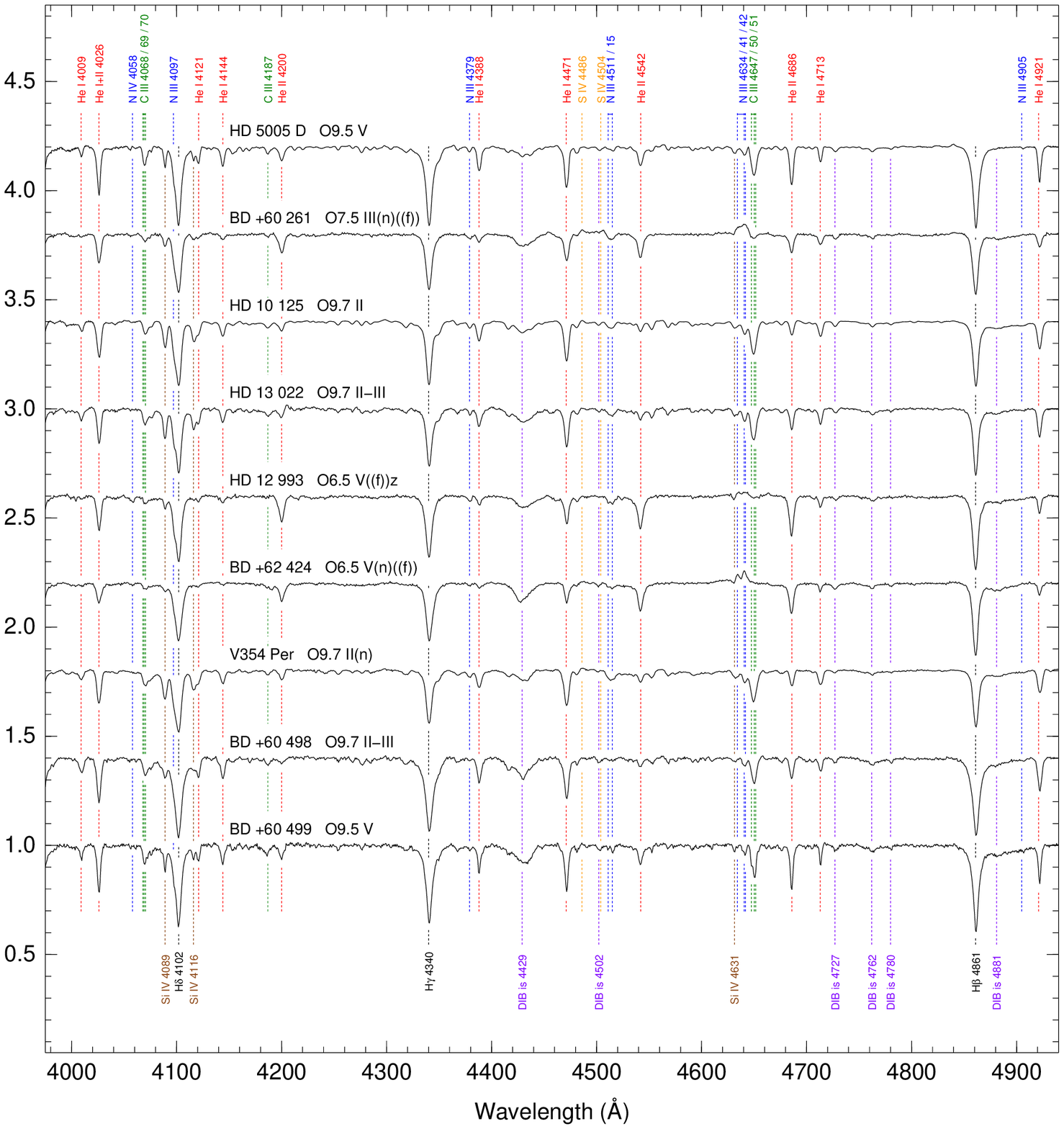}}
\caption{(continued).}
\label{Norh}   
\end{figure*}

\addtocounter{figure}{-1}

\begin{figure*}
%\centerline{\includegraphics*[width=\linewidth]{Nori.ps}}
\centerline{\includegraphics*[width=\linewidth]{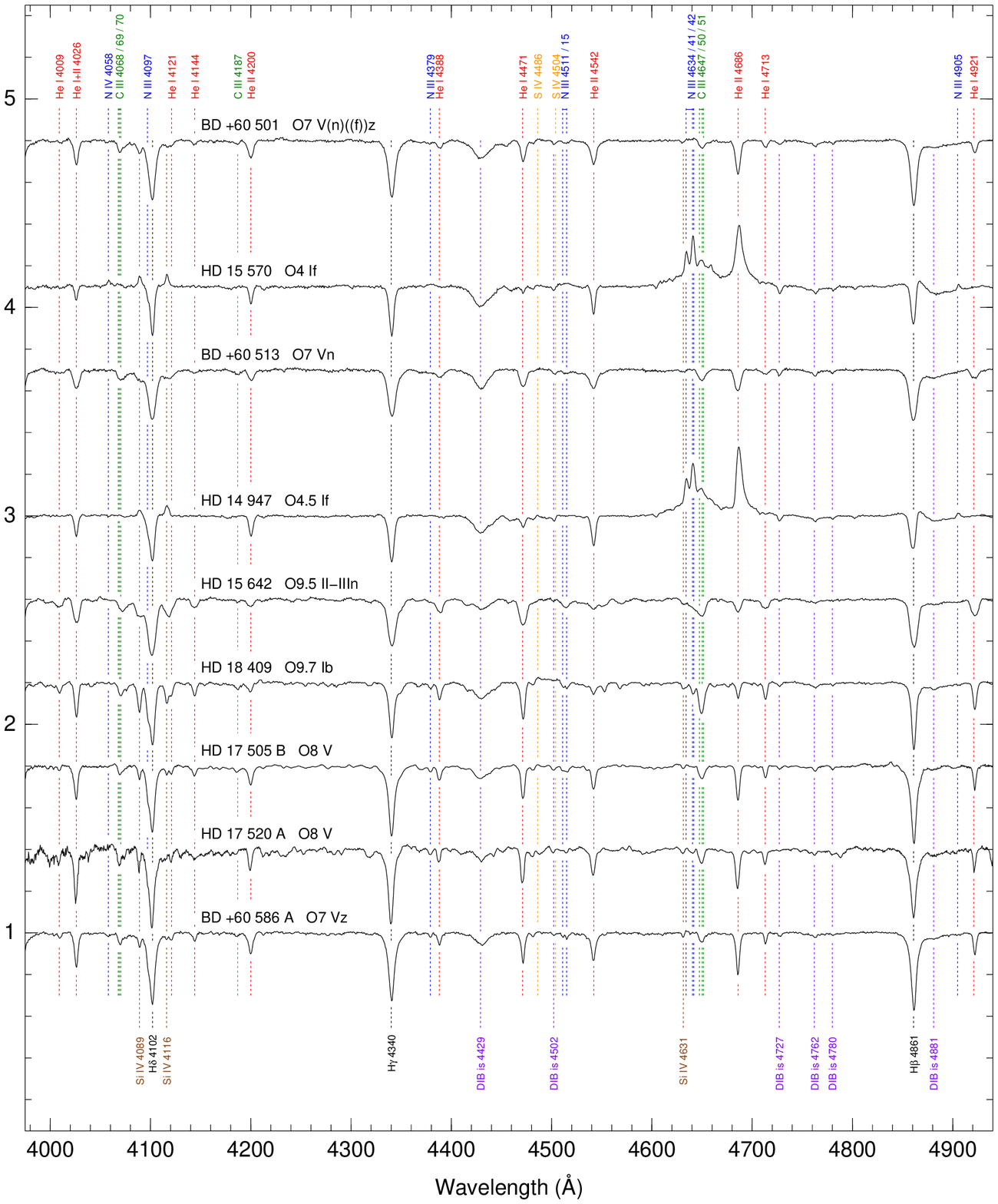}}
\caption{(continued).}
\label{Nori}   
\end{figure*}

\addtocounter{figure}{-1}

\begin{figure*}
%\centerline{\includegraphics*[width=\linewidth]{Norj.ps}}
\centerline{\includegraphics*[width=\linewidth]{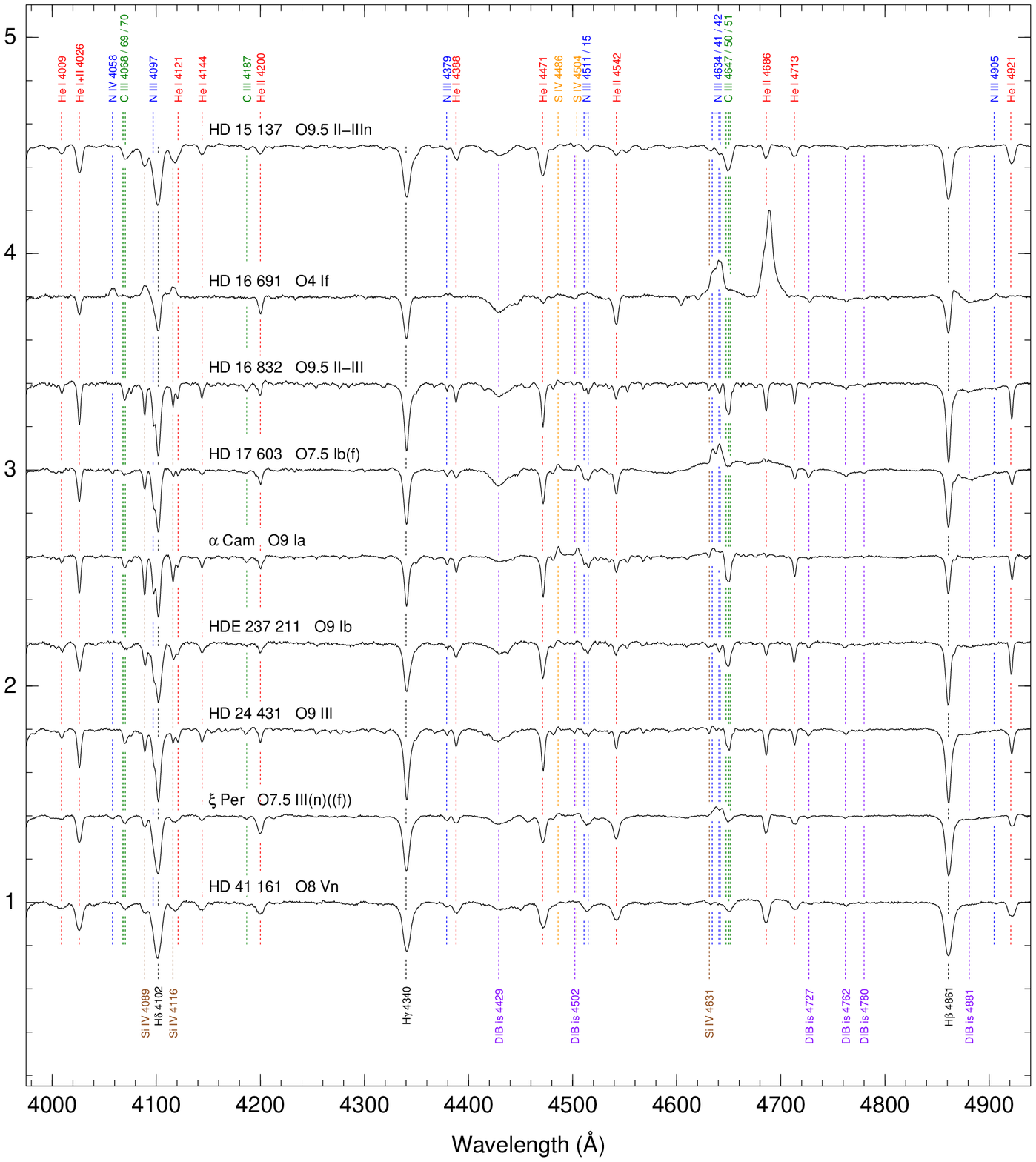}}
\caption{(continued).}
\label{Norj}   
\end{figure*}

\addtocounter{figure}{-1}

\begin{figure*}
%\centerline{\includegraphics*[width=\linewidth]{Nork.ps}}
\centerline{\includegraphics*[width=\linewidth]{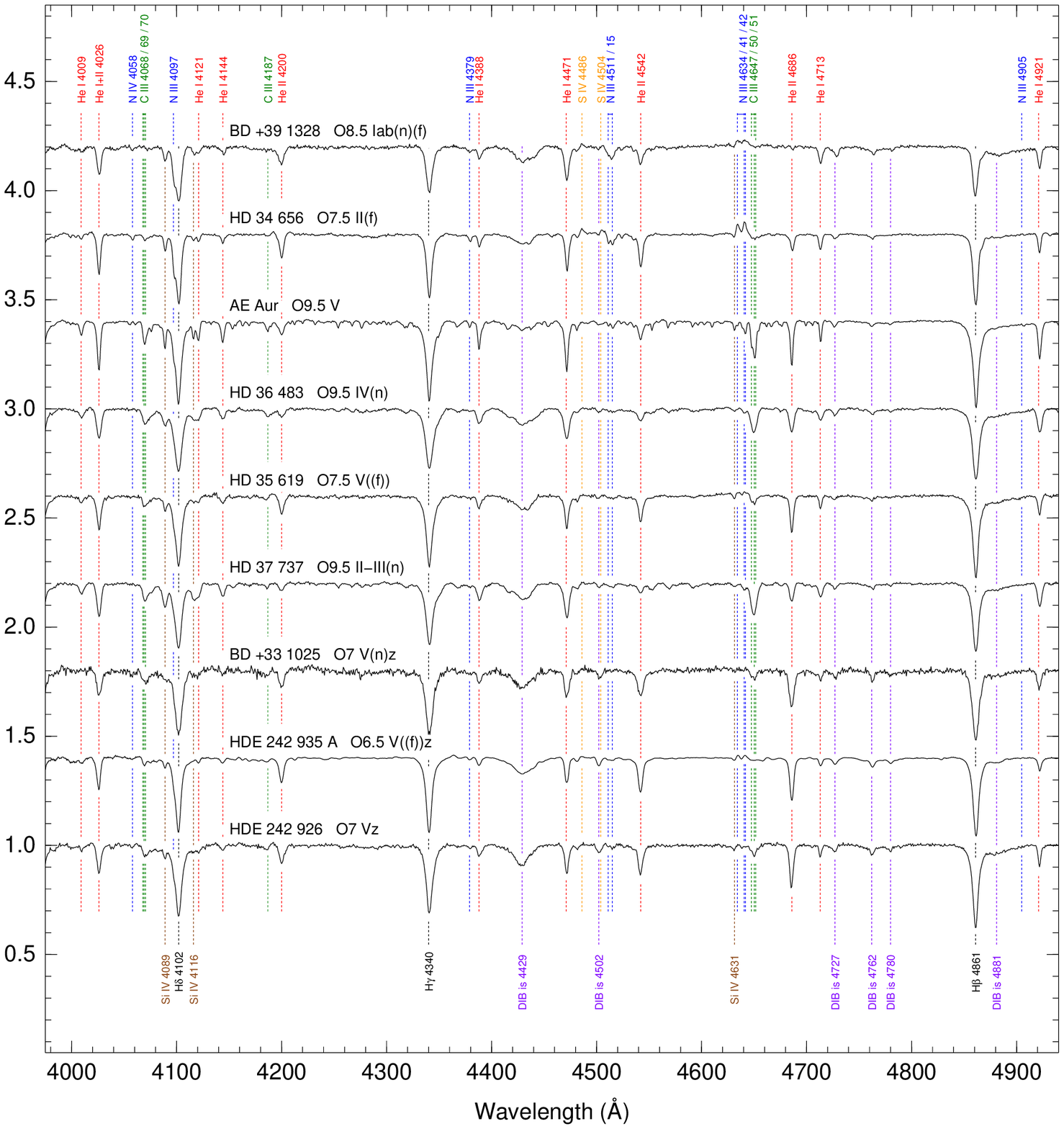}}
\caption{(continued).}
\label{Nork}   
\end{figure*}

\addtocounter{figure}{-1}

\begin{figure*}
%\centerline{\includegraphics*[width=\linewidth]{Norl.ps}}
\centerline{\includegraphics*[width=\linewidth]{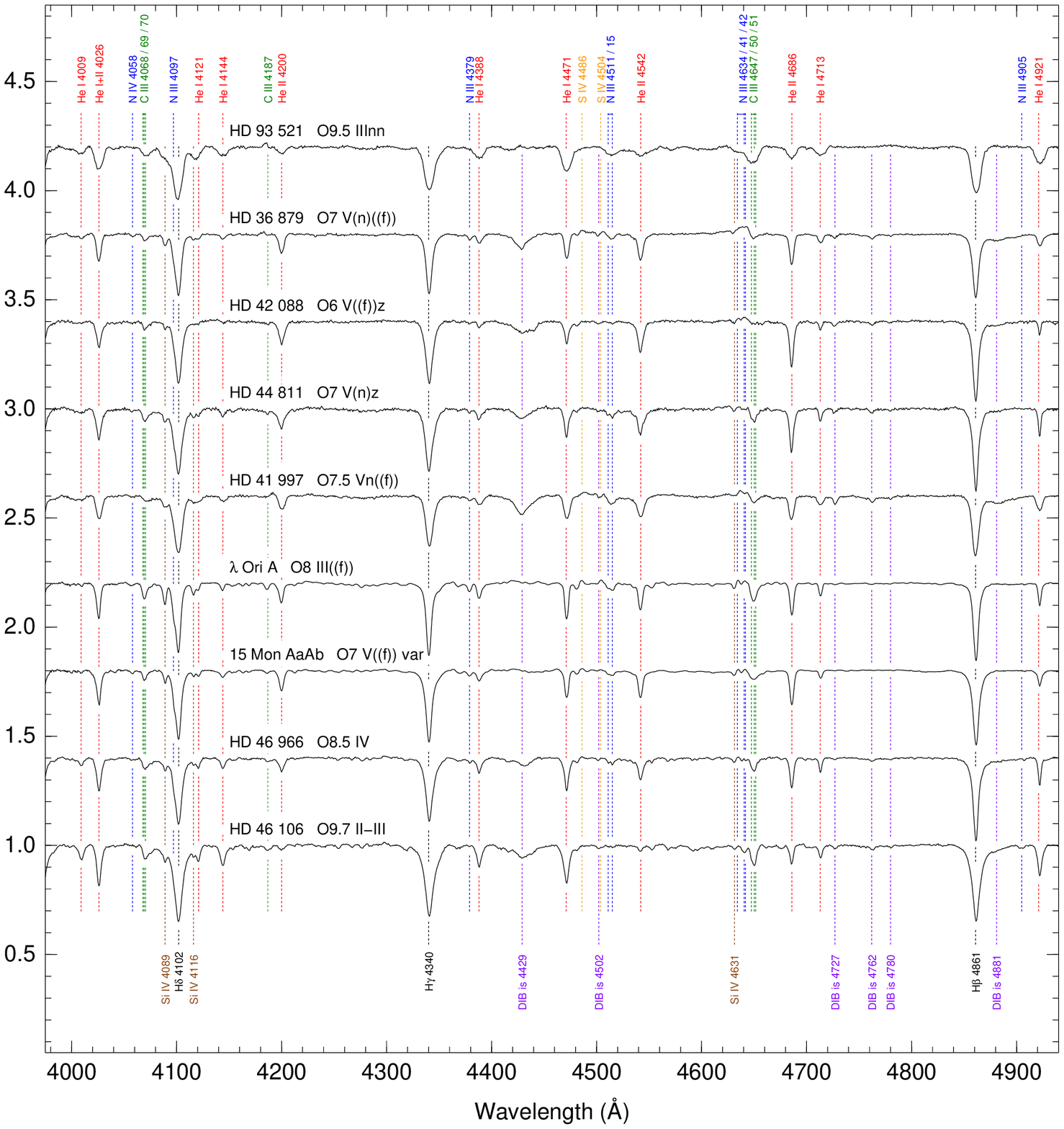}}
\caption{(continued).}
\label{Norl}   
\end{figure*}

\addtocounter{figure}{-1}

\begin{figure*}
%\centerline{\includegraphics*[width=\linewidth]{Norm.ps}}
\centerline{\includegraphics*[width=\linewidth]{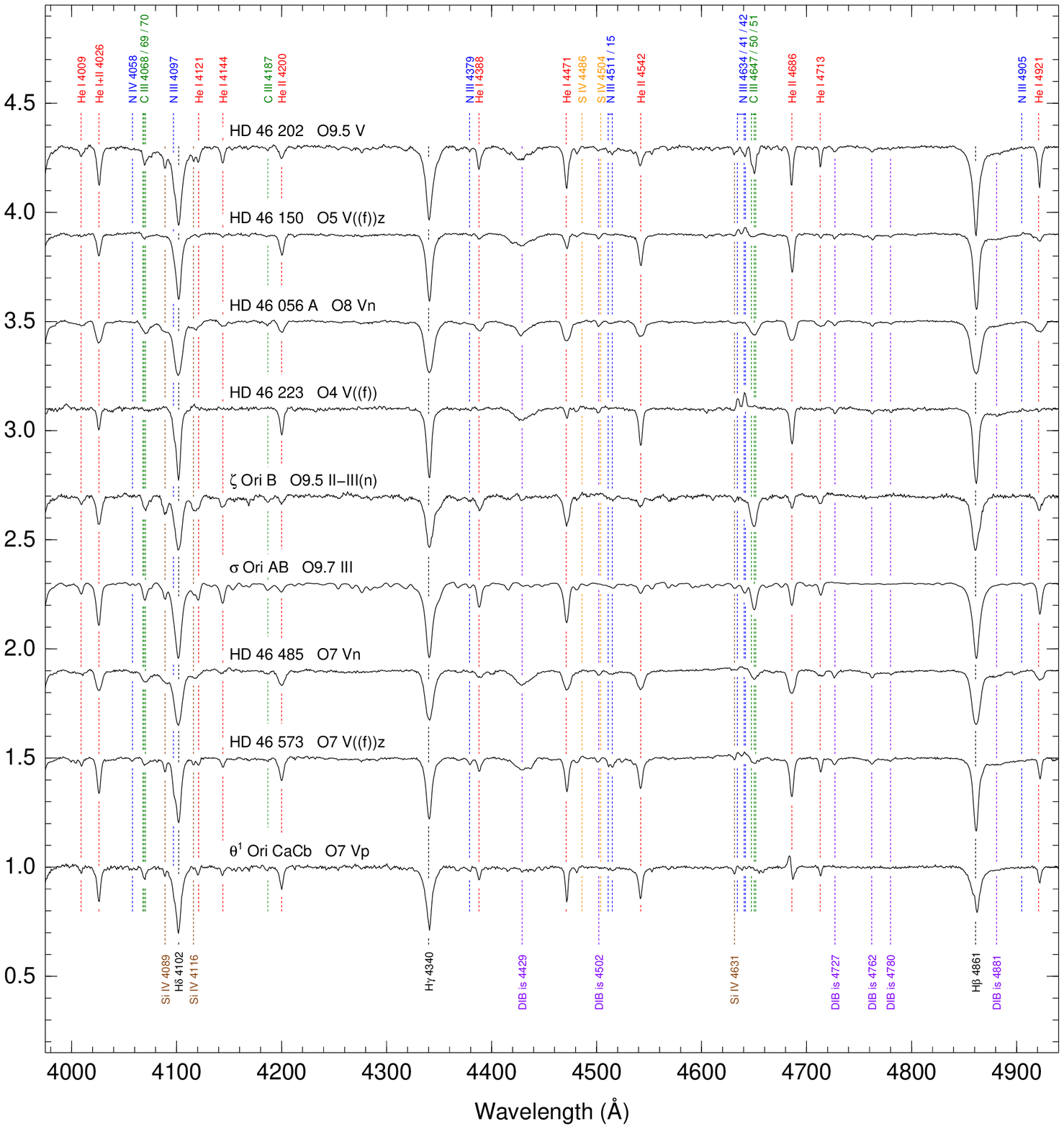}}
\caption{(continued).}
\label{Norm}   
\end{figure*}

\addtocounter{figure}{-1}

\begin{figure*}
%\centerline{\includegraphics*[width=\linewidth]{Norn.ps}}
\centerline{\includegraphics*[width=\linewidth]{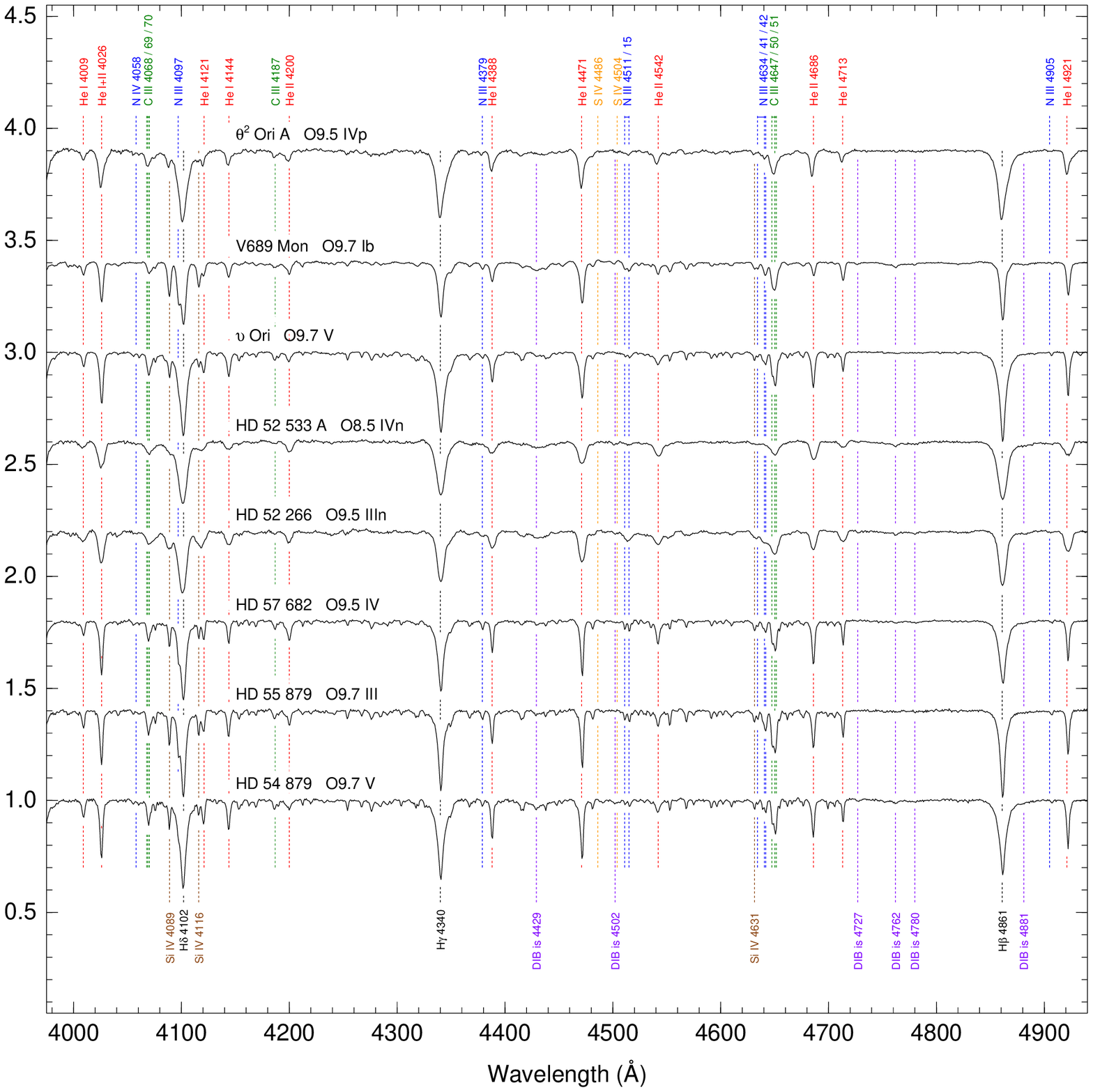}}
\caption{(continued).}
\label{Norn}   
\end{figure*}	

\setlength{\textheight}{280mm}
\setlength{\topmargin}{-15.0mm}
\input{maintable}

\end{document}

%% file: standards.tex
\begin{tabular}{llllllll}
\\
\hline
 & \multicolumn{1}{c}{V} & \multicolumn{1}{c}{IV} & \multicolumn{1}{c}{III} & \multicolumn{1}{c}{II} & \multicolumn{1}{c}{Ib} & \multicolumn{1}{c}{Iab/I} & \multicolumn{1}{c}{Ia} \\
\hline
O2   &                      &                        &                       &                         &                   & {\it HD 93\,129 AaAb} &                         \\
\hline
O3   & {\it HD 64\,568}     &                        & \nodata               &                         &                   &      Cyg OB2-7        &                         \\
\hline
O3.5 & {\it HD 93\,128}     &                        & {\it Pismis 24-17}    &                         &                   & {\it Pismis 24-1 AB}  &                         \\
\hline
O4   & {\bf HD 46\,223}     &                        & {\bf HD 168\,076 AB}  &                         &                   &      HD 15\,570       &                         \\
     & {\it HD 96\,715}     &                        & {\it HD 93\,250}      &                         &                   &      HD 16\,691       &                         \\
     &                      &                        &                       &                         &                   &      HD 190\,429 A    &                         \\
\hline
O4.5 &      HD 15\,629      &                        &      Cyg OB2-8 C      &                         &                   &      HD 14\,947       &                         \\
     & {\it HDE 303\,308}   &                        &                       &                         &                   &      Cyg OB2-9        &                         \\
\hline
O5   & {\bf HD 46\,150}     &                        & {\bf HD 168\,112}     &                         &                   & {\it CPD -47 2963}    &                         \\
     & {\it HDE 319\,699}   &                        & {\it HD 93\,403}      &                         &                   &                       &                         \\
     &                      &                        & {\it HD 93\,843}      &                         &                   &                       &                         \\
\hline
O5.5 & {\it HD 93\,204}     &                        & \nodata               &                         &                   &      Cyg OB2-11       &                         \\
\hline
O6   &      HD 42\,088      & {\it HD 101\,190}      & \nodata               &      HDE 229\,196       & \nodata           & \nodata               & {\bf HD 169\,582}       \\
     & {\it HDE 303\,311}   &                        &                       &                         &                   &                       &                         \\
\hline
O6.5 & {\it HD 91\,572}     & {\it HDE 322\,417}     &      HD 190\,864      & {\bf HD 157\,857}       & \nodata           & \nodata               & {\it HD 163\,758}       \\
     &      HD 12\,993      &                        & {\it HD 96\,946}      &                         &                   &                       &                         \\
     &                      &                        & {\it HD 152\,723}     &                         &                   &                       &                         \\
     &                      &                        & {\it HD 156\,738}     &                         &                   &                       &                         \\
\hline
O7   & {\it HD 93\,146}     & \nodata                &      Cyg OB2-4        & {\it HD 94\,963}        & {\it HD 69\,464}  & \nodata               & \nodata                 \\
     &      HDE 242\,926    &                        &                       & {\it HD 151\,515}       &      HD 193\,514  &                       &                         \\
     & {\it HD 91\,824}     &                        &                       &                         &                   &                       &                         \\
     & {\it HD 93\,222}     &                        &                       &                         &                   &                       &                         \\
     & {\bf 15 Mon AaAb}    &                        &                       &                         &                   &                       &                         \\
\hline
O7.5 & {\it HDE 319\,703 A} & \nodata                & {\it HD 163\,800}     &      HD 34\,656         &      HD 17\,603   &      HD 192\,639      & \nodata                 \\
     & {\it HD 152\,590}    &                        &                       & {\bf HD 171\,589}       & {\it HD 156\,154} & {\bf 9 Sge}           &                         \\
\hline
O8   &      HD 191\,978     & {\it HD 97\,166}       & {\it HDE 319\,702}    & {\it HD 162\,978}       & {\bf BD -11 4586} &      HD 225\,160      & {\it HD 151\,804}       \\
     & {\it HD 97\,848}     &                        & {\bf $\lambda$ Ori A} &                         &                   &                       &                         \\
\hline
O8.5 & {\bf HD 46\,149}     & {\bf HD 46\,966}       & {\it HD 114\,737}     & {\it HD 75\,211}        & {\it HD 125\,241} & \nodata               & {\it HDE 303\,492}      \\
     & {\it HD 57\,236}     &                        &      HD 218\,195 A    &                         &                   &                       &                         \\
     &      HD 14\,633      &                        &                       &                         &                   &                       &                         \\
\hline
O9   &      10 Lac          & {\it CPD -41 7733}     &      HD 24\,431       & {\it $\tau$ CMa}        &      19 Cep       &      HD 202\,124      &      $\alpha$ Cam       \\
     &      HD 216\,898     & {\it HD 93\,028}       & {\it HD 93\,249}      &      HD 207\,198        &                   & {\it HD 148\,546}     &                         \\
     &                      &                        &      HD 193\,443 AB   & {\it HD 71\,304}        &                   & {\it HD 152\,249}     &                         \\
\hline
O9.5 &      AE Aur          &      HD 192\,001       & {\it HD 96\,264}      & {\bf $\delta$ Ori AaAb} & {\it HD 76\,968}  &      HD 188\,209      & \nodata                 \\
     & {\bf HD 46\,202}     & {\it HD 93\,027}       &                       &                         &                   & {\it HD 154\,368}     &                         \\
     &      HD 12\,323      & {\it HD 155\,889}      &                       &                         &                   & {\it HD 123\,008}     &                         \\
     &                      & {\it HD 96\,622}       &                       &                         &                   &                       &                         \\
\hline
O9.7 & {\bf $\upsilon$ Ori} &      HD 207\,538       &      HD 189\,957      & {\it HD 68\,450}        & {\bf V689 Mon}    &      HD 225\,146      &      HD 195\,592        \\
     &                      &                        & {\it HD 154\,643}     & {\it HD 152\,405}       &                   & {\it HD 75\,222}      & {\bf HD 173\,010}       \\
     &                      &                        &                       &      HD 10\,125         &                   & {\it $\mu$ Nor}       & {\it HD 105\,056}       \\
     &                      &                        &                       &                         &                   &                       & {\it HD 152\,424}       \\
\hline
B0   & {\it $\tau$ Sco}     & \nodata                & {\bf HD 48\,434}      & \nodata                 & \nodata           & \nodata               & {\bf $\varepsilon$ Ori} \\
     &                      &                        &                       &                         &                   &                       & {\it HD 122\,879}       \\
\hline
B0.2 &      HD 2083         & {\bf $\phi$$^{1}$ Ori} &      HD 6675          & \nodata                 & \nodata           & \nodata               & \nodata                 \\
\hline
B0.5 & {\bf HD 36\,960}     & \nodata                &      1 Cas            & \nodata                 & \nodata           & \nodata               & {\bf $\kappa$ Ori}      \\
\hline
Notes & \multicolumn{7}{l}{Normal, {\it italic}, and {\bf bold} typefaces are used for stars with $\delta > +20\degr$, $\delta < -20\degr$, and the equatorial in-between region, respectively.} \\
      & \multicolumn{7}{l}{$\upsilon$ Ori, a previous B0 V standard, is now an O9.7 V.} \\
      & \multicolumn{7}{l}{$\tau$ Sco, a previous B0.2 V standard, is now a B0 V.} \\
      & \multicolumn{7}{l}{HD 189\,957, a previous O9.5 III standard is now an O9.7 III.}
\end{tabular}

%% file: maintable.tex
\begin{deluxetable}{llllllllllll}
\rotate
\tablecaption{Spectral classifications}
\tablewidth{0pt}
\tabletypesize{\scriptsize}
\tablehead{\colhead{Name} & \colhead{GOSSS ID} & \colhead{RA (J2000)} & \colhead{dec (J2000)} & \colhead{SC} & \colhead{LC} & \colhead{Qual.} & \colhead{Second.} & \colhead{Altern. classification} & \colhead{Ref.} & \colhead{Sect.} & \colhead{Flag}}
\startdata
$\zeta$ Oph             & GOS 006.28$+$23.59\_01 & 16:37:09.530 & $-$10:34:01.75 & O9.5  & IV      & nn          & \nodata    & \nodata                                   & \nodata & \ref{sec:Nor}   & ch      \\
HD 164\,438             & GOS 010.35$+$01.79\_01 & 18:01:52.279 & $-$19:06:22.07 & O9    & III     & \nodata     & \nodata    & \nodata                                   & \nodata & \ref{sec:Nor}   & \nodata \\
HD 167\,659             & GOS 012.20$-$01.27\_01 & 18:16:58.562 & $-$18:58:05.20 & O7    & II-III  & (f)         & \nodata    & \nodata                                   & \nodata & \ref{sec:Nor}   & ch      \\
HD 167\,771             & GOS 012.70$-$01.13\_01 & 18:17:28.556 & $-$18:27:48.43 & O7    & III     & (f)         & O8 III     & \nodata                                   & \nodata & \ref{sec:SB2}   & ch      \\
HD 157\,857             & GOS 012.97$+$13.31\_01 & 17:26:17.332 & $-$10:59:34.79 & O6.5  & II      & (f)         & \nodata    & \nodata                                   & \nodata & \ref{sec:Nor}   & \nodata \\
HD 167\,633             & GOS 014.34$-$00.07\_01 & 18:16:49.656 & $-$16:31:04.30 & O6.5  & V       & ((f))       & \nodata    & \nodata                                   & \nodata & \ref{sec:Nor}   & ch      \\
HD 165\,319             & GOS 015.12$+$03.33\_01 & 18:05:58.838 & $-$14:11:53.01 & O9.7  & Ib      & \nodata     & \nodata    & \nodata                                   & \nodata & \ref{sec:Nor}   & new     \\
HD 175\,754             & GOS 016.39$-$09.92\_01 & 18:57:35.709 & $-$19:09:11.25 & O8    & II      & (n)(f)p     & \nodata    & \nodata                                   & \nodata & \ref{sec:Onfp}  & ch      \\
HD 168\,075             & GOS 016.94$+$00.84\_01 & 18:18:36.043 & $-$13:47:36.46 & O7    & V       & (n)((f))z   & \nodata    & O6.5 V((f)) + B0-1 V                      & S09     & \ref{sec:SB2}   & ch      \\
HD 168\,076 AB          & GOS 016.94$+$00.84\_02 & 18:18:36.421 & $-$13:48:02.38 & O4    & III     & (f)         & \nodata    & O3.5 V((f+)) + O7.5 V                     & S09     & \ref{sec:Nor}   & ch      \\
BD -13 4927             & GOS 016.98$+$00.85\_01 & 18:18:40.091 & $-$13:45:18.58 & O7    & II      & (f)         & \nodata    & \nodata                                   & \nodata & \ref{sec:Nor}   & ch      \\
MY Ser                  & GOS 018.25$+$01.68\_01 & 18:18:05.895 & $-$12:14:33.30 & O8    & Ia      & f(n)        & O4/5       & O8 I + O5-8 V + O5-8 V                    & L87     & \ref{sec:Onfp}  & ch      \\
BD -12 4979             & GOS 018.25$+$01.69\_01 & 18:18:03.112 & $-$12:14:34.28 & O9.5  & III-IV  & \nodata     & \nodata    & \nodata                                   & \nodata & \ref{sec:Nor}   & new     \\
HD 168\,112             & GOS 018.44$+$01.62\_01 & 18:18:40.868 & $-$12:06:23.38 & O5    & III     & (f)         & \nodata    & \nodata                                   & \nodata & \ref{sec:Nor}   & \nodata \\
HD 171\,589             & GOS 018.65$-$03.09\_01 & 18:36:12.640 & $-$14:06:55.82 & O7.5  & II      & (f)         & \nodata    & \nodata                                   & \nodata & \ref{sec:Nor}   & ch      \\
HD 166\,734             & GOS 018.92$+$03.63\_01 & 18:12:24.656 & $-$10:43:53.03 & O7.5  & Iab     & f           & \nodata    & O7 Ib(f) + O8-9 I                         & W73     & \ref{sec:SB2}   & ch      \\
BD -11 4586             & GOS 019.08$+$02.14\_01 & 18:18:03.344 & $-$11:17:38.83 & O8    & Ib      & (f)         & \nodata    & \nodata                                   & \nodata & \ref{sec:Nor}   & \nodata \\
HD 169\,582             & GOS 021.33$+$01.20\_01 & 18:25:43.147 & $-$09:45:11.02 & O6    & Ia      & f           & \nodata    & \nodata                                   & \nodata & \ref{sec:Nor}   & ch      \\
HD 173\,010             & GOS 023.73$-$02.49\_01 & 18:43:29.710 & $-$09:19:12.60 & O9.7  & Ia      & \nodata     & \nodata    & \nodata                                   & \nodata & \ref{sec:Nor}   & \nodata \\
HD 173\,783             & GOS 024.18$-$03.34\_01 & 18:47:24.183 & $-$09:18:29.50 & O9    & Iab     & \nodata     & \nodata    & \nodata                                   & \nodata & \ref{sec:Nor}   & ch      \\
V442 Sct                & GOS 024.53$-$00.85\_01 & 18:39:03.776 & $-$07:51:35.44 & O6.5  & I       & (n)fp       & \nodata    & \nodata                                   & \nodata & \ref{sec:Onfp}  & ch      \\
9 Sge                   & GOS 056.48$-$04.33\_01 & 19:52:21.765 & $+$18:40:18.75 & O7.5  & Iab     & f           & \nodata    & \nodata                                   & \nodata & \ref{sec:Nor}   & \nodata \\
HDE 344\,783            & GOS 059.37$-$00.15\_01 & 19:43:06.790 & $+$23:16:12.40 & O9.7  & III     & \nodata     & \nodata    & \nodata                                   & \nodata & \ref{sec:Nor}   & new     \\
HDE 344\,782            & GOS 059.40$-$00.14\_01 & 19:43:08.900 & $+$23:18:08.00 & O9.5  & V       & \nodata     & \nodata    & \nodata                                   & \nodata & \ref{sec:Nor}   & new     \\
HDE 344\,784 A          & GOS 059.40$-$00.15\_01 & 19:43:10.970 & $+$23:17:45.38 & O6.5  & V       & ((f))       & \nodata    & \nodata                                   & \nodata & \ref{sec:Nor}   & ch      \\
HD 186\,980             & GOS 067.39$+$03.66\_01 & 19:46:15.902 & $+$32:06:58.16 & O7.5  & III     & ((f))       & \nodata    & \nodata                                   & \nodata & \ref{sec:Nor}   & \nodata \\
Cyg X-1                 & GOS 071.34$+$03.07\_01 & 19:58:21.678 & $+$35:12:05.81 & O9.7  & Iab     & p var       & \nodata    & \nodata                                   & \nodata & \ref{sec:Nor}   & ch      \\
HD 190\,864             & GOS 072.47$+$02.02\_01 & 20:05:39.800 & $+$35:36:27.98 & O6.5  & III     & (f)         & \nodata    & \nodata                                   & \nodata & \ref{sec:Nor}   & \nodata \\
HD 190\,429 B           & GOS 072.58$+$02.61\_01 & 20:03:29.418 & $+$36:01:28.61 & O9.5  & II-III  & \nodata     & \nodata    & \nodata                                   & \nodata & \ref{sec:Nor}   & new     \\
HD 190\,429 A           & GOS 072.59$+$02.61\_01 & 20:03:29.399 & $+$36:01:30.01 & O4    & I       & f           & \nodata    & \nodata                                   & \nodata & \ref{sec:Nor}   & ch      \\
HD 191\,201 A           & GOS 072.75$+$01.78\_01 & 20:07:23.684 & $+$35:43:05.91 & O9.5  & III     & \nodata     & B0 IV      & \nodata                                   & \nodata & \ref{sec:SB2}   & ch      \\
HD 191\,201 B           & GOS 072.75$+$01.78\_02 & 20:07:23.766 & $+$35:43:06.01 & O9.7  & III     & \nodata     & \nodata    & \nodata                                   & \nodata & \ref{sec:Nor}   & new     \\
HD 191\,612             & GOS 072.99$+$01.43\_01 & 20:09:28.608 & $+$35:44:01.31 & O8    & \nodata & f?p var     & \nodata    & \nodata                                   & \nodata & \ref{sec:Of?p}  & ch      \\
HD 192\,639             & GOS 074.90$+$01.48\_01 & 20:14:30.429 & $+$37:21:13.83 & O7.5  & Iab     & f           & \nodata    & \nodata                                   & \nodata & \ref{sec:Nor}   & ch      \\
HDE 228\,766            & GOS 075.19$+$00.96\_01 & 20:17:29.703 & $+$37:18:31.13 & O4    & I       & f           & O8: II:    & \nodata                                   & \nodata & \ref{sec:SB2}   & ch      \\
HD 193\,443 AB          & GOS 076.15$+$01.28\_01 & 20:18:51.707 & $+$38:16:46.50 & O9    & III     & \nodata     & \nodata    & \nodata                                   & \nodata & \ref{sec:Nor}   & \nodata \\
BD +36 4063             & GOS 076.17$-$00.34\_01 & 20:25:40.608 & $+$37:22:27.08 & ON9.7 & Ib      & \nodata     & \nodata    & \nodata                                   & \nodata & \ref{sec:ON/OC} & ch      \\
HDE 228\,841            & GOS 076.60$+$01.68\_01 & 20:18:29.692 & $+$38:52:39.76 & O6.5  & V       & n((f))      & \nodata    & \nodata                                   & \nodata & \ref{sec:Nor}   & \nodata \\
HD 193\,514             & GOS 077.00$+$01.80\_01 & 20:19:08.498 & $+$39:16:24.23 & O7    & Ib      & (f)         & \nodata    & \nodata                                   & \nodata & \ref{sec:Nor}   & \nodata \\
V2011 Cyg               & GOS 077.12$+$03.40\_01 & 20:12:33.121 & $+$40:16:05.45 & O4.5  & V       & n(f)        & \nodata    & \nodata                                   & \nodata & \ref{sec:Nor}   & ch      \\
Y Cyg                   & GOS 077.25$-$06.23\_01 & 20:52:03.577 & $+$34:39:27.51 & O9.5  & IV      & \nodata     & O9.5 IV    & \nodata                                   & \nodata & \ref{sec:SB2}   & ch      \\
HDE 229\,232            & GOS 077.40$+$00.93\_01 & 20:23:59.183 & $+$39:06:15.27 & O4    & V       & n((f))      & \nodata    & \nodata                                   & \nodata & \ref{sec:Nor}   & \nodata \\
HD 189\,957             & GOS 077.43$+$06.17\_01 & 20:01:00.005 & $+$42:00:30.83 & O9.7  & III     & \nodata     & \nodata    & \nodata                                   & \nodata & \ref{sec:Nor}   & ch      \\
HD 191\,978             & GOS 077.87$+$04.25\_01 & 20:10:58.281 & $+$41:21:09.91 & O8    & V       & z           & \nodata    & \nodata                                   & \nodata & \ref{sec:Nor}   & ch      \\
HD 193\,322 AaAb        & GOS 078.10$+$02.78\_01 & 20:18:06.990 & $+$40:43:55.46 & O9    & IV      & (n)         & \nodata    & \nodata                                   & \nodata & \ref{sec:Nor}   & ch      \\
HD 201\,345             & GOS 078.44$-$09.54\_01 & 21:07:55.416 & $+$33:23:49.25 & ON9.5 & IV      & \nodata     & \nodata    & \nodata                                   & \nodata & \ref{sec:ON/OC} & ch      \\
HD 192\,001             & GOS 078.53$+$04.66\_01 & 20:11:01.706 & $+$42:07:36.39 & O9.5  & IV      & \nodata     & \nodata    & \nodata                                   & \nodata & \ref{sec:Nor}   & \nodata \\
HD 191\,423             & GOS 078.64$+$05.37\_01 & 20:08:07.113 & $+$42:36:21.98 & ON9   & II-III  & nn          & \nodata    & \nodata                                   & \nodata & \ref{sec:ON/OC} & ch      \\
HDE 229\,196            & GOS 078.76$+$02.07\_01 & 20:23:10.787 & $+$40:52:29.85 & O6    & II      & (f)         & \nodata    & \nodata                                   & \nodata & \ref{sec:Nor}   & ch      \\
Cyg OB2-5 A             & GOS 080.12$+$00.91\_01 & 20:32:22.422 & $+$41:18:18.91 & O7    & Ia      & fpe         & \nodata    & \nodata                                   & \nodata & \ref{sec:SB2}   & ch      \\
V2185 Cyg               & GOS 080.14$+$00.74\_01 & 20:33:09.600 & $+$41:13:00.60 & O9.5  & III     & n           & \nodata    & \nodata                                   & \nodata & \ref{sec:Nor}   & new     \\
Cyg OB2-22 A            & GOS 080.14$+$00.75\_01 & 20:33:08.800 & $+$41:13:18.50 & O3    & I       & f*          & \nodata    & \nodata                                   & \nodata & \ref{sec:Nor}   & \nodata \\
Cyg OB2-22 B            & GOS 080.14$+$00.75\_02 & 20:33:08.876 & $+$41:13:17.22 & O6    & V       & ((f))       & \nodata    & \nodata                                   & \nodata & \ref{sec:Nor}   & \nodata \\
Cyg OB2-9               & GOS 080.17$+$00.76\_01 & 20:33:10.734 & $+$41:15:08.25 & O4.5  & I       & fc          & \nodata    & \nodata                                   & \nodata & \ref{sec:Ofc}   & ch      \\
NSV 13\,148             & GOS 080.21$+$00.76\_01 & 20:33:17.480 & $+$41:17:09.30 & O8    & V       & (n)         & \nodata    & \nodata                                   & \nodata & \ref{sec:Nor}   & new     \\
Cyg OB2-8 A             & GOS 080.22$+$00.79\_01 & 20:33:15.078 & $+$41:18:50.51 & O5    & III     & (fc)        & \nodata    & O6 + O5.5; see note                       & D04     & \ref{sec:Ofc}   & ch      \\
Cyg OB2-8 B             & GOS 080.22$+$00.79\_02 & 20:33:14.756 & $+$41:18:41.79 & O6    & II      & (f)         & \nodata    & \nodata                                   & \nodata & \ref{sec:Nor}   & new     \\
Cyg OB2-4               & GOS 080.22$+$01.02\_01 & 20:32:13.823 & $+$41:27:11.99 & O7    & III     & ((f))       & \nodata    & \nodata                                   & \nodata & \ref{sec:Nor}   & \nodata \\
Cyg OB2-8 C             & GOS 080.23$+$00.78\_01 & 20:33:17.977 & $+$41:18:31.19 & O4.5  & III     & (fc)        & \nodata    & \nodata                                   & \nodata & \ref{sec:Ofc}   & ch      \\
Cyg OB2-8 D             & GOS 080.23$+$00.79\_01 & 20:33:16.340 & $+$41:19:01.80 & O9    & V       & (n)         & \nodata    & \nodata                                   & \nodata & \ref{sec:Nor}   & new     \\
Cyg OB2-7               & GOS 080.24$+$00.80\_01 & 20:33:14.112 & $+$41:20:21.88 & O3    & I       & f*          & \nodata    & \nodata                                   & \nodata & \ref{sec:Nor}   & \nodata \\
Cyg OB2-11              & GOS 080.57$+$00.83\_01 & 20:34:08.514 & $+$41:36:59.42 & O5.5  & I       & fc          & \nodata    & \nodata                                   & \nodata & \ref{sec:Ofc}   & ch      \\
HD 188\,209             & GOS 080.99$+$10.09\_01 & 19:51:59.068 & $+$47:01:38.44 & O9.5  & Iab     & \nodata     & \nodata    & \nodata                                   & \nodata & \ref{sec:Nor}   & \nodata \\
HD 191\,781             & GOS 081.18$+$06.61\_01 & 20:09:50.581 & $+$45:24:10.44 & ON9.7 & Iab     & \nodata     & \nodata    & \nodata                                   & \nodata & \ref{sec:ON/OC} & \nodata \\
HD 195\,592             & GOS 082.36$+$02.96\_01 & 20:30:34.970 & $+$44:18:54.87 & O9.7  & Ia      & \nodata     & \nodata    & O9.7 I + B                                & D10     & \ref{sec:SB2}   & \nodata \\
HD 199\,579             & GOS 085.70$-$00.30\_01 & 20:56:34.779 & $+$44:55:29.01 & O6.5  & V       & ((f))z      & \nodata    & O6 V((f)) + B1-2 V                        & W01     & \ref{sec:SB2}   & ch      \\
HD 202\,124             & GOS 087.29$-$02.66\_01 & 21:12:28.389 & $+$44:31:54.14 & O9    & Iab     & \nodata     & \nodata    & \nodata                                   & \nodata & \ref{sec:Nor}   & ch      \\
68 Cyg                  & GOS 087.61$-$03.84\_01 & 21:18:27.187 & $+$43:56:45.40 & O7.5  & III     & n((f))      & \nodata    & \nodata                                   & \nodata & \ref{sec:Nor}   & ch      \\
10 Lac                  & GOS 096.65$-$16.98\_01 & 22:39:15.679 & $+$39:03:01.01 & O9    & V       & \nodata     & \nodata    & \nodata                                   & \nodata & \ref{sec:Nor}   & \nodata \\
HD 206\,183             & GOS 098.89$+$03.40\_01 & 21:38:26.284 & $+$56:58:25.45 & O9.5  & IV-V    & \nodata     & \nodata    & \nodata                                   & \nodata & \ref{sec:Nor}   & new     \\
HD 204\,827 AaAb        & GOS 099.17$+$05.55\_01 & 21:28:57.763 & $+$58:44:23.20 & O9.7  & III     & \nodata     & \nodata    & \nodata                                   & \nodata & \ref{sec:Nor}   & new     \\
HD 206\,267 AaAb        & GOS 099.29$+$03.74\_01 & 21:38:57.618 & $+$57:29:20.55 & O6.5  & V       & ((f))       & O9/B0 V    & \nodata                                   & \nodata & \ref{sec:SB2}   & ch      \\
HD 210\,809             & GOS 099.85$-$03.13\_01 & 22:11:38.601 & $+$52:25:47.95 & O9    & Iab     & \nodata     & \nodata    & \nodata                                   & \nodata & \ref{sec:Nor}   & \nodata \\
HD 207\,538             & GOS 101.60$+$04.67\_01 & 21:47:39.790 & $+$59:42:01.35 & O9.7  & IV      & \nodata     & \nodata    & \nodata                                   & \nodata & \ref{sec:Nor}   & new     \\
LZ Cep                  & GOS 102.01$+$02.18\_01 & 22:02:04.576 & $+$58:00:01.33 & O9    & IV      & (n) var     & B1: V:     & \nodata                                   & \nodata & \ref{sec:SB2}   & ch      \\
HD 207\,198             & GOS 103.14$+$06.99\_01 & 21:44:53.278 & $+$62:27:38.05 & O9    & II      & \nodata     & \nodata    & \nodata                                   & \nodata & \ref{sec:Nor}   & \nodata \\
$\lambda$ Cep           & GOS 103.83$+$02.61\_01 & 22:11:30.584 & $+$59:24:52.25 & O6.5  & I       & (n)fp       & \nodata    & \nodata                                   & \nodata & \ref{sec:Onfp}  & ch      \\
19 Cep                  & GOS 104.87$+$05.39\_01 & 22:05:08.791 & $+$62:16:47.35 & O9    & Ib      & \nodata     & \nodata    & \nodata                                   & \nodata & \ref{sec:Nor}   & ch      \\
DH Cep                  & GOS 107.07$-$00.90\_01 & 22:46:54.111 & $+$58:05:03.55 & O5    & V       & ((f))       & O6 V ((f)) & \nodata                                   & \nodata & \ref{sec:SB2}   & ch      \\
HD 218\,915             & GOS 108.06$-$06.89\_01 & 23:11:06.948 & $+$53:03:29.64 & O9.5  & Iab     & \nodata     & \nodata    & \nodata                                   & \nodata & \ref{sec:Nor}   & \nodata \\
HD 218\,195 A           & GOS 109.32$-$01.79\_01 & 23:05:12.928 & $+$58:14:29.34 & O8.5  & III     & \nodata     & \nodata    & \nodata                                   & \nodata & \ref{sec:Nor}   & ch      \\
HD 216\,532             & GOS 109.65$+$02.68\_01 & 22:52:30.555 & $+$62:26:25.92 & O8.5  & V       & (n)         & \nodata    & \nodata                                   & \nodata & \ref{sec:Nor}   & ch      \\
HD 216\,898             & GOS 109.93$+$02.39\_01 & 22:55:42.460 & $+$62:18:22.83 & O9    & V       & \nodata     & \nodata    & \nodata                                   & \nodata & \ref{sec:Nor}   & ch      \\
HD 217\,086             & GOS 110.22$+$02.72\_01 & 22:56:47.194 & $+$62:43:37.60 & O7    & V       & nn((f))     & \nodata    & \nodata                                   & \nodata & \ref{sec:Nor}   & ch      \\
BD +60 2522             & GOS 112.23$+$00.22\_01 & 23:20:44.519 & $+$61:11:40.53 & O6.5  & \nodata & (n)fp       & \nodata    & \nodata                                   & \nodata & \ref{sec:Onfp}  & ch      \\
HD 225\,146             & GOS 117.23$-$01.24\_01 & 00:03:57.504 & $+$61:06:13.07 & O9.7  & Iab     & \nodata     & \nodata    & \nodata                                   & \nodata & \ref{sec:Nor}   & ch      \\
HD 225\,160             & GOS 117.44$-$00.14\_01 & 00:04:03.796 & $+$62:13:18.99 & O8    & Iab     & f           & \nodata    & \nodata                                   & \nodata & \ref{sec:Nor}   & ch      \\
AO Cas                  & GOS 117.59$-$11.09\_01 & 00:17:43.059 & $+$51:25:59.12 & O9.5  & II      & (n)         & O8 V       & \nodata                                   & \nodata & \ref{sec:Onfp}  & ch      \\
HD 108                  & GOS 117.93$+$01.25\_01 & 00:06:03.386 & $+$63:40:46.75 & O8    & \nodata & fp var      & \nodata    & \nodata                                   & \nodata & \ref{sec:Of?p}  & ch      \\
HD 5005\,A              & GOS 123.12$-$06.24\_01 & 00:52:49.199 & $+$56:37:39.59 & O4    & V       & ((fc))      & \nodata    & \nodata                                   & \nodata & \ref{sec:Ofc}   & ch      \\
HD 5005\,C              & GOS 123.12$-$06.24\_02 & 00:52:49.541 & $+$56:37:36.91 & O8.5  & V       & (n)         & \nodata    & \nodata                                   & \nodata & \ref{sec:Nor}   & ch      \\
HD 5005\,B              & GOS 123.12$-$06.24\_03 & 00:52:49.383 & $+$56:37:39.80 & O9.7  & II-III  & \nodata     & \nodata    & \nodata                                   & \nodata & \ref{sec:Nor}   & new     \\
HD 5005\,D              & GOS 123.12$-$06.25\_01 & 00:52:48.940 & $+$56:37:30.92 & O9.5  & V       & \nodata     & \nodata    & \nodata                                   & \nodata & \ref{sec:Nor}   & new     \\
BD +60 261              & GOS 127.87$-$01.35\_01 & 01:32:32.720 & $+$61:07:45.84 & O7.5  & III     & (n)((f))    & \nodata    & \nodata                                   & \nodata & \ref{sec:Nor}   & \nodata \\
HD 10\,125              & GOS 128.29$+$01.82\_01 & 01:40:52.762 & $+$64:10:23.13 & O9.7  & II      & \nodata     & \nodata    & \nodata                                   & \nodata & \ref{sec:Nor}   & \nodata \\
HD 13\,022              & GOS 132.91$-$02.57\_01 & 02:09:30.067 & $+$58:47:01.58 & O9.7  & II-III  & \nodata     & \nodata    & \nodata                                   & \nodata & \ref{sec:Nor}   & ch      \\
HD 12\,323              & GOS 132.91$-$05.87\_01 & 02:02:30.126 & $+$55:37:26.38 & ON9.5 & V       & \nodata     & \nodata    & \nodata                                   & \nodata & \ref{sec:ON/OC} & ch      \\
HD 12\,993              & GOS 133.11$-$03.40\_01 & 02:09:02.473 & $+$57:55:55.93 & O6.5  & V       & ((f))z      & \nodata    & \nodata                                   & \nodata & \ref{sec:Nor}   & ch      \\
HD 13\,268              & GOS 133.96$-$04.99\_01 & 02:11:29.700 & $+$56:09:31.70 & ON8.5 & III     & n           & \nodata    & \nodata                                   & \nodata & \ref{sec:ON/OC} & new     \\
HD 14\,442              & GOS 134.21$-$01.32\_01 & 02:22:10.701 & $+$59:32:58.92 & O5    & \nodata & n(f)p       & \nodata    & \nodata                                   & \nodata & \ref{sec:Onfp}  & \nodata \\
BD +62 424              & GOS 134.53$+$02.46\_01 & 02:36:18.221 & $+$62:56:53.35 & O6.5  & V       & (n)((f))    & \nodata    & \nodata                                   & \nodata & \ref{sec:Nor}   & ch      \\
V354 Per                & GOS 134.58$-$04.96\_01 & 02:15:45.938 & $+$55:59:46.73 & O9.7  & II      & (n)         & \nodata    & \nodata                                   & \nodata & \ref{sec:Nor}   & ch      \\
BD +60 497              & GOS 134.58$+$01.04\_01 & 02:31:57.087 & $+$61:36:43.95 & O6.5  & V       & ((f))       & O8/B0 V    & \nodata                                   & \nodata & \ref{sec:SB2}   & ch      \\
BD +60 498              & GOS 134.63$+$00.99\_01 & 02:32:10.855 & $+$61:33:07.95 & O9.7  & II-III  & \nodata     & \nodata    & \nodata                                   & \nodata & \ref{sec:Nor}   & new     \\
BD +60 499              & GOS 134.64$+$01.00\_01 & 02:32:16.752 & $+$61:33:15.07 & O9.5  & V       & \nodata     & \nodata    & \nodata                                   & \nodata & \ref{sec:Nor}   & \nodata \\
BD +60 501              & GOS 134.71$+$00.94\_01 & 02:32:36.272 & $+$61:28:25.60 & O7    & V       & (n)((f))z   & \nodata    & \nodata                                   & \nodata & \ref{sec:Nor}   & ch      \\
HD 15\,558 A            & GOS 134.72$+$00.92\_01 & 02:32:42.536 & $+$61:27:21.56 & O4.5  & III     & (fc)        & \nodata    & O5.5 III(f) + O7 V                        & D06     & \ref{sec:Ofc}   & ch      \\
HD 15\,570              & GOS 134.77$+$00.86\_01 & 02:32:49.422 & $+$61:22:42.07 & O4    & I       & f           & \nodata    & \nodata                                   & \nodata & \ref{sec:Nor}   & ch      \\
HD 15\,629              & GOS 134.77$+$01.01\_01 & 02:33:20.586 & $+$61:31:18.18 & O4.5  & V       & ((fc))      & \nodata    & \nodata                                   & \nodata & \ref{sec:Ofc}   & ch      \\
BD +60 513              & GOS 134.90$+$00.92\_01 & 02:34:02.530 & $+$61:23:10.87 & O7    & V       & n           & \nodata    & \nodata                                   & \nodata & \ref{sec:Nor}   & \nodata \\
HD 14\,947              & GOS 134.99$-$01.74\_01 & 02:26:46.992 & $+$58:52:33.11 & O4.5  & I       & f           & \nodata    & \nodata                                   & \nodata & \ref{sec:Nor}   & ch      \\
HD 14\,434              & GOS 135.08$-$03.82\_01 & 02:21:52.413 & $+$56:54:18.03 & O5.5  & V       & nn((f))p    & \nodata    & \nodata                                   & \nodata & \ref{sec:Onfp}  & ch      \\
HD 16\,429 A            & GOS 135.68$+$01.15\_01 & 02:40:44.951 & $+$61:16:56.04 & O9    & II-III  & (n)         & \nodata    & O9.5 II + O8 III-IV + B0 V?               & M03     & \ref{sec:SB2}   & ch      \\
HD 15\,642              & GOS 137.09$-$04.73\_01 & 02:32:56.383 & $+$55:19:39.07 & O9.5  & II-III  & n           & \nodata    & \nodata                                   & \nodata & \ref{sec:Nor}   & ch      \\
HD 18\,409              & GOS 137.12$+$03.46\_01 & 03:00:29.719 & $+$62:43:19.05 & O9.7  & Ib      & \nodata     & \nodata    & \nodata                                   & \nodata & \ref{sec:Nor}   & \nodata \\
HD 17\,505 A            & GOS 137.19$+$00.90\_01 & 02:51:07.971 & $+$60:25:03.88 & O6.5  & III     & n((f))      & \nodata    & O6.5 III((f)) + O7.5 V((f)) + O7.5 V((f)) & H06     & \ref{sec:SB2}   & ch      \\
HD 17\,505 B            & GOS 137.19$+$00.90\_02 & 02:51:08.263 & $+$60:25:03.78 & O8    & V       & \nodata     & \nodata    & \nodata                                   & \nodata & \ref{sec:Nor}   & new     \\
HD 17\,520 A            & GOS 137.22$+$00.88\_01 & 02:51:14.434 & $+$60:23:09.97 & O8    & V       & \nodata     & \nodata    & \nodata                                   & \nodata & \ref{sec:Nor}   & ch      \\
HD 17\,520 B            & GOS 137.22$+$00.88\_02 & 02:51:14.397 & $+$60:23:10.12 & O9:   & V       & e           & \nodata    & \nodata                                   & \nodata & \ref{sec:Oe}    & new     \\
BD +60 586 A            & GOS 137.42$+$01.28\_01 & 02:54:10.672 & $+$60:39:03.59 & O7    & V       & z           & \nodata    & \nodata                                   & \nodata & \ref{sec:Nor}   & ch      \\
HD 15\,137              & GOS 137.46$-$07.58\_01 & 02:27:59.811 & $+$52:32:57.60 & O9.5  & II-III  & n           & \nodata    & \nodata                                   & \nodata & \ref{sec:Nor}   & ch      \\
HD 16\,691              & GOS 137.73$-$02.73\_01 & 02:42:52.028 & $+$56:54:16.45 & O4    & I       & f           & \nodata    & \nodata                                   & \nodata & \ref{sec:Nor}   & ch      \\
HD 16\,832              & GOS 138.00$-$02.88\_01 & 02:44:12.717 & $+$56:39:27.23 & O9.5  & II-III  & \nodata     & \nodata    & \nodata                                   & \nodata & \ref{sec:Nor}   & ch      \\
HD 18\,326              & GOS 138.03$+$01.50\_01 & 02:59:23.171 & $+$60:33:59.50 & O6.5  & V       & (n)((f))    & O9/B0 V:   & \nodata                                   & \nodata & \ref{sec:SB2}   & ch      \\
HD 17\,603              & GOS 138.77$-$02.08\_01 & 02:51:47.798 & $+$57:02:54.46 & O7.5  & Ib      & (f)         & \nodata    & \nodata                                   & \nodata & \ref{sec:Nor}   & \nodata \\
CC Cas                  & GOS 140.12$+$01.54\_01 & 03:14:05.333 & $+$59:33:48.50 & O8.5  & III     & (n)((f))    & \nodata    & O8.5 III + B0 V                           & H94     & \ref{sec:SB2}   & ch      \\
HD 14\,633              & GOS 140.78$-$18.20\_01 & 02:22:54.293 & $+$41:28:47.72 & ON8.5 & V       & \nodata     & \nodata    & \nodata                                   & \nodata & \ref{sec:ON/OC} & ch      \\
$\alpha$ Cam            & GOS 144.07$+$14.04\_01 & 04:54:03.011 & $+$66:20:33.58 & O9    & Ia      & \nodata     & \nodata    & \nodata                                   & \nodata & \ref{sec:Nor}   & ch      \\
HDE 237\,211            & GOS 147.14$+$02.97\_01 & 04:03:15.652 & $+$56:32:24.85 & O9    & Ib      & \nodata     & \nodata    & \nodata                                   & \nodata & \ref{sec:Nor}   & ch      \\
HD 24\,431              & GOS 148.84$-$00.71\_01 & 03:55:38.420 & $+$52:38:28.75 & O9    & III     & \nodata     & \nodata    & \nodata                                   & \nodata & \ref{sec:Nor}   & \nodata \\
NGC 1624-2              & GOS 155.36$+$02.61\_01 & 04:40:37.266 & $+$50:27:40.96 & O7    & \nodata & f?p         & \nodata    & \nodata                                   & \nodata & \ref{sec:Of?p}  & new     \\
$\xi$ Per               & GOS 160.37$-$13.11\_01 & 03:58:57.900 & $+$35:47:27.72 & O7.5  & III     & (n)((f))    & \nodata    & \nodata                                   & \nodata & \ref{sec:Nor}   & \nodata \\
X Per                   & GOS 163.08$-$17.14\_01 & 03:55:23.078 & $+$31:02:45.04 & O9.5: & \nodata & npe         & \nodata    & \nodata                                   & \nodata & \ref{sec:Oe}    & ch      \\
HD 41\,161              & GOS 164.97$+$12.89\_01 & 06:05:52.456 & $+$48:14:57.41 & O8    & V       & n           & \nodata    & \nodata                                   & \nodata & \ref{sec:Nor}   & \nodata \\
BD +39 1328             & GOS 169.11$+$03.60\_01 & 05:32:13.845 & $+$40:03:57.88 & O8.5  & Iab     & (n)(f)      & \nodata    & \nodata                                   & \nodata & \ref{sec:Nor}   & ch      \\
HD 34\,656              & GOS 170.04$+$00.27\_01 & 05:20:43.080 & $+$37:26:19.23 & O7.5  & II      & (f)         & \nodata    & \nodata                                   & \nodata & \ref{sec:Nor}   & ch      \\
AE Aur                  & GOS 172.08$-$02.26\_01 & 05:16:18.149 & $+$34:18:44.34 & O9.5  & V       & \nodata     & \nodata    & \nodata                                   & \nodata & \ref{sec:Nor}   & \nodata \\
HD 36\,483              & GOS 172.29$+$01.88\_01 & 05:33:41.154 & $+$36:27:34.97 & O9.5  & IV      & (n)         & \nodata    & \nodata                                   & \nodata & \ref{sec:Nor}   & ch      \\
LY Aur A                & GOS 172.76$+$00.61\_01 & 05:29:42.647 & $+$35:22:30.07 & O9.5  & II      & \nodata     & O9 III     & \nodata                                   & \nodata & \ref{sec:SB2}   & ch      \\
HD 35\,619              & GOS 173.04$-$00.09\_01 & 05:27:36.146 & $+$34:45:18.97 & O7.5  & V       & ((f))       & \nodata    & \nodata                                   & \nodata & \ref{sec:Nor}   & ch      \\
HD 37\,737              & GOS 173.46$+$03.24\_01 & 05:42:31.160 & $+$36:12:00.50 & O9.5  & II-III  & (n)         & \nodata    & \nodata                                   & \nodata & \ref{sec:Nor}   & ch      \\
HDE 242\,908            & GOS 173.47$-$01.66\_01 & 05:22:29.302 & $+$33:30:50.43 & O4.5  & V       & (n)((fc))   & \nodata    & \nodata                                   & \nodata & \ref{sec:Ofc}   & ch      \\
BD +33 1025             & GOS 173.56$-$01.66\_01 & 05:22:44.001 & $+$33:26:26.65 & O7    & V       & (n)z        & \nodata    & \nodata                                   & \nodata & \ref{sec:Nor}   & new     \\
HDE 242\,935 A          & GOS 173.58$-$01.67\_01 & 05:22:46.542 & $+$33:25:11.51 & O6.5  & V       & ((f))z      & \nodata    & \nodata                                   & \nodata & \ref{sec:Nor}   & ch      \\
HDE 242\,926            & GOS 173.65$-$01.74\_01 & 05:22:40.099 & $+$33:19:09.37 & O7    & V       & z           & \nodata    & \nodata                                   & \nodata & \ref{sec:Nor}   & ch      \\
HD 37\,366 A            & GOS 177.63$-$00.11\_01 & 05:39:24.799 & $+$30:53:26.75 & O9.5  & IV      & \nodata     & \nodata    & O9.5 V + B0-1 V                           & B07     & \ref{sec:SB2}   & ch      \\
HD 93\,521              & GOS 183.14$+$62.15\_01 & 10:48:23.511 & $+$37:34:13.09 & O9.5  & III     & nn          & \nodata    & \nodata                                   & \nodata & \ref{sec:Nor}   & ch      \\
HD 36\,879              & GOS 185.22$-$05.89\_01 & 05:35:40.527 & $+$21:24:11.72 & O7    & V       & (n)((f))    & \nodata    & \nodata                                   & \nodata & \ref{sec:Nor}   & ch      \\
HD 42\,088              & GOS 190.04$+$00.48\_01 & 06:09:39.574 & $+$20:29:15.46 & O6    & V       & ((f))z      & \nodata    & \nodata                                   & \nodata & \ref{sec:Nor}   & ch      \\
HD 44\,811              & GOS 192.40$+$03.21\_01 & 06:24:38.354 & $+$19:42:15.83 & O7    & V       & (n)z        & \nodata    & \nodata                                   & \nodata & \ref{sec:Nor}   & ch      \\
V1382 Ori               & GOS 194.07$-$05.88\_01 & 05:54:44.731 & $+$13:51:17.06 & O6    & V:      & [n]pe var   & \nodata    & \nodata                                   & \nodata & \ref{sec:Oe}    & \nodata \\
HD 41\,997              & GOS 194.15$-$01.98\_01 & 06:08:55.821 & $+$15:42:18.18 & O7.5  & V       & n((f))      & \nodata    & \nodata                                   & \nodata & \ref{sec:Nor}   & ch      \\
$\lambda$ Ori A         & GOS 195.05$-$12.00\_01 & 05:35:08.277 & $+$09:56:02.96 & O8    & III     & ((f))       & \nodata    & \nodata                                   & \nodata & \ref{sec:Nor}   & \nodata \\
HD 45\,314              & GOS 196.96$+$01.52\_01 & 06:27:15.777 & $+$14:53:21.22 & O9:   & \nodata & npe         & \nodata    & \nodata                                   & \nodata & \ref{sec:Oe}    & ch      \\
HD 60\,848              & GOS 202.51$+$17.52\_01 & 07:37:05.731 & $+$16:54:15.29 & O8:   & V:      & pe          & \nodata    & \nodata                                   & \nodata & \ref{sec:Oe}    & ch      \\
15 Mon AaAb             & GOS 202.94$+$02.20\_01 & 06:40:58.656 & $+$09:53:44.71 & O7    & V       & ((f)) var   & \nodata    & \nodata                                   & \nodata & \ref{sec:Nor}   & ch      \\
$\delta$ Ori AaAb       & GOS 203.86$-$17.74\_01 & 05:32:00.401 & $-$00:17:56.73 & O9.5  & II      & Nwk         & \nodata    & O9.5 II + B0.5 III                        & H02     & \ref{sec:ON/OC} & ch      \\
HD 46\,966              & GOS 205.81$-$00.55\_01 & 06:36:25.887 & $+$06:04:59.47 & O8.5  & IV      & \nodata     & \nodata    & \nodata                                   & \nodata & \ref{sec:Nor}   & ch      \\
HD 47\,129              & GOS 205.87$-$00.31\_01 & 06:37:24.042 & $+$06:08:07.38 & O8    & \nodata & fp var      & \nodata    & O8 III/I + O7.5 III                       & L08     & \ref{sec:Onfp}  & ch      \\
HD 46\,106              & GOS 206.20$-$02.09\_01 & 06:31:38.395 & $+$05:01:36.38 & O9.7  & II-III  & \nodata     & \nodata    & \nodata                                   & \nodata & \ref{sec:Nor}   & new     \\
HD 48\,099              & GOS 206.21$+$00.80\_01 & 06:41:59.231 & $+$06:20:43.54 & O6.5  & V       & (n)((f))    & \nodata    & O5.5 V ((f)) + O9 V                       & M10     & \ref{sec:SB2}   & ch      \\
HD 46\,149              & GOS 206.22$-$02.04\_01 & 06:31:52.533 & $+$05:01:59.19 & O8.5  & V       & \nodata     & \nodata    & O8 V + B0-1 V                             & M09     & \ref{sec:SB2}   & \nodata \\
HD 46\,202              & GOS 206.31$-$02.00\_01 & 06:32:10.471 & $+$04:57:59.79 & O9.5  & V       & \nodata     & \nodata    & \nodata                                   & \nodata & \ref{sec:Nor}   & ch      \\
HD 46\,150              & GOS 206.31$-$02.07\_01 & 06:31:55.519 & $+$04:56:34.27 & O5    & V       & ((f))z      & \nodata    & \nodata                                   & \nodata & \ref{sec:Nor}   & ch      \\
HD 46\,056 A            & GOS 206.34$-$02.25\_01 & 06:31:20.862 & $+$04:50:03.85 & O8    & V       & n           & \nodata    & \nodata                                   & \nodata & \ref{sec:Nor}   & ch      \\
HD 46\,223              & GOS 206.44$-$02.07\_01 & 06:32:09.306 & $+$04:49:24.73 & O4    & V       & ((f))       & \nodata    & \nodata                                   & \nodata & \ref{sec:Nor}   & ch      \\
$\zeta$ Ori A           & GOS 206.45$-$16.59\_01 & 05:40:45.527 & $-$01:56:33.26 & O9.5  & Ib      & Nwk var     & \nodata    & \nodata                                   & \nodata & \ref{sec:ON/OC} & ch      \\
$\zeta$ Ori B           & GOS 206.45$-$16.59\_02 & 05:40:45.571 & $-$01:56:35.59 & O9.5  & II-III  & (n)         & \nodata    & \nodata                                   & \nodata & \ref{sec:Nor}   & new     \\
$\sigma$ Ori AB         & GOS 206.82$-$17.34\_01 & 05:38:44.768 & $-$02:36:00.25 & O9.7  & III     & \nodata     & \nodata    & \nodata                                   & \nodata & \ref{sec:Nor}   & ch      \\
HD 46\,485              & GOS 206.90$-$01.84\_01 & 06:33:50.957 & $+$04:31:31.61 & O7    & V       & n           & \nodata    & \nodata                                   & \nodata & \ref{sec:Nor}   & ch      \\
HD 46\,573              & GOS 208.73$-$02.63\_01 & 06:34:23.568 & $+$02:32:02.94 & O7    & V       & ((f))z      & \nodata    & \nodata                                   & \nodata & \ref{sec:Nor}   & ch      \\
$\theta$$^{1}$ Ori CaCb & GOS 209.01$-$19.38\_01 & 05:35:16.463 & $-$05:23:23.18 & O7    & V       & p           & \nodata    & \nodata                                   & \nodata & \ref{sec:Nor}   & \nodata \\
$\theta$$^{2}$ Ori A    & GOS 209.05$-$19.37\_01 & 05:35:22.900 & $-$05:24:57.79 & O9.5  & IV      & p           & \nodata    & \nodata                                   & \nodata & \ref{sec:Nor}   & ch      \\
$\iota$ Ori             & GOS 209.52$-$19.58\_01 & 05:35:25.981 & $-$05:54:35.64 & O9    & III     & var         & \nodata    & O9 III + B1 III                           & S87     & \ref{sec:SB2}   & ch      \\
V689 Mon                & GOS 210.03$-$02.11\_01 & 06:38:38.187 & $+$01:36:48.66 & O9.7  & Ib      & \nodata     & \nodata    & \nodata                                   & \nodata & \ref{sec:Nor}   & \nodata \\
HD 48\,279 A            & GOS 210.41$-$01.17\_01 & 06:42:40.548 & $+$01:42:58.23 & O8.5  & V       & Nstr var?   & \nodata    & \nodata                                   & \nodata & \ref{sec:ON/OC} & ch      \\
$\upsilon$ Ori          & GOS 210.44$-$20.99\_01 & 05:31:55.860 & $-$07:18:05.53 & O9.7  & V       & \nodata     & \nodata    & \nodata                                   & \nodata & \ref{sec:Nor}   & new     \\
HD 52\,533 A            & GOS 216.85$+$00.80\_01 & 07:01:27.048 & $-$03:07:03.28 & O8.5  & IV      & n           & \nodata    & \nodata                                   & \nodata & \ref{sec:Nor}   & ch      \\
HD 52\,266              & GOS 219.13$-$00.68\_01 & 07:00:21.077 & $-$05:49:35.95 & O9.5  & III     & n           & \nodata    & \nodata                                   & \nodata & \ref{sec:Nor}   & ch      \\
HD 54\,662              & GOS 224.17$-$00.78\_01 & 07:09:20.249 & $-$10:20:47.64 & O7    & V       & ((f))z var? & \nodata    & O6.5 V + O7-9.5 V                         & B07     & \ref{sec:SB2}   & ch      \\
HD 57\,682              & GOS 224.41$+$02.63\_01 & 07:22:02.053 & $-$08:58:45.77 & O9.5  & IV      & \nodata     & \nodata    & \nodata                                   & \nodata & \ref{sec:Nor}   & ch      \\
HD 55\,879              & GOS 224.73$+$00.35\_01 & 07:14:28.253 & $-$10:18:58.50 & O9.7  & III     & \nodata     & \nodata    & \nodata                                   & \nodata & \ref{sec:Nor}   & ch      \\
HD 54\,879              & GOS 225.55$-$01.28\_01 & 07:10:08.149 & $-$11:48:09.86 & O9.7  & V       & \nodata     & \nodata    & \nodata                                   & \nodata & \ref{sec:Nor}   & ch      \\
HD 53\,975              & GOS 225.68$-$02.32\_01 & 07:06:35.964 & $-$12:23:38.23 & O7.5  & V       & z           & \nodata    & O7.5 V + B2-3 V                           & G94     & \ref{sec:SB2}   & ch      \\
\enddata
\tablecomments{{\it GOSSS ID} is the identification for each star with ``GOS'' standing for ``Galactic O Star''. {\it Ref.} is the reference for the alternative classification. {\it Sect.} is the section where the star is discussed. {\it Flag} can be either ``ch'' (O-type classification change from \citealt{Maizetal04b}) or ``new'' (star not present or not O type in \citealt{Maizetal04b}). At the original resolution of our CAHA spectra, Cyg OB2-8 A appears as a SB2 with spectral types O5.5 III (fc) + O5.5 III (fc).}
\tablerefs{B07: \citet{Boyaetal07a}, D04: \citet{DeBeetal04}, D06: \citet{DeBeetal06a}, D10: \citet{DeBeetal10}, G94: \citet{Giesetal94}, H94: \citet{Hilletal94}, H02: \citet{Harvetal02}, H06: \citet{Hilletal06}, L87: \citet{Leitetal87}, L08: \citet{Lindetal08}, M03: \citet{McSw03}, M09: \citet{Mahyetal09}, M10: \citet{Mahyetal10}, S87: \citet{Sticetal87}, S09: \citet{Sanaetal09}, W73: \citet{Walb73a}, W01: \citet{Willetal01}.}
\label{spectralclas}
\end{deluxetable}